\PassOptionsToPackage{usenames,dvipsnames}{xcolor}
\documentclass[10pt]{article}
\usepackage{geometry}
\usepackage[english]{babel}
\usepackage[latin1]{inputenc}
\usepackage{natbib}
\usepackage{enumerate}
\usepackage[dvipsnames,usenames]{xcolor}
\usepackage{pstricks}
\usepackage{graphicx}
\usepackage{amsmath}
\usepackage{amsfonts}
\usepackage{amssymb}
\usepackage{float}
\usepackage{cancel}
\usepackage{placeins}

\title{Does rare, noise-induced, bypass transition in plane Couette flow bypass instantons ?}
\author{Joran Rolland\thanks{Univ. Lille, CNRS, ONERA, Arts et M\'etiers Institute of Technology, Centrale Lille,
UMR 9014 - LMFL - Laboratoire de M\'ecanique des Fluides de Lille - Kamp\'e de F\'eriet, F-59000 Lille, France.}
\thanks{joran.rolland@centralelille.fr}}
\date{\today}

\begin{document}

\maketitle

\begin{abstract}
This text presents the study of rare noise induced transitions
from stable laminar flow to transitional turbulence
in plane Couette flow, that we will term \emph{build up}.
We wish to study forced paths that go all the way from laminar to turbulent flow
and to focus the investigation on whether these paths share the properties of noise induced transitions in simpler systems.
The forcing noise has a red spectrum without any component in the natural, large scale, linear receptivity range of the flow.
As we decreased the forcing energy injection rate, the transitions became rare.
The rare paths from laminar to turbulent flow are computed using Adaptive Multilevel Splitting (AMS), a rare event simulation method,
and are validated against Direct Numerical Simulations (DNS) at moderately small energy injection rate.
On the computed trajectories, the flow manages to non-linearly redistribute energy from the small
forced scales to the unforced large scales so that the reactive trajectories
display forced streamwise velocity tubes at the natural scale of
velocity streaks. As the trajectory proceeds, these tubes gradually grow in amplitude until
they cross the separatrix between laminar and turbulent flow.
Streamwise vortices manifest themselves only after velocity tubes have reached near turbulent amplitude,
displaying a two stage process reminiscent of the ``backward'' path from turbulence to laminar flow.
We checked that these were not time reversed turbulence collapse paths.
As the domain size is increased from a Minimal Flow Unit (MFU) type flow at $L_x\times L_z=6\times 4$
(in half gap units) to a large domain $L_x\times L_z=36\times 24$,
spatial localisation then extension of the generated coherent streaks and vortices in the spanwise direction is observed in the reactive paths.
The paths systematically computed in MFU display many of the characteristics of instantons,
that often structure noise induced transitions: such as concentration of trajectories,
exponentially increasing waiting times before transition, and Gumbel distribution of trajectory durations.
However, bisections started from successive states on the reactive trajectories
indicate that for all sizes and energy injection rates investigated, the trajectory
lack two key ingredients of instantons. Firstly, they do not visit the neighbourhood of the nearest saddle point
and do not display the natural relaxation path from that saddle to transitional wall turbulence.
This discrepancy is observed for all system sizes.
Secondly, the reactive paths do not concentrate more and more around the same trajectory as energy injection rate is decreased, but instead
gradually move in phase space.
They might reconnect with instantons at very small energy injection rate and exceedingly long waiting times.
They would explain why classical instanton calculations have proved to be tremendously difficult in wall flows.
\end{abstract}

\section{Introduction}

\subsection{Bistability in physics and its description by statistical physics}\label{intstat}

Many physical systems, after they underwent a subcritical bifurcation or a first order transition (or a similar transition in more complex systems),
 display two or more metastable states.
This situation means that the physical system can switch from one metastable state to another if it undergoes some statistically steady forcing,
usually by crossing a boundary of the unforced dynamics separating the basin of attraction of one state from another.
Conversely, this physical system will relax toward either states
if it is prepared with an initial condition on the corresponding side of the separatrix, when no forcing noise is applied.
Traditional bistable systems are found in kinetic chemistry \cite{onsager1938initial,hanggi1990reaction,van2003novel,lopes2019analysis},
where the two states are the reactants and the products or different configurations or conformations \cite{lopes2019analysis} of a complex molecule,
while the force driving transitions is the thermal noise.
Much insight on bistability has been gained from formal noise driven systems
\cite{metzner2006illustration,cerou2011multiple,bouchet2012non,rolland2015statistical,rolland2016computing,lucente2022coupling}.
More recently, a lot of effort has also been directed toward
the study of bistable climate systems
\cite{lucarini2017edge,lucarini2019transitions,baars2021application,herbert2020atmospheric}
and bistable turbulent flows \cite{podvin2017precursor} of aerodynamical \cite{grandemange2013turbulent,kim1988investigation}
or geophysical interest \cite{Berhanu2007,prl_jet}.
In these systems, the metastable states are usually large scale coherent circulations, like jets, or vortices,
while the force driving transition is intrinsic to the system and related to the fluctuating small scales of turbulence.

There are several points of view to describe multistability arising from a deterministic system to which a random forcing
is added.
In this text we describe multistability in flows where turbulence can be sustained even in the absence of forcing: there
are some differences with the case where turbulence is a consequence of said forcing \cite{prl_jet}.
Strictly speaking, the flow to which noise is added becomes a random dynamical system, to which the notions of stable and unstable
attractors of autonomous dynamical systems cannot be applied \cite{arnold1995random}.
What is instead attracting the dynamics is a more complex object that represents the probable random fluctuations around either multistable states,
with excursions from one multistable state to another.
The application of such descriptions to small degrees of freedom systems has already been performed \cite{chekroun2011stochastic}.
However, using such a point of view to describe multistable turbulent flows, with many degrees of freedoms is extremely complex and outside the scope of this text.
Instead, we choose to follow an approach inspired by statistical physics, which has become popular as it is much more tractable.
This requires the variance of the noise added to the deterministic part of the dynamics to be small enough.
In that case, the properties of the unforced original system strongly determine those of the stochastically forced system.
The first of these properties are the attractors of the unforced systems (or what is closest to attractors).
Under the forcing of vanishing variance, the flow spends most of its time fluctuating near these attractors:
the bistable states can then be defined as the narrow regions of phase space within one or two standard deviation of the attractors.

From that point of view of statistical physics, a central object of study for metastable systems is the succession of states visited by the system when it transits from
one bistable state to another.
In kinetic chemistry, this part of the dynamics is termed a \emph{reactive trajectory} \cite{hanggi1990reaction} or a transition path \cite{van2003novel}.
In kinetic chemistry, and more generally, in a wide class of stochastic systems,
as the thermal noise is decreased and the transition becomes rare,
the transition paths tend to concentrate around a specific path termed the instanton.
Again, the phase space of the unforced deterministic system strongly influences the path followed by the instanton.
The instanton generally consists in a fluctuation path under noise from the first attractor of the unforced dynamics toward a specific transition state
which is a saddle of the deterministic part of the dynamics,
followed by a deterministic relaxation path from that saddle to the second attractor \cite{touchette2009large} (Fig.~\ref{skconcl} (a)).
This path is important because it highlights the physical mechanisms of the transition,
while the properties of the transition state
control the parametric dependence of the transition rate \cite{hanggi1990reaction,touchette2009large,bouchet2016generalisation}.
While the instanton always visits the transition state, complex trajectories can be found in non-gradient systems \cite{wan2015model}.
The rarity and the concentration of trajectories is quantitatively
studied in the case where noise variance is a constant times $\epsilon$, a vanishing parameter.
This rarity is quantified by the the transition rate, which is the inverse of the mean first passage time before a transition occurs $T$.
The mean first passage time goes to infinity exponentially with $\epsilon$, as
\begin{equation}
\lim_{\epsilon\rightarrow 0}\epsilon \log(T)=\mathcal{I}(\cancel{\epsilon})\,,
\, T\underset{\epsilon\rightarrow 0}{\asymp}\exp\left(\frac{\mathcal{I}}{\epsilon} \right)\,.\label{exldp}
\end{equation}
We introduced the rate function $\mathcal{I}$. This quantity is finite (the limit is defined), not everywhere zero and is independent on $\epsilon$.
This rate function is directly controlled by the properties of the escaped multistable state and the saddle.
In kinetic chemistry, for relatively low temperatures, this relation was identified as Arrhenius law.
It was then formalised in the context of stochastic gradient systems, systems where the deterministic part of the dynamics is the gradient of some potential $V$,
as Eyring--Kramers theory \cite{hanggi1990reaction}.
In that case $\mathcal{I}$ is the different of potential between the first metastable state and the transition state: the saddle.
Such a formula can also be obtained in non gradient systems \cite{bouchet2016generalisation}.
In that case, these exists a quasipotential that controls part of the dynamics, and $\mathcal{I}$
is the difference of quasipotential between the first metastable state and the saddle.
In general, the asymptotic study of exponential dependences in probabilities or rate of probabilities (Eq.~(\ref{exldp}))
is termed Large Deviations, and encompasses the case of rare reactive trajectories.
In the case of rare paths, this Large Deviation principle is named after Freidlin and Wentzell \cite{freidlin1998random}.
The difficulty of computing the transition path directly from the dynamics, either because of its complexity or
because direct simulation times are too long for molecular dynamics, has led to the development and application
of rare event simulation methods for kinetic chemistry \cite{van2003novel,lopes2019analysis}.
Another approach for these computations consists in exploiting the fact that the instanton minimises the
so-called Onsager--Machlup action \cite{onsager1953fluctuations}
in optimisation procedures \cite{grafke2019numerical,wan2013minimum,wan2015model,mam_pois,borner2023saddle}.
Note that the noise covariance matrix is involved in the action, so that the type of forcing can influence the
fluctuation path and the transition rate.
The transition state, however, is a property of the deterministic part of the dynamics and thus has a degree of universality.

To conclude this section, we note that the addition of a noise makes bistability possible in systems where several stable fixed points can exist.
In the case of low noise variance, statistical physics can greatly help the study of the problem.
Among other things, it tells us that the properties of the noise induced bistability is
strongly influenced  by the phase space of the unforced system: the stable fixed points become the multistable states
and the saddle between them will play a key role in the noise induced transitions.
This means
that it is very relevant to mention them when discussing reactive trajectories.
It can also be very enlightening to present those path on top of the phase space of the unforced system.

\subsection{Towards the use of rare events simulation methods in transitional wall turbulence}

The success of rare event simulation methods in chemistry has encouraged their use
in fluid dynamics. For that purpose, they have been applied to spatially extended systems of gradual complexity,
starting from one dimensional stochastically forced systems \cite{rolland_pre18,baars2021application},
moving to two dimensional stochastically forced turbulence \cite{prl_jet} and recently reaching
three dimensional deterministic transitional turbulence \cite{rolland2022collapse,gome2022extreme}.
The progress has been milestoned in this manner because of the size of the leap
between stochastic differential equation and fully developed turbulent flows.
This means a drastic increase of the number of degrees of freedom (DoF) involved,
a complete change of the force driving the metastability,
and sometimes a strong increase in complexity of the transition state between the two metastable states.
While the extended systems studied were firstly chosen for their physical interest, the order of studies was also
motivated by the need to tackle these changes separately from one another at first.

This was the case of the most recent works, concerning the collapse of turbulence in plane Couette flow \cite{rolland2022collapse,gome2022extreme}.
Plane Couette flow (PCF) is the flow between two parallel moving walls (Fig.~\ref{sksize}, left). In PCF, like in
other wall flows such as plane Poiseuille flow \cite{wan2013minimum,wan2015model,mam_pois}, Couette--Poiseuille flow \cite{liu2021decay},
Hagen--Poiseuille pipe flow \cite{willis2009turbulent} or boundary layers \cite{spangler1968effects,rigas2021nonlinear},
wall turbulence can coexist at Reynolds numbers (or positions in the case of boundary layers)
for which the laminar baseflow is linearly stable (see \cite{romanov1973stability} for the case of PCF).
As a consequence, transitional turbulence can collapse under its own fluctuations,
while the flow can go from laminar to turbulent under a forcing \cite{spangler1968effects,rigas2021nonlinear}
or if it is given an initial condition ``close enough'' to turbulence \cite{schmiegel1997fractal}.
The study of the collapse of turbulence using rare event methods gave the occasion to investigate the physical mechanism of laminar hole opening
as well as the technical solutions to study rare multistability events
in a three dimensional flow not stochastically forced \cite{rolland2022collapse,gome2022extreme}.
Conversely, the study of the forcing of laminar flow toward turbulence, under a noise or another, would
give the occasion to investigate the crossing of a
very complex separatrix \footnote{We use the term separatrix for lack of a better word. strictly speaking,
an unforced wall flow starting from any initial condition eventually reaches the laminar state:
 what changes is the relaxation time before it does so.
 This duration dramatically increases as an hypersurface is crossed \cite{schmiegel1997fractal}.}.

As such, this has brought the question of the separatrix between laminar and turbulent flow at the forefront of the study of transition.
The first descriptions by Waleffe and collaborators of the mechanisms maintaining transitional wall turbulence
was the self sustaining process (SSP) of wall turbulence \cite{hamilton1995regeneration,waleffe1997self}.
This process represented wall turbulence as a stable limit cycle, which was therefore separated from the laminar baseflow in phase space
by an actual separatrix with a saddle on it.
The SSP describes how streamwise velocity streaks and streamwise vortices regenerate one another.
It can take place provided the Reynolds number is large enough and the domain size is larger than
the so called Minimal Flow Unit (MFU) \cite{jimenez1991minimal}.
Refinements of that picture indicated that this process was not periodic nor an attractor (thus leaving the room for collapse of turbulence)
but nevertheless confirmed the idea that there was a separatrix, on which saddles could be found.
The structure of the separatrix, which is a key matter for the forced path from laminar to turbulent flow, was often studied using
methods of dichotomy or bisections \cite{nusse1989procedure}.
This has led to the study of edge states, saddles on that separatrix,
in wall flows \cite{toh2003periodic,schneider2007turbulence,schneider2008laminar,schneider2010localized,willis2009turbulent} as well as in
climate models \cite{lucarini2017edge}.
Note that additional unstable limit cycles and fixed points can be found on the turbulent side of the separatrix \cite{rolland_Gib08}. 
In wall flows, these unstable fixed points and limit cycles are often reminiscent of the turbulent flow, in
that they display velocity streaks and streamwise vortices.
However, they display less small scale fluctuations and are often very spatially regular.
This regularity can be characterised using the symmetry groups of the flow \cite{rolland_Gib08,schneider2010localized}.
Note finally that, as the separatrix and the unstable states on it are a property of the flow in the configuration of interest
and not of initial conditions or forcing type, they also represent an universal feature of transition to turbulence.

In wall flows, the complexity of this boundary is such that the saddle
on it is not the point of the separatrix which is has the smallest amplitude of departure to the laminar baseflow.
Experimental and numerical studies of initial conditions, generated using various specific rules, of minimal energy that evolve into
turbulence have highlighted that.
This leaves the question of whether the most probable forced path would visit this saddle at very small but finite forcing noise variance.
Alternate investigations have thus been aimed  at identifying a path forced and/or triggered by a finite amplitude initial condition
that would drive the flow from laminar to turbulent.
Linear studies of optimal initial conditions of infinitesimal amplitude leading to
maximal amplification of energy identified lift-up: the extraction of energy from the laminar base flow from streamwise vortices.
The lift-up mechanism was thus identified in transient growth studies that highlight the fact that
energy growth can happen while all linear modes are stable if the linearised Navier--Stokes equations lead
to a non normal operator \cite{farrell1993optimal,schmid1994optimal,trefethen1993hydrodynamic}.
Such an optimal (in shape) initial condition whose amplitude places it on the laminar side of the boundary would remain on the laminar side
of said boundary.
However, for initial conditions starting on the turbulent side of the boundary,
lift-up is an important part of the self-sustaining process \cite{waleffe1997self}.
Refinement of optimisation of initial condition energy amplification
technics can be used to compute the optimal finite amplitude initial condition leading to maximal non-linear energy growth
\cite{cherubini2011minimal,rabin2012triggering,monokrousos2011nonequilibrium}.
This initial condition is termed a minimal seed.
Such computations are still performed on the laminar side of the separatrix, however a slight modification of the seed,
in the right direction, can generate an initial condition which has very little energy
but can nevertheless travel to turbulence extremely efficiently.
Note that by discretising this procedure in phase space,
the instanton linking different stable fixed points can be computed \cite{lecoanet2018connection}.
There is in practice no explicit stochastic forcing in that case, but the resulting trajectory
still undergoes a fluctuation toward the relevant saddle, followed by a relaxation on which forcing is absent.

The question of whether a random forcing of a wall flow can generate a minimal seed,
a flow organisation whose shape and amplitude is precisely defined,
remains open.
Indeed, the study of forced wall flows gives little weight to effects like the Orr mechanism, involved in the growth of minimal seeds.
In parallel with studies of optimal initial conditions leading to transient growth, the
study of the response of the linearised Navier--Stokes equations to a forcing with a wide range of spatial spectra, which are white in time,
identified a reacting flow structure \cite{farrell1993stochastic}.
The weakly non linear behaviour of flows under such stochastic forcing can also be performed \cite{ducimetiere2022weakly}.
The inclusion of the strong non-linearities may very well strongly change this picture because they introduce
the aforementioned complex separatrix.
The experimental study of the response of boundary layers to harmonic forcing of different spectrum has a long history
and shows a strong dependence on the forcing frequency
and amplitude of the point of appearance of wall turbulence \cite{spangler1968effects}.
Numerical equivalent of such studies of the response to an harmonic forcing have only been recently undertaken \cite{rigas2021nonlinear}.
Meanwhile, to refocus on our subject of interest, studying the response to a stochastic forcing of vanishing variance requires a
rare event simulation method.
This has been performed in the case of a model of Hagen--Poiseuille pipe flow \cite{rolland_pre18}.
In that case, we have termed the forcing of the flow model from laminar to turbulent \emph{build up}.
However the separatrix in that model
is nowhere near as complex as the one found in actual Hagen--Poiseuille pipe flow.
A direct instanton computation through action minimisation in plane Poiseuille flow can be extremely complex, and only
instantons linking the laminar baseflow to non linear Tolmien--Schlichting type waves have been performed so far \cite{wan2013minimum,wan2015model}.

\subsection{Outline of the article}

This motivates our present study.
On the one hand, we would like to propose a method to study transition that can emulate experiments \cite{spangler1968effects,liu2021decay}.
On the other hand, we would like to study multistability forced by noise in the case of a very complex separatrix.
As to the rare event simulation method, we use a method alternate to action minimisation termed
Adaptive Multilevel Splitting (AMS) \cite{rolland_CG07,cerou2011multiple},
to compute the rare transitions from laminar baseflow to turbulence in plane Couette flow,
in domains of increasing sizes at transitional Reynolds numbers.
By decreasing the variance of the forcing, proportional to the energy injection rate, we will investigate the possible concentration
of reactive trajectories and whether this trajectories are structured by an instanton.
Using several noise spectra, we will test whether forcing directly the most responsive structure is a \emph{sine qua non} condition
to travel to turbulence, while increasing the system size will highlight possible spatial localisation of the growing  turbulence.
We thus present this study in the following manner.
We first present the forced Navier--Stokes equations (\S~\ref{sfns}), then the version of Adaptive Multilevel Splitting that
is used to compute the trajectories (\S~\ref{sams}), including the details of the set up of AMS computations
and additional diagnostics used to study the trajectories such as bisections.
We describe the computed paths in section~\ref{spath}, first visually in relevant examples (\S~\ref{svisup}),
then statistically, on the whole collection of computed path (\S~\ref{sstatp}).
We finally study the effect of the energy injection rate, on the transition rate (\S~\ref{sea})
and on the trajectories duration (\S~\ref{setau}).
We will discuss these results in view of the literature in the conclusion (\S~\ref{sconcl}).

\section{Method}\label{smet}

\subsection{Forced Navier--Stokes equations}\label{sfns}

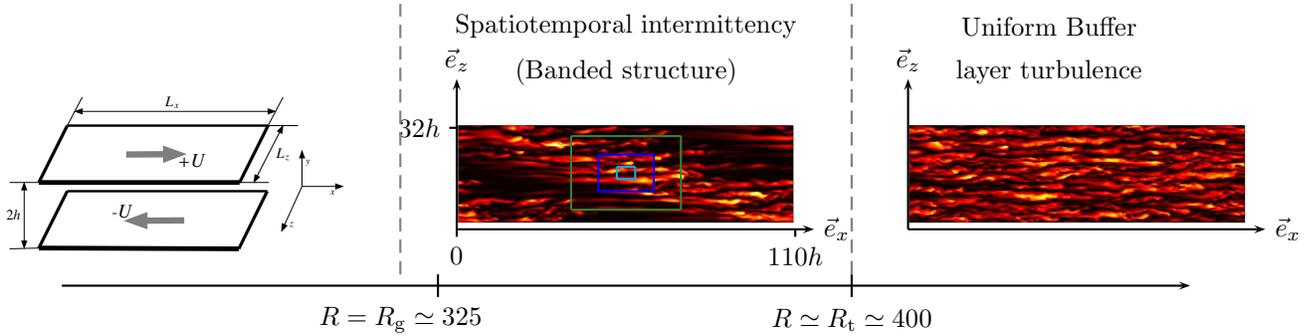
\begin{figure*}[!htbp]
\begin{center}
\begin{pspicture}(16.5,4.5)
\rput(2.25,2.25){\includegraphics[width=4.5cm]{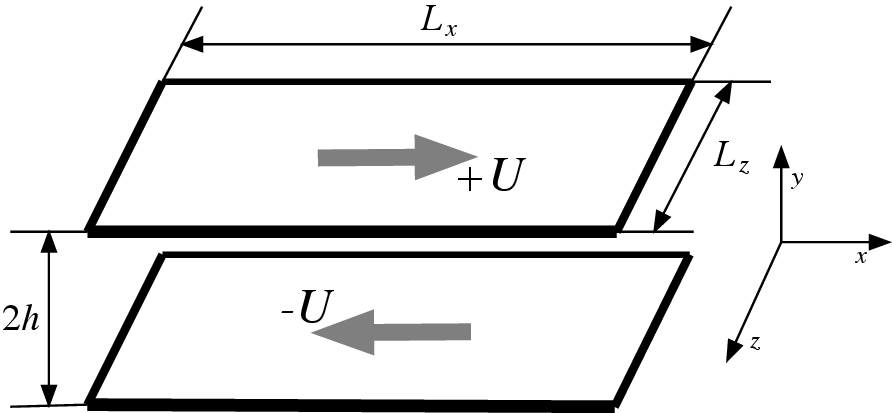}}
\rput(8.25,2.25){\includegraphics[width=4.5cm,clip]{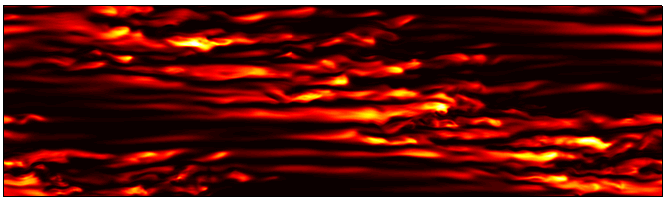}}
\psline{}(6,1.5)(6,1.4)
\psline[linecolor=OliveGreen](8.98,1.76)(8.98,2.74)
\psline[linecolor=OliveGreen](8.98,1.76)(7.52,1.76)
\psline[linecolor=OliveGreen](7.52,2.74)(8.98,2.74)
\psline[linecolor=OliveGreen](7.52,2.74)(7.52,1.76)
\psline[linecolor=blue](8.6175,2.005)(8.6175,2.495)
\psline[linecolor=blue](8.6175,2.005)(7.8825,2.005)
\psline[linecolor=blue](7.8825,2.495)(8.6175,2.495)
\psline[linecolor=blue](7.8825,2.495)(7.8825,2.005)
\psline[linecolor=cyan](8.3725,2.1683)(8.3725,2.3317)
\psline[linecolor=cyan](8.3725,2.1683)(8.1275,2.1683)
\psline[linecolor=cyan](8.1275,2.3317)(8.3725,2.3317)
\psline[linecolor=cyan](8.1275,2.3317)(8.1275,2.1683)
\psline{}(10.5,1.5)(10.5,1.4)
\psline{}(6,2.85)(5.9,2.85)
\rput(6,1.15){0}
\rput(10.5,1.15){$110h$}
\rput(5.5,2.85){$32h$}
\psline{->}(6,1.5)(6,3.5) \rput(6,3.75){$\vec e_z$}
\psline{->}(6,1.5)(10.75,1.5)\rput(11.05,1.5){$\vec e_x$}
\rput(14.25,2.25){\includegraphics[width=4.5cm,clip]{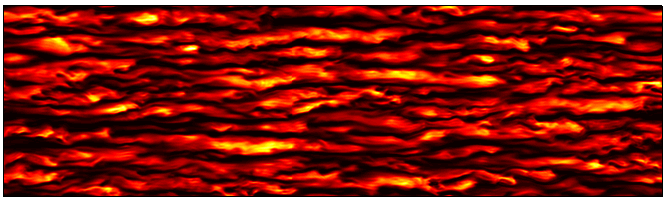}}
\psline{->}(12,1.5)(12,3.5) \rput(12,3.75){$\vec e_z$}
\psline{->}(12,1.5)(16.75,1.5) \rput(17.05,1.5){$\vec e_x$}
\psline{->}(0.75,0.75)(15.75,0.75)
\psline(5.75,0.6)(5.75,0.9)
\psline(11.25,0.6)(11.25,0.9)
\rput(5.25,0.3){$R=R_{\rm g}\simeq 325$}
\rput(11.25,0.3){$R\simeq R_{\rm t}\simeq 400$}
\psline[linecolor=gray,linestyle=dashed](5.25,0.9)(5.25,4.5)
\psline[linecolor=gray,linestyle=dashed](11.25,0.9)(11.25,4.5)
\rput(8.25,4.2){Spatiotemporal intermittency}
\rput(8.25,3.6){(Banded structure)}
\rput(13.875,4.2){ Uniform Buffer }
\rput(13.875,3.6){layer turbulence}
\end{pspicture}
\end{center}
\caption{A picture of transition to turbulence in plane Couette flow indicating the type of flow obtained as the Reynolds number is increased.
Plane Couette flow is sketched on the left: we indicate the two walls moving at velocities $\pm U$, the domain size $L_x$, $L_z$ and $h$
as well as the coordinate system.
For Reynolds numbers larger than a $R_{\rm g}\simeq 325$, laminar-turbulent coexistence can be sustained,
as indicated by the colour levels of kinetic energy $\frac{1}{2}\mathbf{u}^2$
in the horizontal midplane $y=0$ obtained by means of direct numerical simulations
in a domain of size $L_x\times L_z=110\times 32$ at $R=370$ (centre).
The coloured frames indicate the domain sizes considered here (cyan: $L_x\times L_z=6\times 4$,
blue $L_x\times L_z=18\times 12$, green $L_x\times L_z=36\times 24$).
As the Reynolds number is further increased, wall turbulence can invade the whole domain, as illustrated
by the colour levels of kinetic energy $\frac{1}{2}\mathbf{u}^2$
in the horizontal midplane $y=0$ obtained by means of direct numerical simulations
in a domain of size $L_x\times L_z=110\times 32$ at $R=370$ (right).
The visualisations are performed using numerical results originally performed for \cite{rolland2015mechanical}, and the whole figure
was originally created for \cite{rolland2022collapse}.
}
\label{sksize}
\end{figure*}
\subsubsection{Governing equations}

We consider plane Couette flow, the flow between two parallel walls separated by a distance $2h$
moving at velocities $\pm U$ (sketched in figure~\ref{sksize}, left).
We use our standard notations \cite{rolland2022collapse} so that $\mathbf{e_x}$ is the streamwise direction,
$\mathbf{e_y}$ is the wall normal direction and $\mathbf{e_z}$ is the spanwise direction.
Velocities are made dimensionless with $U$, lengths are made dimensionless with $h$ and times are made dimensionless with $h/U$.
The main dimensionless control parameter is the Reynolds number $R=\frac{hU}{\nu}$ where $\nu$ is the kinematic viscosity.
The flow is also controlled by the dimensionless domain sizes $L_x$ and $L_z$.

The forced incompressible Navier--Stokes equations for the field $\textbf{u}$,
the departure to the laminar base flow $y \textbf{e}_x$, and $p$ the dimensionless pressure, read
\begin{equation}
\frac{\partial u_l}{\partial t}+u_j\frac{\partial u_l}{\partial x_j}+y\frac{\partial u_l}{\partial x}+\delta_{l,x}u_y
=-\frac{\partial p}{\partial x_l}+\frac{1}{R}\left( \frac{\partial^2u_l}{\partial x^2}
+\frac{\partial^2u_l}{\partial y^2}+\frac{\partial^2u_l}{\partial z^2}\right)
+\sqrt{\frac{2}{\beta}}f_l(\mathbf{x},t)
\,,\,
 \frac{\partial u_l}{\partial x_l}=0\,.\label{nsf}
\end{equation}
Using tensorial notations, we denote each component of $\mathbf{u}$ by $u_l$, where subscript $l$ and $j$
stands for the component
$\textbf{e}_x$, $\textbf{e}_y$ and $\textbf{e}_z$ (respectively $l,j=1,2,3$).
The forcing $\mathbf{f}$ has zero ensemble average and has in plane spatial correlations.
We will see that its energy injection rate is controlled by $\beta$.
Before giving further details on $\mathbf{f}$, we need to present the numerical discretisation that
we use. The Navier--Stokes equations are discretised in space using $N_x$ and $N_z$ dealiased Fourier modes in
the $x$ and $z$ direction (so that $\frac32 N_x$ and $\frac{3}{2} N_z$ modes are used in total)
and $N_y$ Chebyshev Polynomials in the $y$ direction.
This discretisation and the time integration are performed using a code based on {\sc channelflow},
by J. Gibson \cite{rolland_Gib08}. We will use a fixed time step $\Delta t$ for more control on our numerical procedure.
For each system size, Reynolds number and resolution, the value of $\Delta t$ is chosen so that the CFL criterion is respected.
We give the numerical resolutions used for our simulations in the three considered domain sizes $L_x\times L_z=6\times 4$, $18\times 12$ and $36\times 24$ in table~\ref{tabres}.
Note that we have considered several resolutions for the smallest domain $L_x\times L_z=6\times 4$
to test numerical convergence with spatial resolution of the simulations.
We will comment on the effect of resolution throughout the text.
We present the domain sizes in figure~\ref{sksize}, centre, using concentric frames.

\begin{table}[!htbp]
\centerline{
\begin{tabular}{|c|c|c|c|c|c|}
\hline $L_x\times L_z$&$\frac32 N_x$&$\frac32 N_z$&$N_y$&$\Delta t$& resolution type \\ \hline
$36\times 24$&192&144&27&0.05& --- \\ \hline
$18\times 12$&96&72&27&0.05&--- \\ \hline
$6\times 4$&32&24&27&0.05& standard resolution \\ \hline
$6\times 4$&64&48&47&0.0125& high resolution \\ \hline
$6\times 4$&64&48&27&0.05&--- \\ \hline
$6\times 4$&128&96&35&0.025&--- \\ \hline
\end{tabular}
}
\caption{
Table listing the domain sizes $L_x\times L_z$ considered in simulations,
along with the number of dealiased Fourier modes in the streamwise and spanwise directions $\frac32 N_x$,
$\frac32 N_z$, the number of Chebyshev modes  $N_y$ and the time step $\Delta t$.
Two specific resolutions are distinguished in the smaller domain of size $L_x\times L_z=6\times 4$.}
\label{tabres}
\end{table}

\begin{figure*}[!htbp]
\centerline{\includegraphics[width=6.5cm]{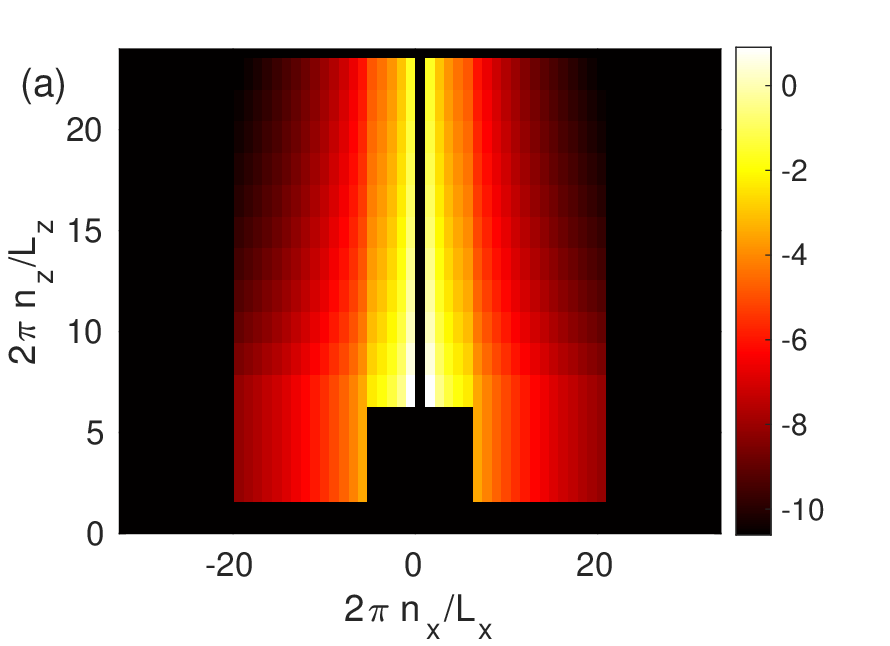}\includegraphics[width=6.5cm]{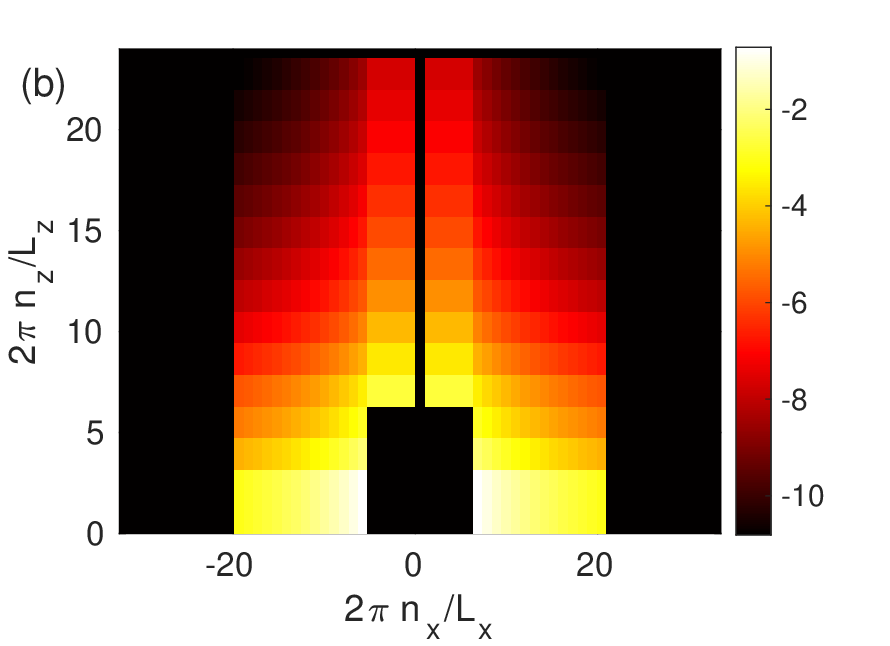}}
\centerline{\includegraphics[width=6.5cm]{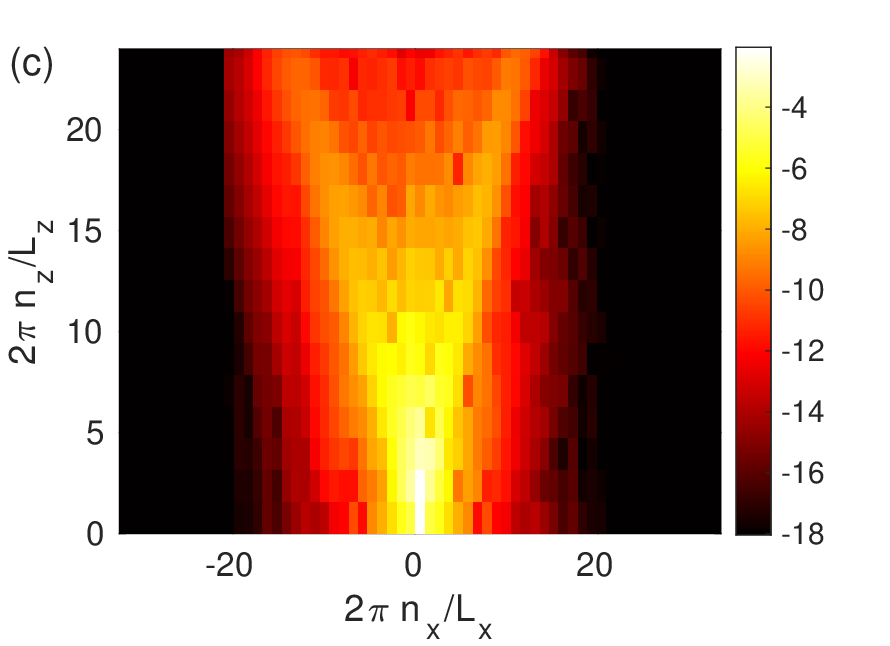}\includegraphics[width=6.5cm]{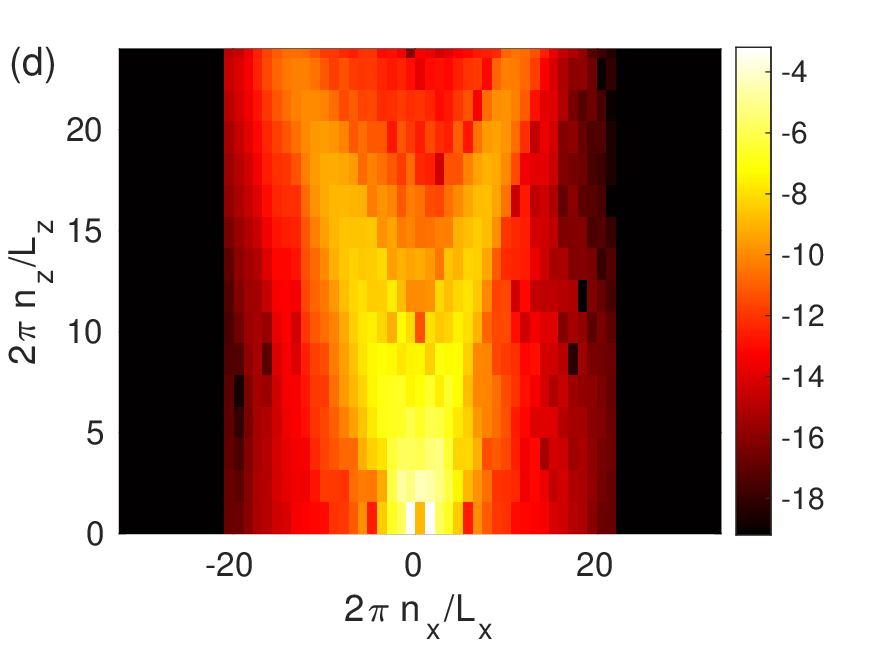}}
\caption{
Colour levels in the wavenumber $2\pi n_x/L_x,2\pi n_z/L_z$ plane of shape factors of divergence free forcing used
for AMS computations in domains of size $L_x\times L_z=6\times 4$ at resolution $N_x=64$, $N_y=48$, $N_z=48$ at Reynolds number $R=500$:
(a) logarithm of the product $((2\pi n_z/L_z)\gamma_{n_x}\gamma_{n_z})^2$ for forcing applied on $u_x$,
(b) logarithm of the product $((2\pi n_x/L_x)\gamma_{n_x}\gamma_{n_z})^2$ for forcing applied on
$u_z$ (the details of $\gamma_x$ and $\gamma_z$ are given in appendix~\ref{stnoise}).
See appendix~\ref{stnoise} for full detail on how this generates the vector potential used to exert a divergence free forcing
on the flow.
From the velocity field obtained in a DNS of a turbulent flow in a domain of  $L_x\times L_z=6\times 4$
at resolution $N_x=64$, $N_y=48$, $N_z=48$ at Reynolds number $R=500$,
(c) logarithm of the modulus of $x-z$ spectrum of $u_x$ in the midplane $y=0$,
(d) logarithm of the modulus of $x-z$ spectrum of $u_z$ in the midplane $y=0$.
}
\label{fact}
\end{figure*}

The forcing $\mathbf{f}$ is quantitatively prescribed by the correlations in space and time of each of its component $f_l$
\begin{equation}
\langle f_l(\textbf{x},t)f_l(\textbf{x}',t')\rangle=
\delta(t-t')C_l(x-x',z-z')\delta(y-y')\,,\\, \hat{C}_l(n_x,n_z)=\Gamma_{l,n_x,n_z}\,.\label{spcor}
\end{equation}
The force $\mathbf{f}$ is white (as indicated by the delta correlations) in time and in the wall normal direction.
Our numerical simulations use a finite number $N_y$ of Chebyshev modes in the wall normal direction.
This means that with a red spectrum in $x$ and $z$, the force $\mathbf{f}$ injects a finite amount of energy.
The correlation function $C_l(x-x',z-z')$ in the streamwise and spanwise direction on component $l$ is defined using its Fourier transform.
Said transform is expressed using a shape factor $\Gamma_{l,n_x,n_z}$, where
$n_x$ stands for the streamwise  wavenumber, $n_z$ stands for the spanwise wavenumber.
We have followed two different options as to the details of the forcing, presented in details in appendix~\ref{stnoise}. One type of noise
is forcing selected components (either a single components or two or all three components).
In that case we have that different components are decorrelated $\langle f_lf_j\rangle=0$ if $l\ne j$.
Another type of forcing is divergence free and forces components $\mathbf{e}_x$ and $\mathbf{e}_z$:
it is the curl of a vector potential along $\mathbf{e}_y$.
In that case components $\mathbf{e}_x$ and $\mathbf{e}_z$ are forced and a
correlation can exists between these two components of the forcing.
Numerically, we chose to add the forcing after the predictor-corrector stage of time integration, in line
with the formulation of an It\^o Process.
We checked \emph{a posteriori} that incompressibility was enforced up to the same numerical precision as in unforced simulations.
Note that in studies of action minimisation in plane Poiseuille flow, both an unconstrained noise white in time and space \cite{wan2013minimum}
and a noise defined as a Wiener process  on the space of divergent free vector fields \cite{wan2015model,mam_pois},
still white in time, were considered.
Linear \cite{farrell1993stochastic} and weakly non linear \cite{ducimetiere2022weakly}
studies of flows used noise with little constraint on spectrum shape and component correlations.
In the large majority of cases, we will not directly force the spatial scales expected in the flow of order $\mathcal{O}(1)$ or larger:
we force below  a corresponding cut-off scale $L_c$.
The value of this scale will be specified in each forcing case.
What we do is that we let the flow non linearly select its scales from energy injected solely at moderately small scales
(some comparison with forcing all large scales have been performed).
Both types of forcing typically give a flat spectrum for the forcing at intermediate wave numbers
and after a wavenumber cut-off $N_r$, we will use a decaying spectrum for the forcing at larger wavenumbers.
We display an example of shape factor $\Gamma_{x,n_x,n_z}$ in the divergence free case in figure~\ref{fact} (a) and $\Gamma_{z,n_x,n_z}$
in figure~\ref{fact} (b), we we chose $L_c=1.0$.
As a comparison, the spectra of $u_x$ and $u_z$ computed in the midplane $y=0$
for transitional turbulence in a MFU type domain at the largest resolution are displayed in figure~\ref{fact} (c)
and figure~\ref{fact} (d) respectively.
We can notice that these two components have their intense modes in the wavenumber range $\left|\frac{2\pi n_x}{L_x}\right|\le 3$,
$\left|\frac{2\pi n_z}{L_z}\right|\le 6$, where the forcing is absent.
The forcing starts to act in a wavenumber range where the natural Fourier modes of the velocity field
are at least one order of magnitude smaller than their maximum.

\subsubsection{Diagnostics of the flow}\label{sdfl}

We define some quantities of interest that are useful to follow the transition.
We define the spatially averaged kinetic energy as
\begin{equation}
E_k=\frac{1}{2L_xL_z}\int_{x=0}^{L_x}\int_{y=-1}^1\int_{z=0}^{L_z}\frac{u_x^2+u_y^2+u_z^2}{2} \,{\rm d}x{\rm d}y{\rm d}z\,.\label{avek}
\end{equation}
The kinetic energy is used to monitor the global state of the flow.
When the flow is turbulent in the whole domain, we find $E_k=\mathcal{O}(0.1)$.
Meanwhile, when the flow is globally laminar, $E_k$ is much smaller and close to zero.
We will also distinguish $E_{k,l}$ the kinetic energy contained in a given component $l=x,y,z$, on the one hand,
and $E_{k,y-z}$ the kinetic energy contained in the spanwise and wall normal components, on the other hand
\begin{equation}
E_{k,l}=\frac{1}{2L_xL_z}\int_{x=0}^{L_x}\int_{y=-1}^1\int_{z=0}^{L_z}\frac{u_l^2}{2} \,{\rm d}x{\rm d}y{\rm d}z\,,
\\ E_{k,y-z}=\frac{1}{2L_xL_z}\int_{x=0}^{L_x}\int_{y=-1}^1\int_{z=0}^{L_z}\frac{u_y^2+u_z^2}{2} \,{\rm d}x{\rm d}y{\rm d}z\,.
\end{equation}
The kinetic energy in the streamwise component $E_{k,x}$ will be the most used.
The energy $E_{k,x}$ roughly quantifies the amount of energy contained in the velocity streaks, while
the energy $E_{k,y-z}$ roughly quantifies the amount of energy contained in the streamwise vortices,
which are the two main coherent structures in transitional wall flows.
These quantities have been able to highlight the processes at play
in the collapse of turbulence \cite{rolland2022collapse},
and are quite close to what can be measured in experiments by means of two dimensional, two components particle
image velocimetry in planes parallel to the moving wall \cite{liu2021decay}.
Note that other choices can be made to follow the amplitude of streamwise vortices.
For instance isocontour of streamwise vorticity can be added to isocontour of streamwise velocity in
visualisations \cite{toh2003periodic,cherubini2011minimal}.
Other choices include colour levels of streamwise vorticity \cite{willis2009turbulent},
or isocontours of the Q-criterion \cite{rigas2021nonlinear}.
In order to have a global measure corresponding to these visualisations,
one could use the square of the streamwise component of vorticity.
We stick to the kinetic energy $E_{K,y-z}$ because of the simplicity of the diagnostic and
because contribution of other types of eddies is relatively small in our case.
We finally used a turbulent fraction $F$ to  observe the possible spatial localisation of turbulence.
This indicator is computed in the following manner: we spatially average the kinetic energy contained in components $u_y$ and $u_z$
in $\tilde{N_z}$ streamwise bands of width $\delta z=L_z/\tilde{N_z}=2.0$, to define
\begin{equation}
e(\tilde{n}_z)=\frac{1}{2L_x \delta z}\int_{x=0}^{L_x}\int_{y=-1}^1\int_{z=\tilde{n}_z\delta z}^{(\tilde{n}_z+1)\delta z}(u_y^2+u_z^2) \,{\rm d}x{\rm d}y{\rm d}z\,,
\label{defef}			
\end{equation}
with $0\le \tilde{n}_z< \tilde{N}_z$.
We then compute the proportion of bands where $e(\tilde{n}_z)>0.0005$, which gives our turbulent fraction $F$.
This definition will be adapted to the spatial localisation we will observe on the trajectories.

We now introduce a parameter alternative to $\beta$ which is possibly better suited to follow the transitions:
the energy injection rate.
This parameter is present in the ensemble average of the spatially average kinetic energy budget.
In order to obtain this budget, we multiply equation~(\ref{nsf}) by $u_l$, sum over the three components and simplify using incompressibility.
We perform a spatial average, to obtain the time derivative of the spatially averaged kinetic energy (Eq.~(\ref{avek})),
and we perform ensemble average with respect to the forcing noise $\mathbf{f}$, in order to remove it from the equation.
In appendix~\ref{abud}, we detail how
we use It\^o's Lemma (see \citep{gardiner2009stochastic} \S~4) to manage this non linear change of variables.
This yields the average energy budget
\begin{equation}
\frac{\partial \langle E_k\rangle}{\partial t}=
\underbrace{-\left\langle\frac{1}{2L_xL_z}\int u_xu_y\,{\rm d}x{\rm d}y{\rm d}z\right\rangle}_{\text{extraction}}
\underbrace{-\left\langle\frac{1}{R 2L_xL_z}\int \frac{\partial u_l}{\partial x_j}
\frac{\partial u_l}{\partial x_j}\,{\rm d}x{\rm d}y{\rm d}z\right\rangle}_{\text{dissipation}}
+\epsilon\,,\label{rolland_bud}
\end{equation}
where we have the energy injection rate $\epsilon$
\begin{equation}
\epsilon=\frac{1}{\beta}\left(N_y\sum_{n_x=-\frac{N_x}{2}+1}^{\frac{N_x}{2}}
\sum_{n_z=-\frac{N_z}{2}+1}^{\frac{N_z}{2}}\sum_{l=1}^3\Gamma_{l,n_x,n_z} \right)\,,\label{defepsilon}
\end{equation}
in the simulation.
Note that equation~(\ref{rolland_bud}) is similar to the Reynolds--Orr equation
which has often been used for non-linear stability analysis of wall flows \cite{drazin2004hydrodynamic,stuart1958non}.
In our case we have an ensemble average of each term and the addition of the energy injection rate.
In this energy budget (Eq.~(\ref{rolland_bud})), the extraction is performed by lift-up, in the streamwise direction, thus involving $u_x$ and $u_y$
and the wall normal gradient of the
laminar baseflow equal to $1$. Dissipation is performed by viscosity.
In the case of an unforced turbulent plane Couette flow, these two terms compensate one another after time average.
For small energy injection rate $\epsilon$, it is generally expected that the flow remains close to laminar
and that equation~(\ref{nsf}) can be treated linearly \cite{farrell1993stochastic} or could be processed using a weakly non linear approach \cite{ducimetiere2022weakly}.
However, this behaviour is conditioned on no transition happening
and is thus not really represented by the actual ensemble average in equation~(\ref{rolland_bud}).
We will see in the rest of this text that even in that case,
transitions can occur.
This means that starting from the initial condition $\langle E_k\rangle (t=0)=0$ even with a small $\epsilon$,
the solution of
equation~(\ref{rolland_bud}) would display a very slow
but steady increase of $\langle E_{k}\rangle(t>0)$ until the flow becomes turbulent.
Let us discuss how transition would be represented in this energy budget.
We would expect that $\epsilon$ would act against dissipation and could maintain some energy extraction
(that would not be sustainable without energy injection, regardless of the spectrum shape).
It is possible that a large part of the energy transfers would not be visible in this equation as they would be erased by the spatial averages.
We would also expect that the larger $\epsilon$ is, the more efficiently transition is forced,
similarly to the dependence on amplitude of perturbations to boundary layers \cite{spangler1968effects,rigas2021nonlinear}.
The energy injection rate is controlled by two parameters whose effect on the build-up can be presented independently.
Indeed, from equation~(\ref{defepsilon}), we deduce that $\beta$
and the spectrum shape are the two independent control parameters of this energy injection rate $\epsilon$.
Thus, on the one hand, decreasing $\beta$ increases $\epsilon$, and on the other hand, increasing the sum of the $\Gamma_{l,n_x,n_z}$ also increases
$\epsilon$: in both cases, we force the flow more intensely. In our definitions (\S~\ref{stnoise}),
these shape factors $\Gamma$  will be normalised and will decrease at large wavenumbers.
Thus the sum will be larger if that decrease starts at larger wave numbers and if that decrease is slow.
Note that we cannot rule out that there could be effect of the precise shape of the spectrum.
These complex effects would be hidden in the way energy is transferred from the forced scales to the most receptive scales.
There can also be effects of a direct forcing of scales versus energy transfer.
As a final remark, equation~(\ref{defepsilon}) reminds that in order to have a finite or even vanishing energy injection
rate in cases where the flow is forced with a noise white in space (see \cite{wan2013minimum}),
one would need to take the limit of $\beta\rightarrow 0$ first and $N_x,N_y,N_z\rightarrow \infty$  next (or use specific
relation between those parameters).
With our spectrum shape, we keep the energy injection rate finite, similarly to what is done in study of linear response \cite{farrell1993stochastic}
and non linear response \cite{ducimetiere2022weakly} to stochastic forcing.
This physical argument leads to the same mathematical condition for well posedness of the Freidlin--Wentzell Large Deviation principle in
two dimensional Navier--Stokes equations as noted by \cite{wan2015model} and demonstrated by \cite{brzezniak2015quasipotential}.

Note that a similar energy budget is also commonly used in the
study of transition to turbulence \cite{kawahara2001periodic,rolland_Gib08,kreilos2012periodic}.
This time, it is obtained for the total spatially averaged kinetic energy $E_{\rm tot}$.
For this matter, we apply It\^o's Lemma for the change of variable from the Navier--Stokes equations for total flow $\mathbf{u}+y\mathbf{e}_x$,
to the equation giving the evolution of spatially averaged total kinetic energy (see again \S~\ref{abud}).
After ensemble averaging, this gives us
\begin{align}
\notag
\frac{dE_{\rm tot}}{dt}=\frac{1}{R}(I-D)+\epsilon\,,\\,
E_{\rm tot}=\int_{x=0}^{L_x}\int_{y=-1}^{1}\int_{z=0}^{L_z}\frac{(u_x+y)^2+u_y^2+u_z^2}{2}\,{\rm d}x{\rm d}y{\rm d}z
\\ \notag I=1+\frac{1}{2L_xL_z}\int_{x=0}^{L_x}\int_{z=0}^{L_z}\left(\left.\frac{\partial u_x}{\partial y}\right|_{y=+1}+\left.\frac{\partial u_x}{\partial y}\right|_{y=-1}\right)\,{\rm d}x{\rm d}z\,,
\\ D=\frac{1}{2L_xL_z}\int_{x=0}^{L_x}\int_{y=-1}^{1}\int_{z=0}^{L_z}\left\lVert \nabla \times(\mathbf{u}+y\mathbf{e}_x) \right\rVert_2^2\,{\rm d}x{\rm d}y{\rm d}z\,,
\label{total_bud}
\end{align}
where $I$ is the total energy injection at the wall and $D$ is the total dissipation,
written using total vorticity because of incompressibility, periodic and no slip boundary conditions.
As such, both are rescaled by their value in the laminar flow where $\mathbf{u}=\mathbf{0}$ and thus $I=D=1$.
This pair of variable is convenient to observe transition in wall flows.
The line $I=D$ has a special importance because all invariant states lie on it (since $\frac{dE_{\rm tot}}{dt}=0$ on them),
while limit cycles lie over it (since the time average over a period of $\frac{dE_{\rm tot}}{dt}$ is $0$ on them).
Meanwhile turbulent flow is close to $I=D=3$ \cite{kawahara2001periodic,rolland_Gib08}.
Note that streamwise vortices contribute strongly to the dissipation. However, it is difficult to
disentangle their contribution from shear.

\subsection{Adaptive Multilevel Splitting}\label{sams}

\subsubsection{The algorithm}\label{detams}

\begin{figure}[!htbp]
\centerline{\textbf{(a)}\includegraphics[width=6.5cm]{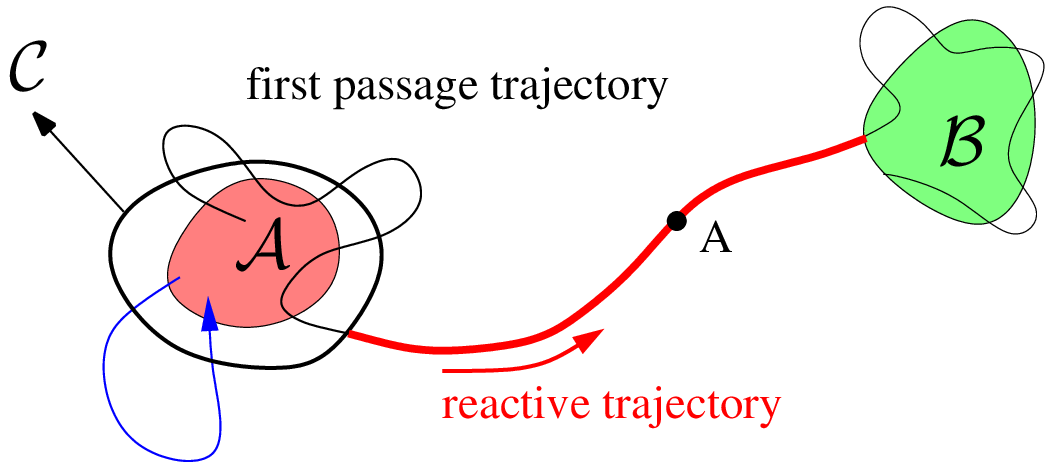}\textbf{(b)}\includegraphics[width=5cm]{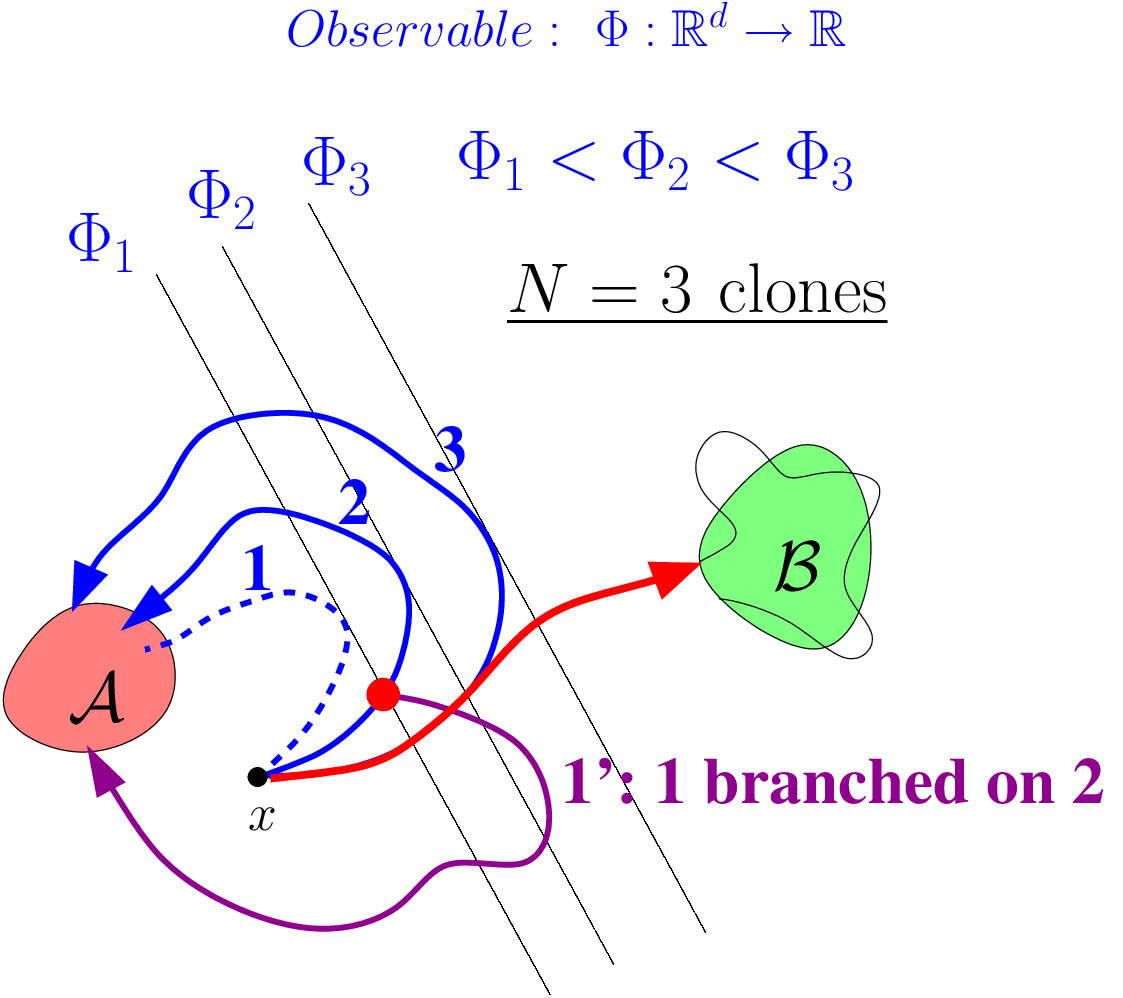}}
\caption{(a) Sketch of two bistable states $\mathcal{A}$ and $\mathcal{B}$ and the hypersurface $\mathcal{C}$ closely surrounding $\mathcal{A}$.
two realisations of the dynamics are sketched: a single excursion in blue and a first passage trajectory in black and red.
The red part of the first passage trajectory is the reactive trajectory (figure originally made for \cite{rolland2015statistical}).
(b) Sketch of the principle of AMS, showing to iterations of the algorithm, with $N=3$ clones,
indicating the starting state $\mathcal{A}$ and its neighbourhood,
the arrival state and its neighbourhood $\mathcal{B}$, three trajectories are ordered by their $\max_t\Phi$.
Trajectory one (dashed blue line) is suppressed and branched on another trajectory
at level $\max_t\Phi_1$ and then ran according to its natural dynamics.
Trajectory $2$ is then suppressed and branched on 3 at level $\max_t\Phi_2$ (figure originally made for \cite{simonnet2016combinatorial}). }
\label{rolland_figrare}
\end{figure}

In order to present the numerical method used to compute rare trajectories,
let us first give a formal phase space description of the rare events we will study in this text.
We sketch the build up trajectories in figure~\ref{rolland_figrare} (a)
and consider $\mathcal{A}$ a neighbourhood of the laminar flow in phase space,
an hypersurface $\mathcal{C}$ that closely surrounds $\mathcal{A}$ and $\mathcal{B}$
a  neighbourhood of the turbulent flow in phase space.
A realisation of the dynamics which starts in $\mathcal{A}$
fluctuates around it (for instance has several excursions out of $\mathcal{C}$,
an hypersurface closely surrounding $\mathcal{A}$) and eventually crosses $\mathcal{C}$ and reaches
$\mathcal{B}$ before coming back to $\mathcal{A}$ is termed a \emph{first passage} (Black then red curve in Fig.~\ref{rolland_figrare} (a)).
Its average duration is termed the mean first passage time $T$. The last part
of the first passage is termed a \emph{reactive trajectory}: this is the part of the dynamics that starts in $\mathcal{A}$,
crosses $\mathcal{C}$ and reaches $\mathcal{B}$ before
$\mathcal{A}$ (Portion of black curve between $\mathcal{A}$ and $\mathcal{C}$ continued by red curve in Fig.~\ref{rolland_figrare} (a)).
Quantitative definitions of sets $\mathcal{A}$, $\mathcal{B}$ and hypersurface $\mathcal{C}$
for our case will be given in section~\ref{dreac}.

We then present the generic formulation for two of the variants of Adaptive Multilevel Splitting (AMS) which were used for the computation presented
in this text
They were initially proposed by C\'erou \& Guyader \cite{rolland_CG07}.
Both variants of AMS use a reaction coordinate (or observable) $\phi(\mathbf{u}(t))=\Phi(t)$.
The reaction coordinate $\phi$ is a mapping from $\mathbb{R}^D$,
where the discretised velocity field $\mathbf{u}$ lives, to $[0,1]$. The reaction coordinate can be viewed as a function of
time $\Phi(t)$ on trajectories.
The reaction coordinate  measures the position of the flow relatively to starting set $\mathcal{A}$,
where is it $0$ and arrival set $\mathcal{B}$ where it is $1$.
We will define our reaction coordinate quantitatively in section~\ref{dreac}.
All variants of AMS run $N$ clone dynamics of the system to compute iteratively from $N-N_c+1$
up to $N$ reactive trajectories going from set $\mathcal{A}$ to the hypersurface $\mathcal{C}$,
closely surrounding set $\mathcal{A}$, then from $\mathcal{C}$ to the set $\mathcal{B}$.
Here $1\le N_c< N$ is the number of trajectories that will be removed at each iteration.
We will label the velocity field and pressure field of each trajectory by $1\le i\le N$
as in $\left(\begin{matrix}\mathbf{u}_i \\ p_i \end{matrix} \right)(t)$.
The reaction coordinate as a function of time for the corresponding trajectory is then denoted by $\Phi_i(t)=\phi(\mathbf{u}_i(t))$.
 The algorithm is sketched in figure~\ref{rolland_figrare} (b)
 and proceeds in the following manner
\begin{itemize}
\item There is a first stage of natural dynamics, where each clone dynamics starts in $\mathcal{A}$ from $\mathbf{u}(t=0)=\mathbf{0}$, $p(t=0)=0$,
is ran with an independent realisation of the noise $\mathbf{f}$ and is let to exit
and reenter $\mathcal{A}$ until it crosses the hypersurface $\mathcal{C}$.
Once this has happened, this dynamics is run until the first time it reaches  either $\mathcal{A}$ or $\mathcal{B}$.
\item The algorithm then runs the steps of mutation selection. At each step,
the clones $1\le i\le N$, are ordered by $\max_t\Phi_i(t)$:
the maximum of the reaction coordinate achieved on the trajectory.
The $N_c$ clones $1\le i'\le N_c$ having the smallest values of $\max_t\Phi_{i'}(t)$ are removed.
In order to keep a constant number of clones, $N_c$ new clones are generated by branching.
We draw $N_c$ clones $i''$ uniformly  from the clones $N_c+1\le i\le N$ such that $\max_t\Phi_i(t)\ge \max_t\Phi_{N_c}(t)$.
We denote by $t_{i''}$ the first time such that clone $i''$ reaches the reaction coordinate value $\max_t\Phi_{i'}(t)$:
$t_{i''}=\min\left\{t'\ge 0\backslash\Phi_{i''}(t')>\max_t\Phi_{i'}(t)\right\}$.
The new clone $i'$ is such that $\mathbf{u}_{i'}(t)=\mathbf{u}_{i''}(t)$,  $p_{i'}(t)=p_{i''}(t)$,  for $0\le t\le t_{i''}$.
For $t>t_{i''}$ the clone $i'$ then follows its natural dynamics under a new, independent realisation of the forcing $\mathbf{f}$ until
it reaches either $\mathcal{A}$ or $\mathcal{B}$.
\end{itemize}
The algorithm stops after iteration number $\kappa \in \mathbb{N}$,
the first iteration where at least $N-N_c+1$ clones trajectories have reached $\mathcal{B}$.
We denote  by $r \in \left[1-\frac{N_c-1}{N} ,1\right]$ the proportion of clones that have reached $\mathcal{B}$ at the last step.
Note that $\kappa$ and $r$ are \emph{a priori} random numbers, different from one AMS run to the other,
whose value depend on the independent noise realisations drawn during simulation of trajectories.
This yields an estimator of the probability $\alpha$ of reaching $\mathcal{B}$ before $\mathcal{A}$
starting from the distribution of first exit points on $\mathcal{C}$,
and the corresponding mean first passage time $T$
\begin{equation}
\notag
\alpha=r\left(1-\frac{N_c}{N} \right)^\kappa\,,\,
\hat{\alpha}=\left\langle \alpha\right\rangle_o\,, \,
\hat{T}=\left\langle\left(\frac{1}{\alpha}-1\right)\tilde{\tau}+\tau\right\rangle_o\,,
\end{equation}
where $\tau$ is the mean duration of reactive trajectories and $\tilde{\tau}$ is the mean duration
of non reactive trajectories, computed in each AMS run and $\langle \cdot\rangle_o$
corresponds to an arithmetic average over $o$ AMS runs \cite{rolland_CG07,rolland_pre18}.
In particular, note that the expression of the estimator of $\alpha$,
is a consequence of the fact that the estimate probability to transit from one branching level to the next is $1-\frac{N_c}{N}$.
Moreover, the estimated probability to go from the last branching level to $\mathcal{B}$ is one.
In this text, we both performed  calculations with $N_c=1$ (systematically leading to $r=1$) and $N_c>1$.
 The algorithm is naturally parallelised over the $N_c$ suppressed clones:
 the speed up using $N_p$ cores has been measured (not shown here) and requires $N_c/N_p$
 to be strictly larger than one to remain close to linear.
Note that other procedures used to compute instantons use intermediate states \cite{lecoanet2018connection}.
In that case, all the intermediate states are found on the fluctuation part of the trajectory.
This is similar to what is found in AMS computations, since the probability to go from the saddle to state $\mathcal{B}$ goes to one
as quantities controlling the rarity of the event like $\epsilon$ go to zero.

As it was done in other studies \cite{simonnet2020multistability,rolland2022collapse},
we save a state at the last step $\kappa$ of the AMS computation.
This state is taken on the
trajectory that has the largest reaction coordinate $\Phi$ among the remaining non
reactive trajectories at the last step of an AMS computation.
On this trajectory, this state is taken at branching time.
We use the same terminology as in \cite{rolland2022collapse},
we term it the last state at the last step and denote it by $\mathbf{u}_{\rm last}$.
Using the notations introduced above, this state is computed using the clone $\mathbf{u}_{i''}$ that replaces $\mathbf{u}_{i'=N_c}$.
We then have $\mathbf{u}_{\rm last}=\mathbf{u}_{i''}(t_{i''}$).
In complex spatially extended systems, it has been verified that this state is
very similar to the saddle crossed by the trajectory \cite{simonnet2020multistability}.
In gradient systems, (with few DoF or with spatial extension \cite{rolland2016computing}),
it can be verified that this gives a finer and finer approximation of the
actual, perfectly defined saddle as the noise variance is decreased (not shown here).
In unforced system, this indicates a turning point for the trajectory \cite{rolland2022collapse}.
This is directly connected to the fact that the probability
to transit from the last branching level to $\mathcal{B}$ is closer and
closer to one as $\epsilon\rightarrow 0$.
This means that the last branching level is then very close to the first
instant of the relaxation path, which is said saddle in forced systems, in the limit $\epsilon\rightarrow 0$.

\subsubsection{Reaction coordinate}\label{dreac}

\begin{figure*}[!htbp]
\centerline{\includegraphics[width=10cm]{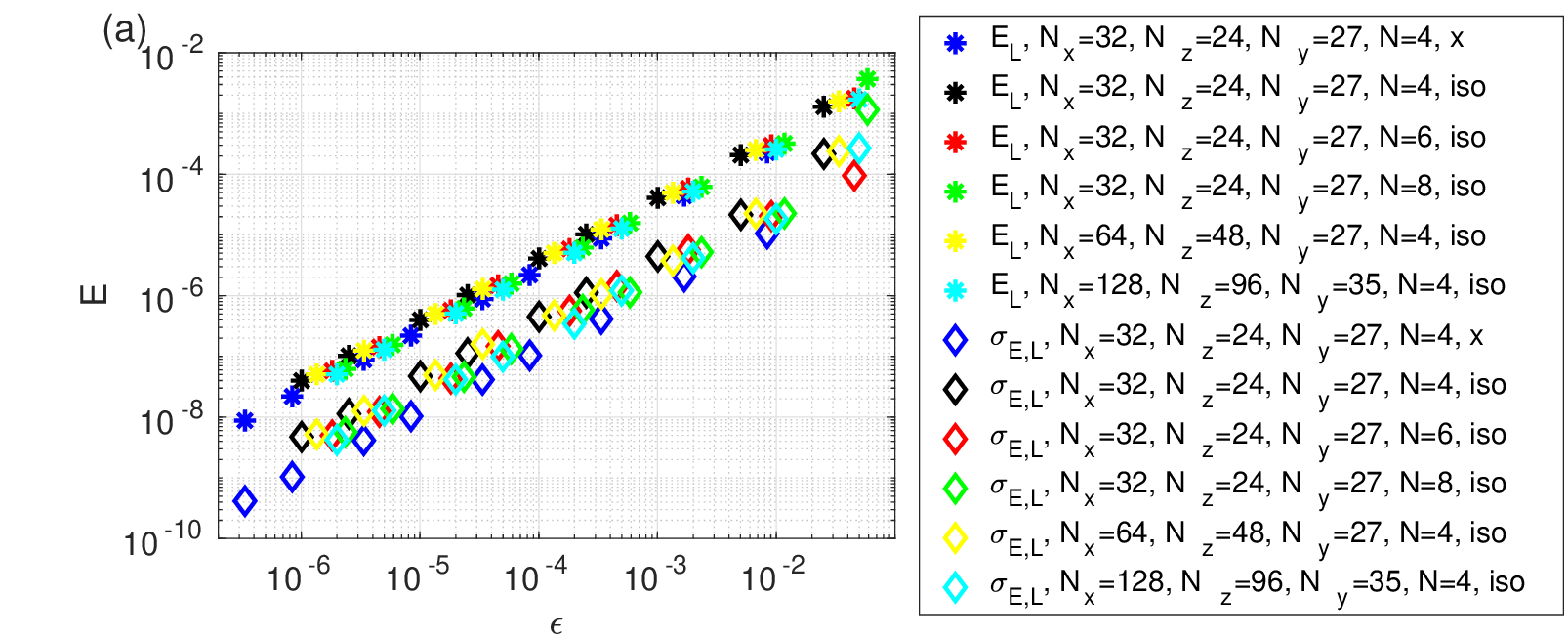}\includegraphics[width=10cm]{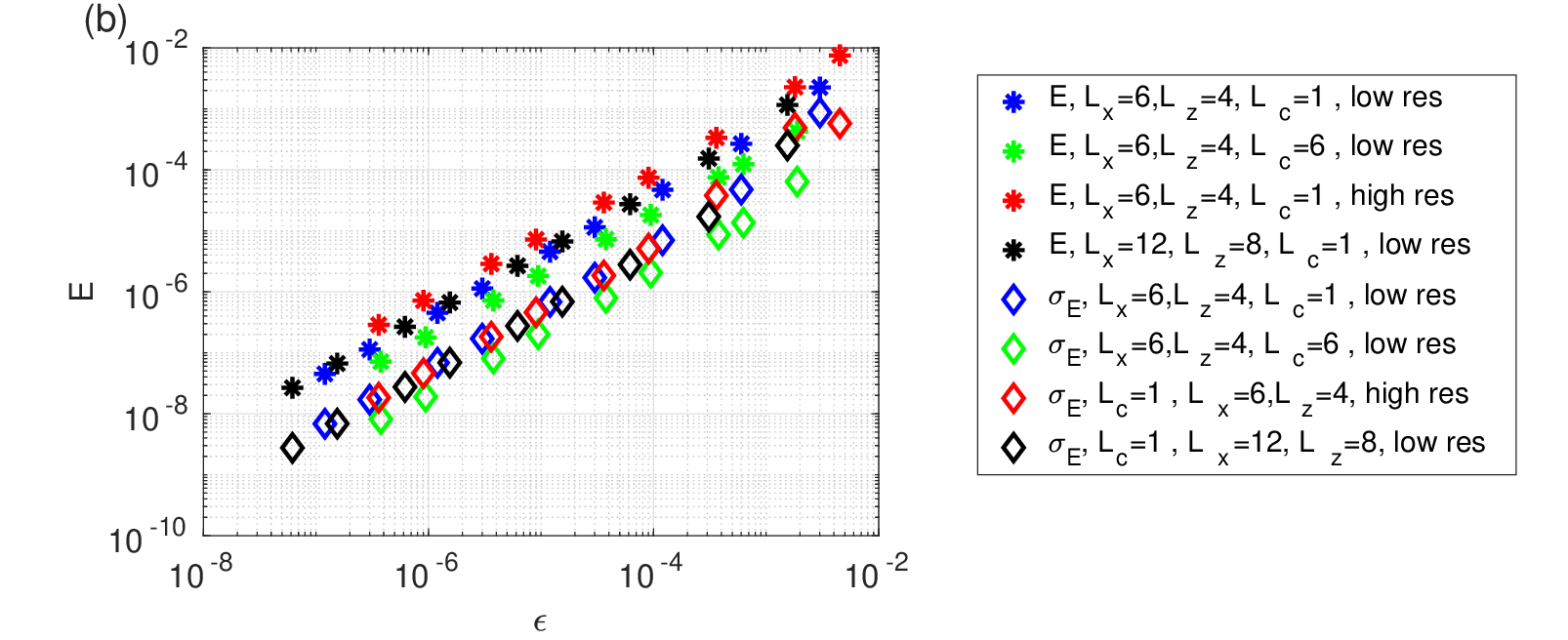}}
\caption{Average kinetic energy conditioned on no build up having occurred and corresponding variance
as a function of the energy injection rate for several shape of the forcing spectrum and/or several resolutions when
(a)  noise on components is applied, (b) when divergence free noise is applied.
}
\label{energy}
\end{figure*}

We will construct a reaction coordinate which is monotonous in the kinetic energy $E_k$.
Indeed, $E_k$
is almost $0$ when the forced flow fluctuates around the first metastable state, the laminar baseflow.
Meanwhile, it takes its largest possible values when the flow is turbulent (the second metastable state).
Finally, $E_k$ grows when the forced flow travels from the laminar state to turbulence.
For simplicity, we used a reaction coordinate affine in spatially averaged kinetic energy
\begin{equation}
\Phi(t)=\frac{E_k(t)-E_{\mathcal{A}}}{E_{\mathcal{B}}}\,.\label{defcoord}
\end{equation}
This reaction coordinate is parameterised by two specific values of kinetic energy $E_{\mathcal{A}}$ and $E_{\mathcal{B}}$ which depend indirectly on
the control parameters used in the AMS calculations: the Reynolds number, the energy injection rate and the shape of the noise spectrum.
The value $E_{\mathcal{A}}$ should be such that $E_k(t)\simeq E_{\mathcal{A}}$ and $\Phi(t)\simeq 0$
when the flow is near the first metastable state and fluctuates near the laminar state under the forcing.
The exact value of $E_{\mathcal{A}}$ is computed using time series of kinetic energy sampled at the control parameters of interest:
$E_{\mathcal{A}}$ is defined as the average of $E_k(t)$ conditioned on no transition occurring.
The  other parameter $E_{\mathcal{B}}$ is such that the reaction coordinate is one when the flow has reached the neighbourhood of the turbulent state.
Equation~(\ref{defcoord}) then implies that $E_{\mathcal{B}}+E_{\mathcal{A}}\simeq E_{k,\text{turb}}$.
The conditional average of the kinetic energy $E_{\mathcal{A}}$ will often be relatively small: the energy injection rate is small enough for our description
of the forced flow as two metastable states makes sense.
As a consequence, we chose $E_{\mathcal{B}}=0.06$ regardless of control parameters,
aside of a few specified cases where we used $E_{\mathcal{B}}=0.06-E_{\mathcal{A}}$.

We furthermore define the hypersurface $\mathcal{C}$ as
\begin{equation}
\mathcal{C}=\left\{\mathbf{u}\backslash \phi(\mathbf{u}) = \frac{\sigma}{E_{\mathcal{B}}} \right\}\,,
\end{equation}
where $\sigma$ is the standard deviation of $E_k(t)$ conditioned on no transition occurring, sampled from the same dataset as $E_{\mathcal{A}}$.
The rationale is that we have a rare occurrence of a reactive trajectory when the flow leaves the neighbourhood of the laminar state, that is to say when
when $E_k\ge E_{\mathcal{A}}+\sigma$.
The narrow region of phase space such that $E_k\le E_{\mathcal{A}}+\sigma$ can then be viewed as the neighbourhood of the first metastable state, as long as $\sigma$ is small enough.

While some AMS computations used \emph{ad hoc} parameters sampled on a case by case basis,
for most values of the control parameters $R$ and $\epsilon$,
We did not resample $E_k(t)$ to compute $E_{\mathcal{A}}$ and $\sigma$.
Instead, we used the parametric dependence of $E_{\mathcal{A}}$ and $\sigma$ on $\beta$ to interpolate values.
Indeed, we can note that provided $\beta$ is large enough, or conversely $\epsilon$ small enough (the two being related, as seen in (Eq.~(\ref{defepsilon})))
for build up events to be rare,
we find that $E_{\mathcal{A}}$ and $\sigma$ are proportional to $\frac{1}{\beta}$.
Moreover, for the two distinct families of forcing, we find that  $E_{\mathcal{A}}$ and $\sigma$
are proportional to $\epsilon$ in good approximation
(Fig.~\ref{energy} (a) for forcing on component
and Fig.~\ref{energy} (b) for divergence free forcing).
This parametric dependence is in line with upper bounds of kinetic energy found for the linearised flow \cite{farrell1993stochastic}.
Note however that the coefficient of this linear dependence depend on the main class of forcing.
The proportionality coefficient in the divergence free case is larger than the one is the case of forcing on components by a factor more than 10.
We also note that within a class of forcing we do not get exactly the same proportionality coefficient between $E_{\mathcal{A}}$ and $\epsilon$.
While $E_{\mathcal{A}}$ and $\sigma$ grow with the energy injection rate, this simplified dependence is an approximation.
Owing to the different spectrum shape and the variety of process bringing energy
from the forcing to the receptive scale (direct forcing and transfer across scales),
the actual dependence is actually more complex, and the relation between $E_{\mathcal{A}}$ and $\epsilon$ is not universal.
We finally note that changing the  numerical resolution does not lead to a change in the way $E_{\mathcal{A}}$ and $\sigma$ depend on $\epsilon$.
Note that in more resolved simulations, we have pushed the cut-off of the spectrum to larger scales,
so that the forcing spectrum is different. Because of that and because the forcing is white in the wall normal direction,
a higher resolved simulation receives more energy
for the same value of $\beta$, as expected from equation~(\ref{defepsilon}).
We have also tested this relation with a larger cut-off size of $L_c=6$ (not shown here), which leads to a forcing that
can directly affect the most energetic scales.
The same linear dependence of $E_{\mathcal{A}}$ on $\frac{1}{\epsilon}$ is found, but in that case the values of $E$ are slightly above those computed
when $L_c=1$ in the forcing, possibly because large scales now do not only receive energy from non linear redistribution.

In order to propose the interpolated values, we adjust the linear dependence of $E_{\mathcal{A}}$
and $\sigma$ on $\beta$ in logarithmic scale for each set of parameters setting the forcing spectrum.
We fit the constants $d_E$, $o_E$, $d_S$ and $o_S$ from data such that
\begin{equation}
\log(E_{\mathcal{A}})=\log(\beta) d_E+o_E\,,\,\log(\sigma) =\log(\beta) d_S+o_S\,.\label{eqlin}
\end{equation}
For this matter, for $M$ values of $\beta_a\in\left[ 4\cdot 10^5 ,10^{10} \right]$, $1\le a \le M$,
we sample $E_a$ and $\sigma_a$ in time series where the flow does not transit
to turbulence (this is extremely rare except maybe for $\beta=4\cdot 10^5$).
The factors $d_E$, $o_E$, $d_S$ and $o_S$ are obtained using the explicit
formula for linear regression by affine functions (Eq.~(\ref{eqlin})) by least squares
using the $M$ sampled pairs of values (Fig.~\ref{energy}).
We always have $d_E\simeq d_S\simeq -1$, owing to the proportionality relation between
$E_{\mathcal{A}}$, $\sigma$ and $\frac{1}{\beta}$ (as reported in table~\ref{sysparam}, \S~\ref{aparam}). Using equation~(\ref{eqlin}),
we can interpolate/extrapolate $\log(E_{\mathcal{A}})$ and $\log(\sigma)$ for most values of $\beta$,
we then exponentiate.

We give the value used to define the sets  $\mathcal{A}$, $\mathcal{C}$ and $\mathcal{B}$ for each set of simulations in appendix~\ref{aparam}.
This comprises of specific cases where we will give $E_{\mathcal{A}}$, $E_{\mathcal{B}}$
and $\sigma$ and generic cases where we give $d_E$, $o_E$, $d_S$ and $o_S$, which mostly concern simulations performed in the largest systems.
We finally note that in all considered case, the values of $E_{\mathcal{A}}$ and $\sigma$ are small enough for the evolution of the
forced flow near the laminar base flow as small fluctuations around one metastable state.

\subsubsection{Direct Numerical simulations of transitions}\label{detdns}

Along with the AMS computations, we perform Direct Numerical Simulations (DNS) of the build up, following a procedure we systematically use
\cite{rolland2015statistical,lucente2022coupling,rolland2022collapse}.
These DNS consist in systematically repeating the first stage of AMS computations, without the additional mutation selection.
Using the same definition of sets  $\mathcal{A}$, $\mathcal{B}$ and hypersurface $\mathcal{C}$,
we start these simulations at $\mathbf{u}=\mathbf{0}$, $p=0$ in the whole domain, let them evolve until they reach $\mathcal{C}$
and then stop them when they reach either $\mathcal{A}$ or $\mathcal{B}$.
The proportion of trajectories that reach $\mathcal{B}$ before $\mathcal{A}$ provides an unbiased estimate of the crossing probability.
Meanwhile the average of the duration of trajectories that reach $\mathcal{B}$ before $\mathcal{A}$ provides an
unbiased estimate of the average duration of reactive trajectories.
These quantities are used to validate AMS computation and detect possible problems up
up to a given degree of rarity of the event.
The estimates rapidly become much more costly than AMS computations as $\beta$ is increased, because
the proportion of trajectories that go straight back to $\mathcal{A}$ becomes almost $1$,
in agreement with the description of the forced flow as having two narrow mulistable states.
For very rare events other tests (mentioned in \S~\ref{aconv}) are used to probe the validity of computations.

\subsubsection{Methodology for bisections}\label{metbis}

We expect that in the relaxation path of noise induced reactive trajectories, the flow crosses the separatrix
between the laminar and the turbulent basins of attractions of the deterministic system.
This supposition would be in line with the statistical physics description of bistability under a vanishing noise.
We will first test this by performing relaxation simulations using states taken on a reactive trajectory.
In practice this consists in using states $\left(\begin{matrix}\mathbf{u} (t_l=l\delta t) \\ p(t_l=l\delta t)\end{matrix}\right)$
sampled every $\delta t>\Delta t$  along a reactive trajectory and using it as an initial condition
for an unforced simulation ($\beta=0$ in equation~(\ref{nsf})) and letting it naturally evolve,
either toward the laminar flow $\mathbf{u} \rightarrow \mathbf{0}$,
or toward turbulence.
We can thus first detect a $t_s$ such that $\left(\begin{matrix}\mathbf{u}(t_s) \\ p(t_s)\end{matrix}\right)$ relaxes toward the laminar flow and
$\left(\begin{matrix}\mathbf{u}(t_{s}+\delta t)\\ p(t_s+\delta t)\end{matrix}\right)$
relaxes toward the turbulent flow.

Starting from these two fields, successive shootings of a bisection procedure are then performed \cite{toh2003periodic}.
During shooting $m\ge 1$, we initialise $\eta_0=\frac12$, $\eta_{\text{lam}}=1$ and $\eta_{\text{turb}}=0$.
For $0\le n\le 36$, we run unforced simulations for $t\ge t_{m,0}$ started from
\begin{equation}
\left(\begin{matrix}\mathbf{u}_{m,n} \\ p_{m,n}\end{matrix}\right)(t_{m,0})
=\eta_n \left(\begin{matrix}\mathbf{u}_{\text{lam},m} \\ p_{\text{lam},m}\end{matrix}\right)
+(1-\eta_{n})\left(\begin{matrix}\mathbf{u}_{\text{turb},m} \\ p_{\text{turb},m}\end{matrix}\right)\,.\end{equation}
During the first shooting $m=1$ we use
$ \left(\begin{matrix}\mathbf{u}_{\text{lam},1} \\ p_{\text{lam},1}\end{matrix}\right)
=\left(\begin{matrix}\mathbf{u}(t_s) \\ p(t_s)\end{matrix}\right)$
and $\left(\begin{matrix}\mathbf{u}_{\text{turb},1} \\ p_{\text{turb},1}\end{matrix}\right)
=\left(\begin{matrix}\mathbf{u}(t_{s}+\delta t)\\ p(t_s+\delta t)\end{matrix}\right)$.
We set $t_{1,0}=t_s+\frac{\delta t}{2}$ for presentation of the time series.
These simulations are run until
\begin{enumerate}[i)]
\item The flow laminarises, which is deemed to happen if the kinetic energy is below the threshold $E_k(t)\le 0.001$.
This threshold is chosen so that the flow has underwent extensive viscous decay when $E_k(t)= 0.001$ is reached,
so that the relaxation toward the saddle is not interrupted and no saddle is excluded. This plays a more important role in the largest domain of size $L_x\times L_z=36\times 24$ where
smaller values of $E_k(t)$ are found because non trivial flow occupies a smaller area in the domain.
In that case we set $\eta_{n+1}=\eta_n-\frac{1}{2^{n+1}}$ and $\eta_{\text{lam}}=\eta_n$.

\centerline{or}
\item The flow first reaches turbulence,  which is deemed to happen if the kinetic energy is above the threshold $E_k(t)\ge 0.07$.
This threshold is chosen large enough so that no saddle is mistaken for turbulent flow, for all domain sizes, but not so
large that it would not be reached by wall turbulence.
This plays a more important role in the smallest domain of size $L_x\times L_z=36\times 24$ where
the saddle has larger values of $E_k(t)$ than for other domain sizes.
In that case we set $\eta_{n+1}=\eta_n+\frac{1}{2^{n+1}}$ and $\eta_{\text{turb}}=\eta_n$.
\end{enumerate}
When the shooting is over at $n=36$, we prepare the initial conditions for the next shooting.
We set
\begin{align}\notag
\left(\begin{matrix}\mathbf{u}_{m,-}\\ p_{m,-}\end{matrix}\right)(t=t_{m,0})
=\eta_{\text{lam}}
\left(\begin{matrix}\mathbf{u}_{\text{lam},m} \\ p_{\text{lam},m}\end{matrix}\right)
+(1-\eta_{\text{lam}})\left(\begin{matrix}\mathbf{u}_{\text{turb},m} \\ p_{\text{turb},m}\end{matrix}\right)
\,,\\
\left(\begin{matrix}\mathbf{u}_{m,+} \\ p_{m,+}\end{matrix}\right)(t=t_{m,0})  =\eta_{\text{turb}}
\left(\begin{matrix}\mathbf{u}_{\text{lam},m} \\ p_{\text{lam},m}\end{matrix}\right)
+(1-\eta_{\text{turb}})\left(\begin{matrix}\mathbf{u}_{\text{turb},m} \\ p_{\text{turb},m}\end{matrix}\right)\,.
\end{align}
We evolve them in parallel until a separation time $t_{m,\text{sep}}$.
Setting $E_{k,m,-}(t)$ and $E_{k,m,+}(t)$ the spatial averaged kinetic energy of these two fields at $t$,
this separation time is the first time multiple of $\delta t$ such that
$\frac{\left| E_{k,m,-}(t_{m,\text{sep}})-E_{k,m,+}(t_{m,\text{sep}}) \right|}{E_{k,m,-}(t_{m,\text{sep}})}>10^{-2}$.
We then define $\left(\begin{matrix}\mathbf{u}_{\text{turb},m+1} \\ p_{\text{turb},m+1}\end{matrix}\right)
=\left(\begin{matrix}\mathbf{u}_{m,-}\\ p_{m,-}\end{matrix}\right)(t=t_{m,\text{sep}})$ and
$\left(\begin{matrix}\mathbf{u}_{\text{turb},m+1} \\ p_{\text{turb},m+1}\end{matrix}\right)
=\left(\begin{matrix}\mathbf{u}_{m,+}\\ p_{m,+}\end{matrix}\right)(t=t_{m,\text{sep}})$.
We also set $t_{m+1,0}=t_{m,\text{sep}}$.
We run shootings until we have reached an unstable fixed point and spent a long enough time near it, or
we have reached an unstable limit cycle and enough periods have elapsed, for a precise computation of its properties.

For the presentation of results, we will concatenate the successive times series of $E_{k,+}$ and $E_{k,-}$
as the result of the successive shootings.
We will also present velocity fields $\mathbf{u}_{\text{lam},m}$
or $\mathbf{u}_{\text{turb},m}$ for $m$ large enough such that they are close enough to either an unstable fixed point or an unstable limit cycle.
In practice we used $m=8$ for all domain sizes except for the largest at $L_x\times L_z=36\times 24$ where we used $m=16$.

On these bisections, we use the same spatial and temporal resolution as in the AMS and DNS computations generating
$\left(\begin{matrix}\mathbf{u}(t_s) \\ p(t_s)\end{matrix}\right)$ and $\left(\begin{matrix}\mathbf{u}(t_s+\delta t) \\ p(t_s+\delta t)\end{matrix}\right)$.
We used $\delta t=1$ for all cases, except in the MFU type system at higher resolution where we used $\delta t=2$.

\section{Paths toward turbulence}\label{spath}

In that section we describe the paths followed by the flow from the neighbourhood
of the laminar state to wall turbulence under forcing, computed by means of AMS.
In the first subsection~\ref{svisup}, We will first describe full velocity fields on selected paths (\S~\ref{Paths}),
in the last state at the last step (\S~\ref{slast}) and after bisections (\S~\ref{sbis}).
In the second subsection~\ref{sstatp}, we will place these observations in the statistical description of all the paths we have computed.

\subsection{Velocity fields in selected examples of transition paths}\label{svisup}

We will first visualise selected paths to examine the physical mechanism
of crossing from laminar to turbulent flow under forcing.
The visualisations will be complemented by examination of the last states at the last step
as well as bisections started from velocity fields taken from the reactive trajectories.
We will investigate whether the trajectories follow the classical instanton phenomenology,
whether they cross a separatrix and in that case, whether they cross it near an unstable fixed point.

\subsubsection{Description of fields on paths}\label{Paths}

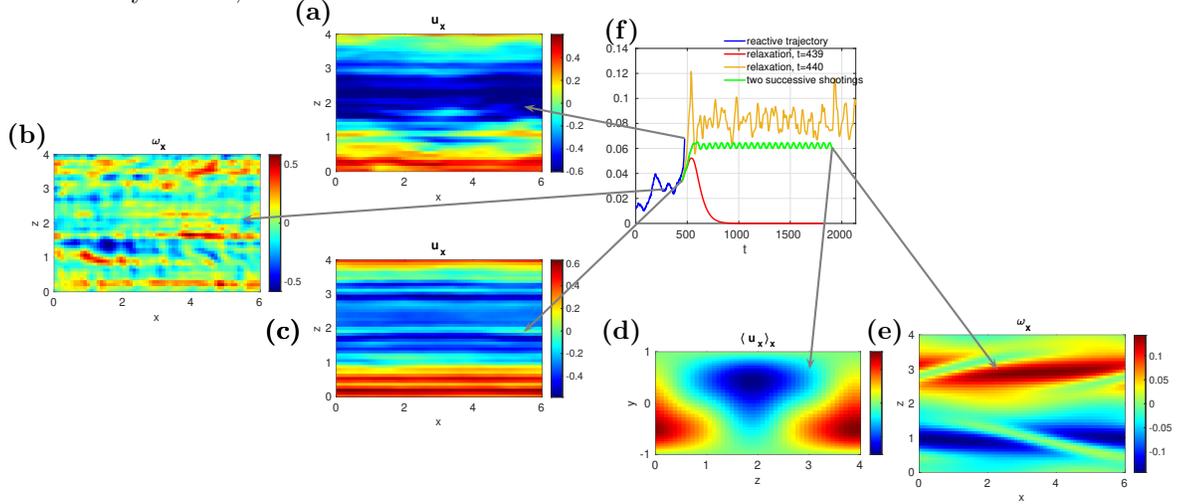
\begin{figure*}[!htbp]
\centerline{
\begin{pspicture}(12,7)
\rput(9.5,5.5){\includegraphics[width=4cm,clip]{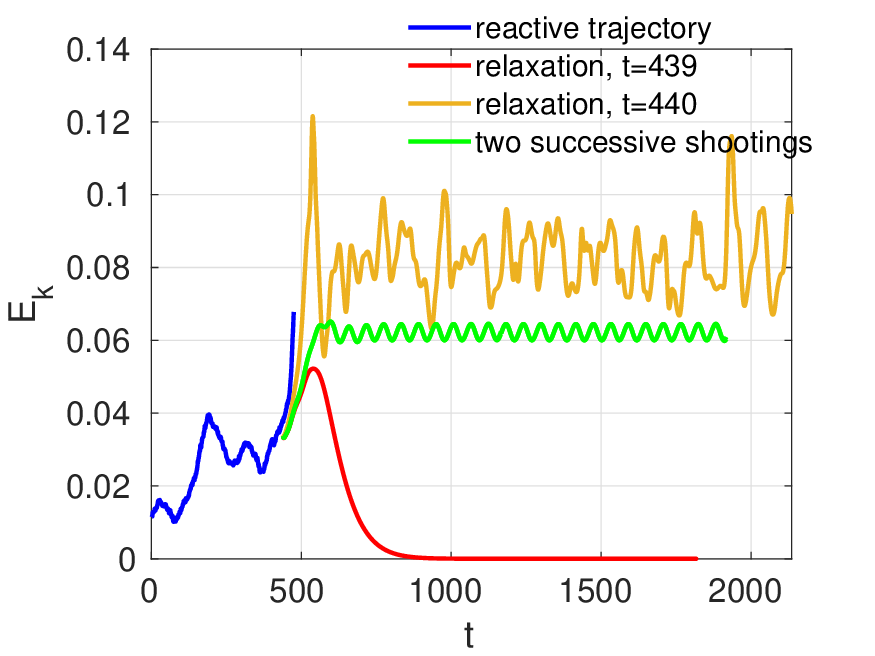}}
\rput(8,7){\textbf{(f)}}
\rput(5.75,6){\includegraphics[width=4cm,clip]{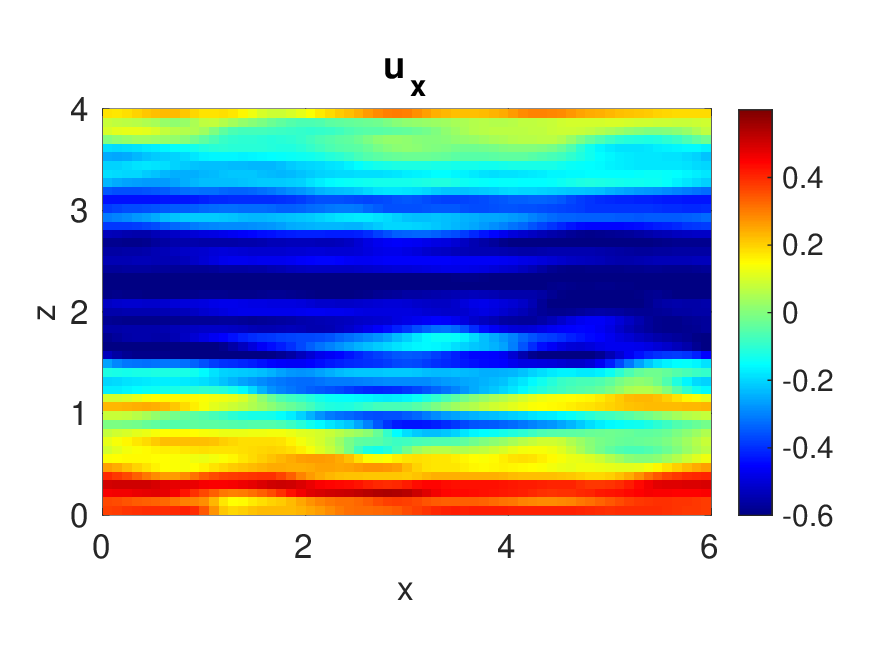}}
\rput(3.9,7.25){\textbf{(a)}}
\rput(5.75,3){\includegraphics[width=4cm,clip]{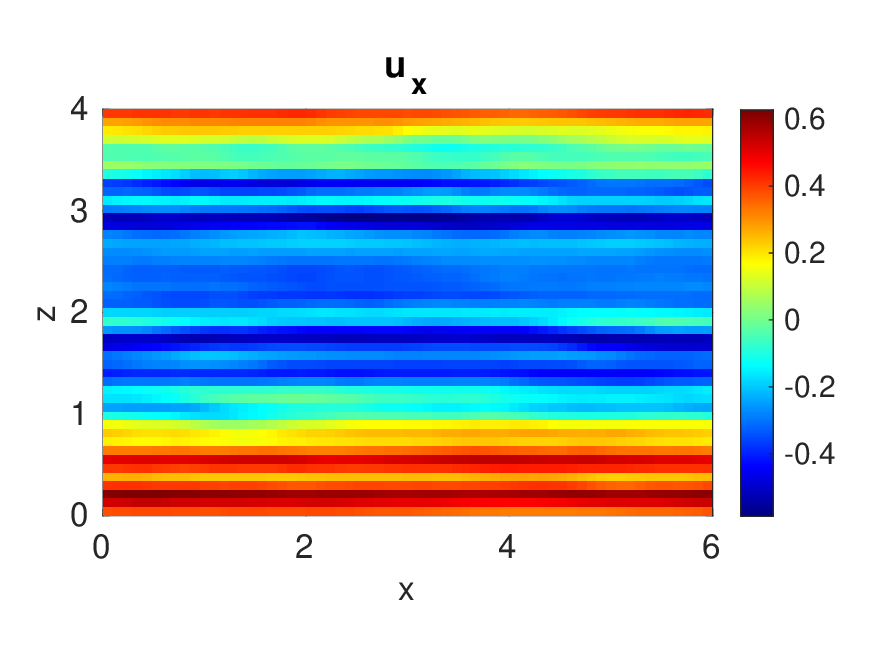}}
\rput(2,4.5){\includegraphics[width=4cm,clip]{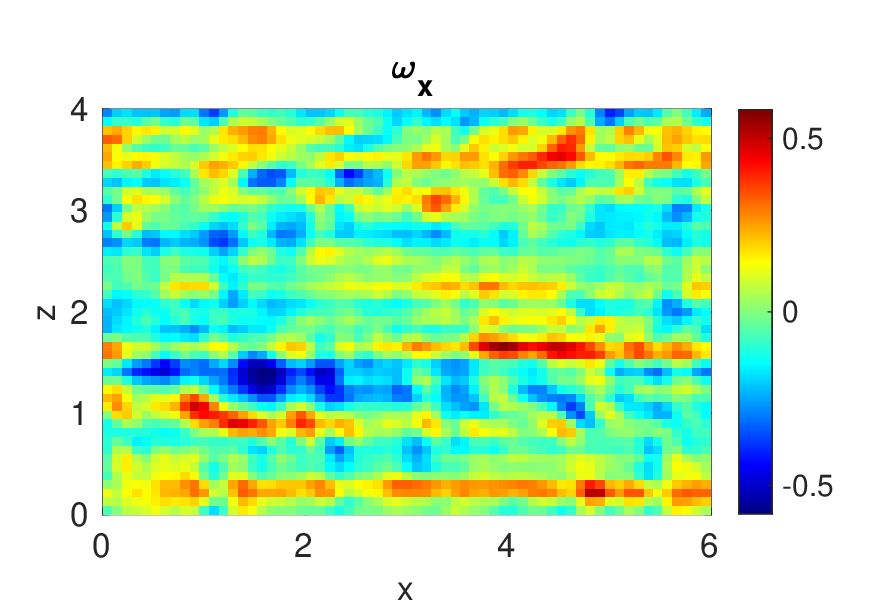}}
\rput(0.1,5.6){\textbf{(b)}}
\rput(3.5,3){\textbf{(c)}}
\rput(10,2){\includegraphics[width=4cm,clip]{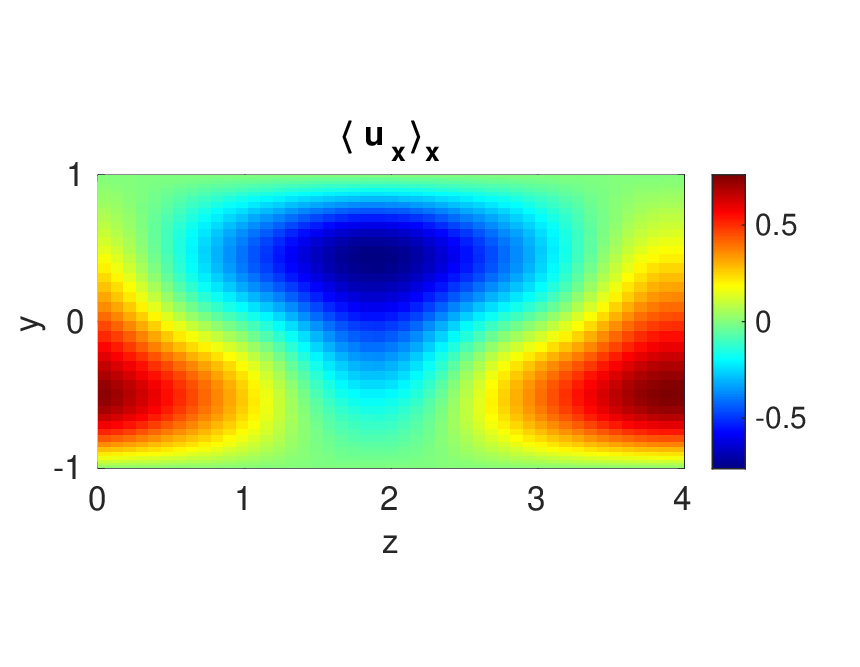}}
\rput(3.5,3){\textbf{(c)}}
\rput(13.5,2){\includegraphics[width=4cm,clip]{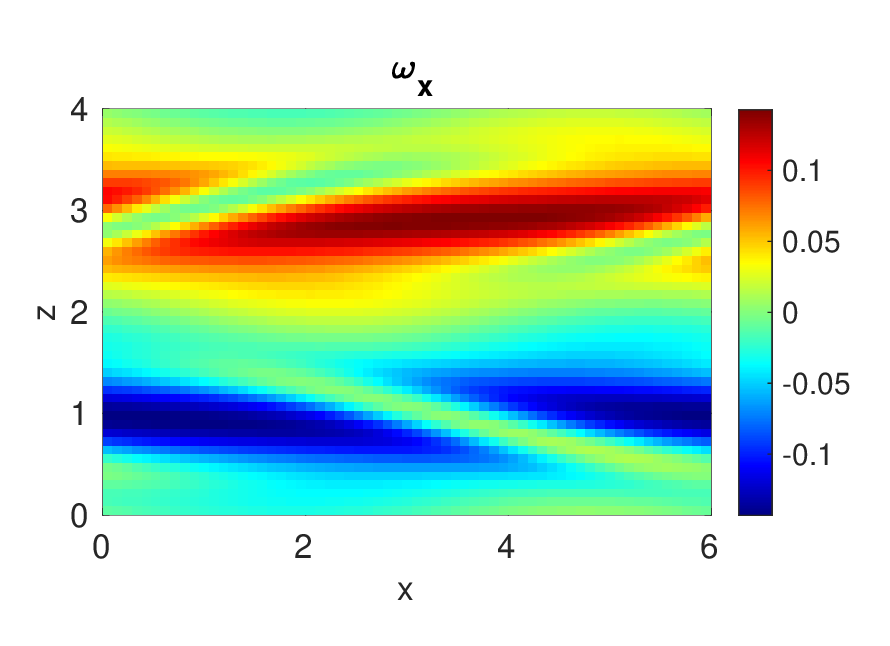}}
\rput(8,3){\textbf{(d)}}
\rput(11.5,3){\textbf{(e)}}
\psline[linecolor=gray]{->}(8.84,5.58)(6.7,6)
\psline[linecolor=gray]{->}(8.80,5.01)(6.7,3)
\psline[linecolor=gray]{->}(8.57,4.9)(2.95,4.5)
\psline[linecolor=gray]{->}(10.8,5.45)(10.5,2.5)
\psline[linecolor=gray]{->}(10.8,5.45)(13,2.5)
\end{pspicture}
}
\caption{View of a reactive trajectory for a system of size $L_x\times L_z=6\times 4$ at Reynolds number $R=500$
computed at high resolution
under a divergence free forcing at $\beta =5.5\cdot 10^4$,
along with the result of relaxations and bisections started from this trajectory.
(a) Colour levels  of the streamwise velocity in the plane $y=0$ when the flow has reached the turbulent state.
(b) Colour levels of the streamwise vorticity in the plane $y=0$ at $t=440$ (first time step relaxing to turbulence).
(c) Colour levels of the streamwise velocity field in the plane $y=0$ at $t=440$.
(d) Streamwise average of the streamwise velocity field in the  $y-z$ plane at a point on the unstable limit cycle reached after from the bisections.
(e) Colour levels of the streamwise vorticity field in the plane $y=0$ at a point on the unstable limit cycle reached after from the bisections.
(f) Time series of the kinetic energy during a reactive trajectory (blue), during two relaxation from states taken on the reactive trajectory,
at $t=439$ (last relaxation to laminar state, red), $t=440$ (first relaxation to turbulent state, orange)
and on the bisections started from these two successive velocity fields (green).}
\label{traj_6_4_hres}
\end{figure*}

\begin{figure*}[!htbp]
\centerline{
\begin{pspicture}(15,8)
\rput(11.5,6.5){\includegraphics[width=8cm]{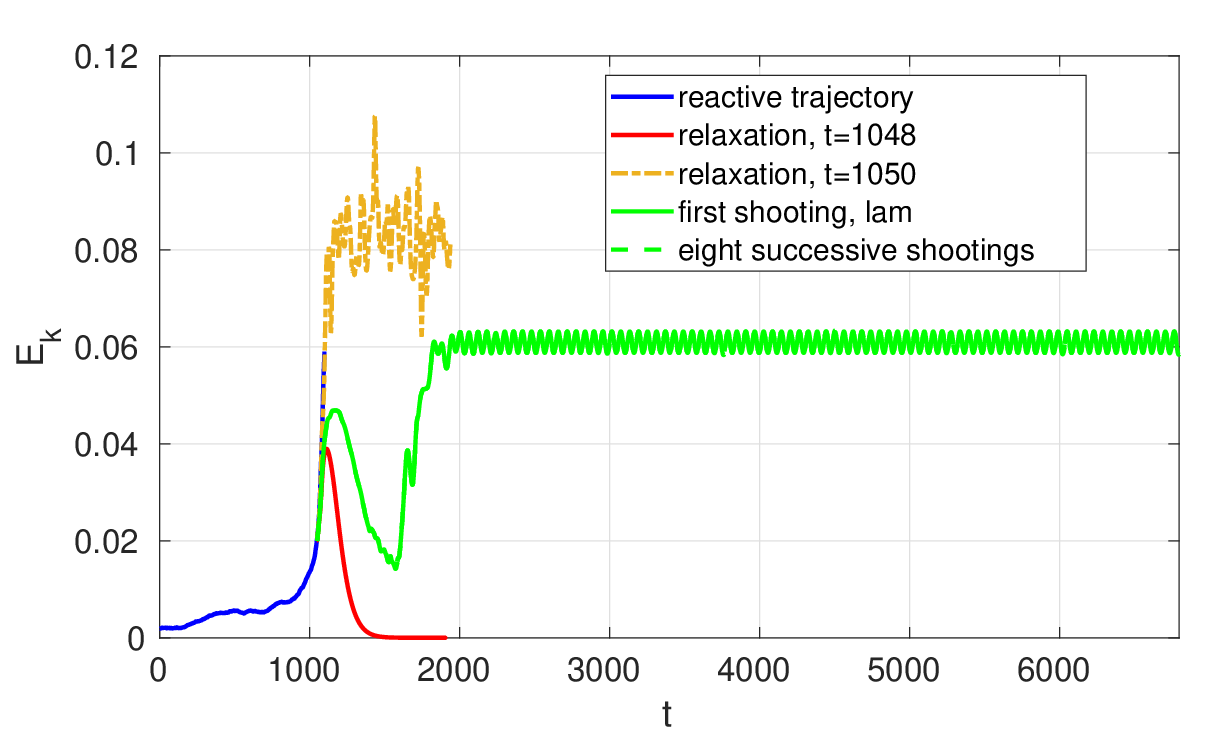}}
\rput(7.75,8.5){\textbf{(e)}}
\rput(5.5,6){\includegraphics[width=4cm,clip]{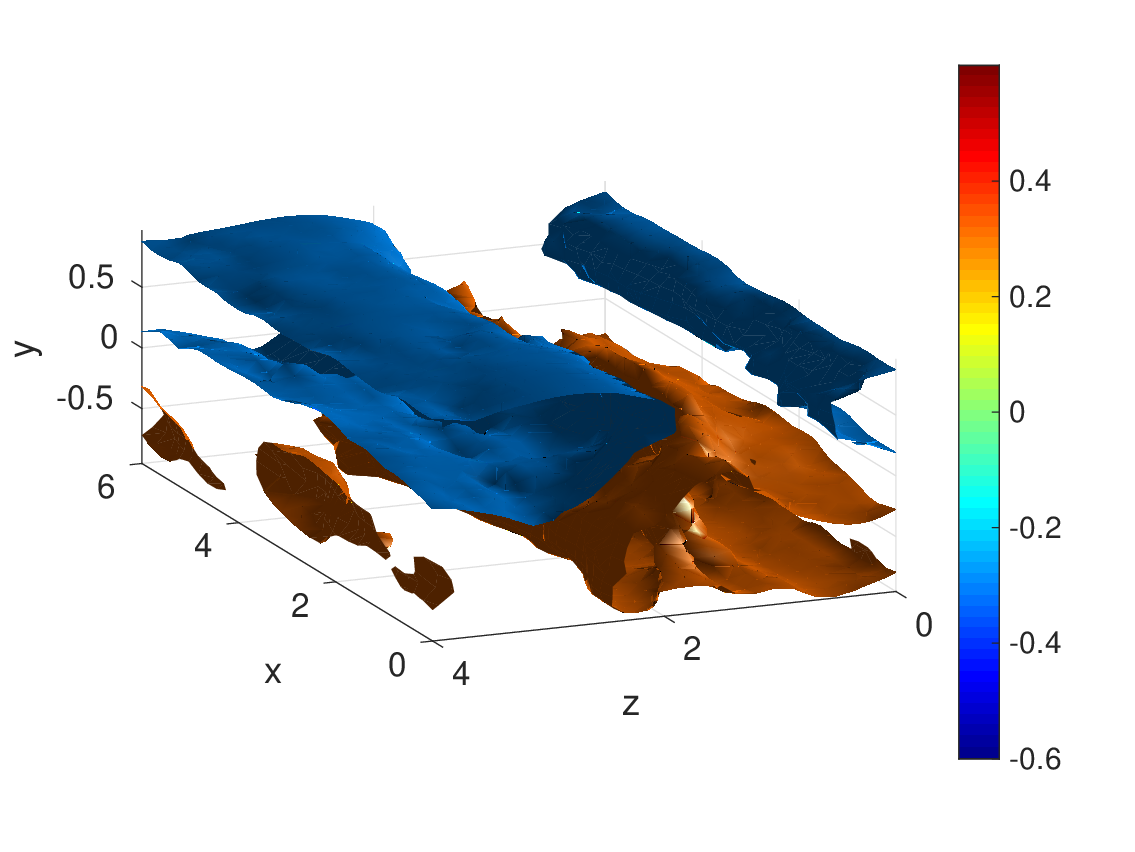}}
\rput(3.75,6.9){\textbf{(a)}}
\rput(5.75,3){\includegraphics[width=4cm,clip]{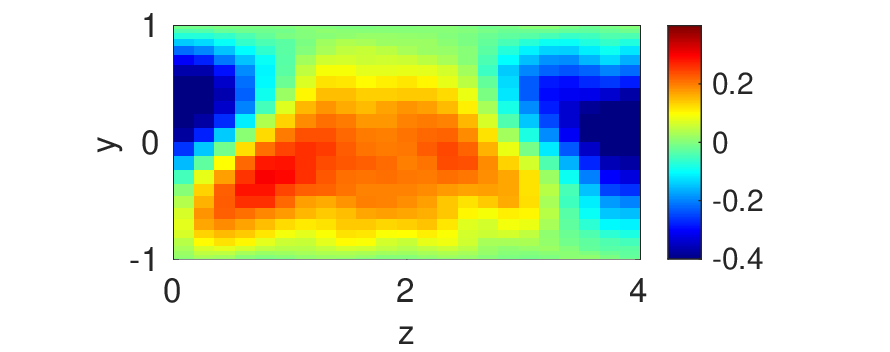}}
\rput(3.75,3.8){\textbf{(b)}}
\rput(10,2){\includegraphics[width=4cm,clip]{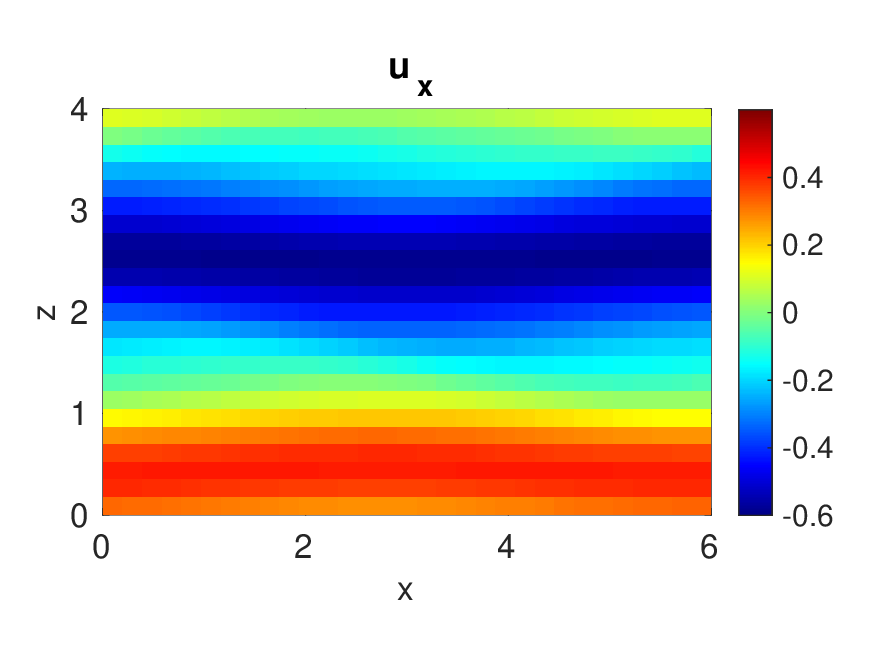}}
\rput(8,2.9){\textbf{(c)}}
\rput(15,2){\includegraphics[width=4cm,clip]{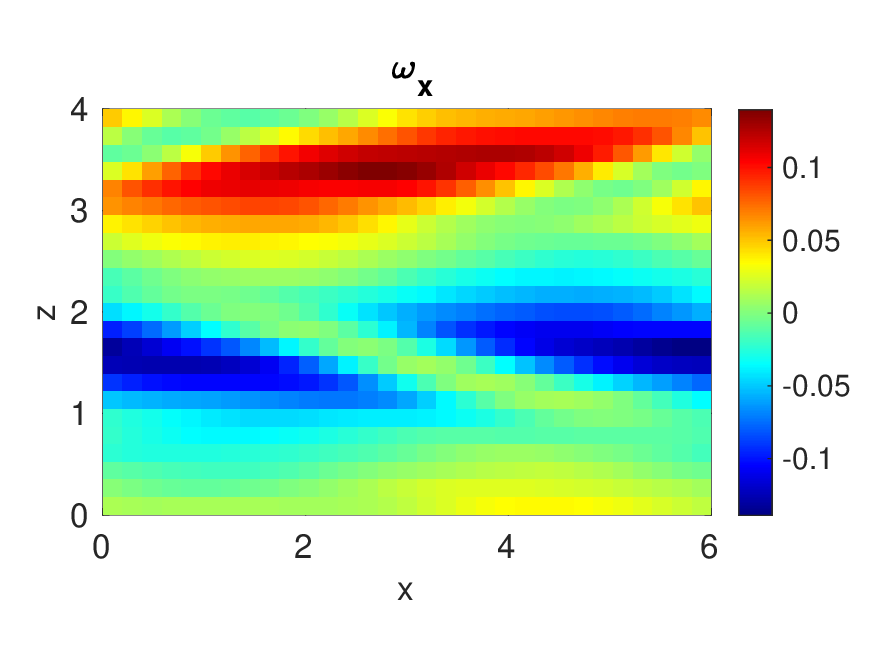}}
\rput(13,2.9){\textbf{(d)}}
\psline[linecolor=gray]{->}(9.68,7.2)(6.7,6)
\psline[linecolor=gray]{->}(9.58,5.4)(7,2.65)
\psline[linecolor=gray]{->}(12.2,6.7)(10.5,2.5)
\psline[linecolor=gray]{->}(12.2,6.7)(14.5,2.5)
\end{pspicture}
}
\caption{
View of a reactive trajectory, relaxations and bisections started from said trajectory,
for a system of size $L_x\times L_z=6\times 4$ at Reynolds number $R=500$
under a forcing on component $x$ at $\beta=1.1\cdot 10^5$.
(a) Isovalues $u_x=-0.3$ (blue) and $u_x=0.3$ (red) of streamwise velocity in the whole domain when the flow has reached the turbulent state.
(b) Colour levels of the streamwise average of the streamwise velocity field in the vertical plane $z-y$
 for the state relaxing to laminar flow used in the bisection.
(c) Colour levels of the streamwise velocity field in the $y=0$ plane when bisections have reached the neighbourhood of the unstable limit cycle.
(d) Colour levels of the streamwise vorticity field in the $y=0$ plane when bisections have reached the neighbourhood of the unstable limit cycle.
(e) Time series of the kinetic energy during a reactive trajectory (blue), during two relaxation from states taken on the reactive trajectory,
at $t=1048$ (last relaxation to laminar state, red), $t=1050$ (first relaxation to turbulent state, orange)
and on the dichotomies started from these two successive velocity fields (green).}
\label{traj_6_4_wcx}
\end{figure*}
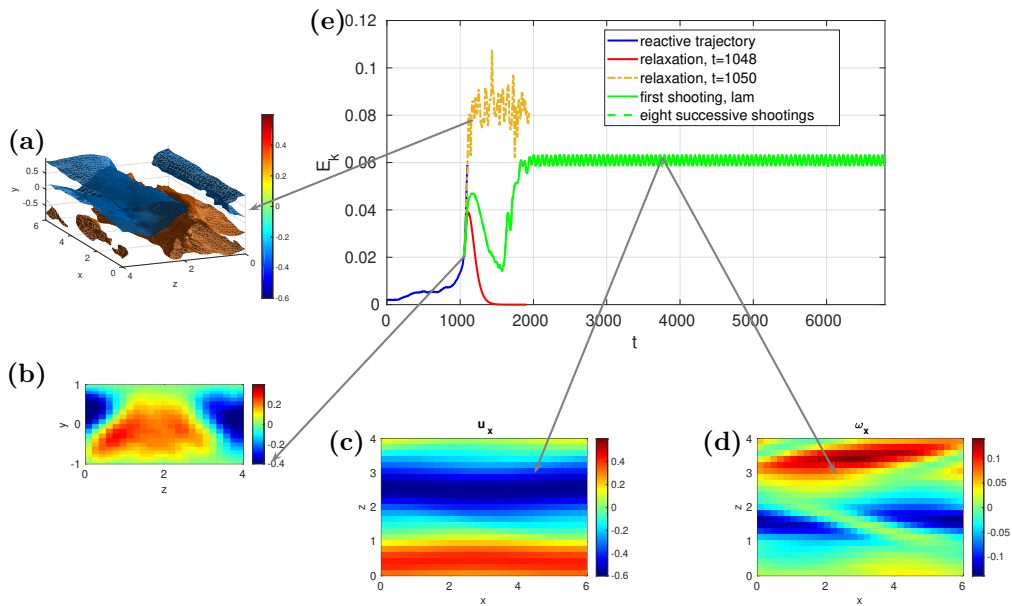

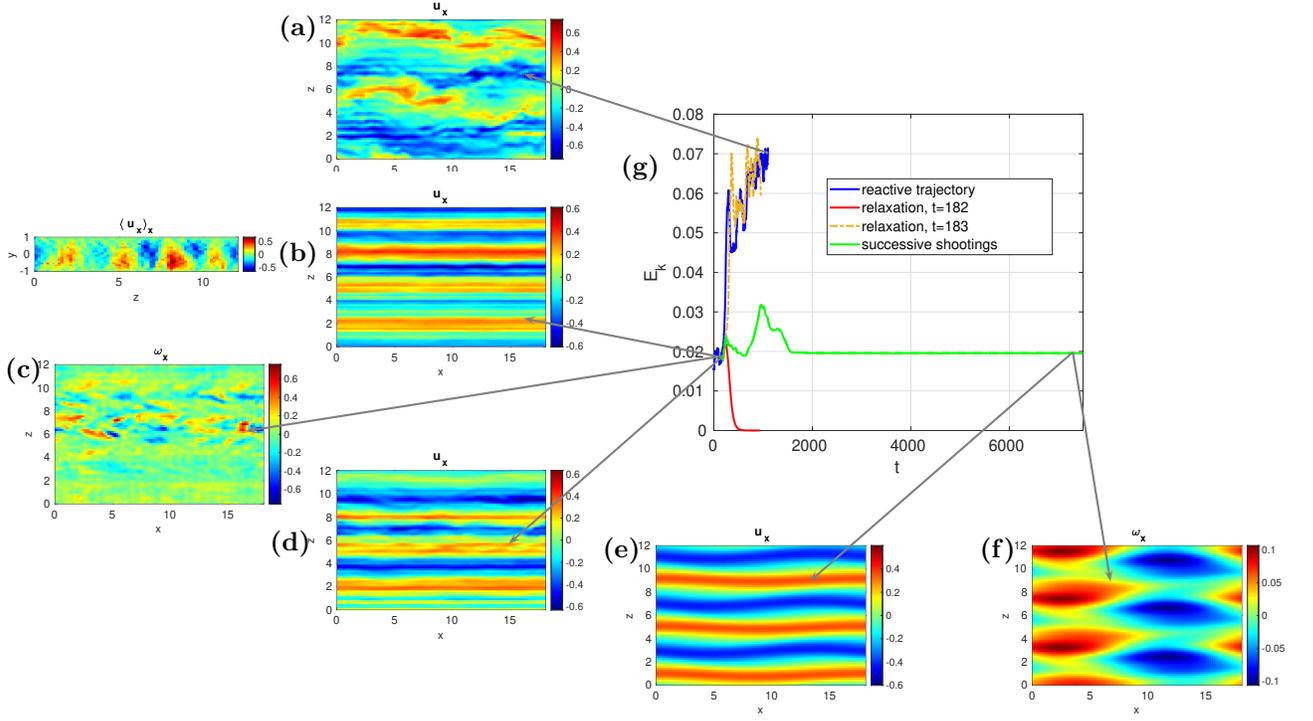
\begin{figure*}[!htbp]
\centerline{
\begin{pspicture}(15,10)
\rput(11.5,6.5){\includegraphics[width=6.5cm]{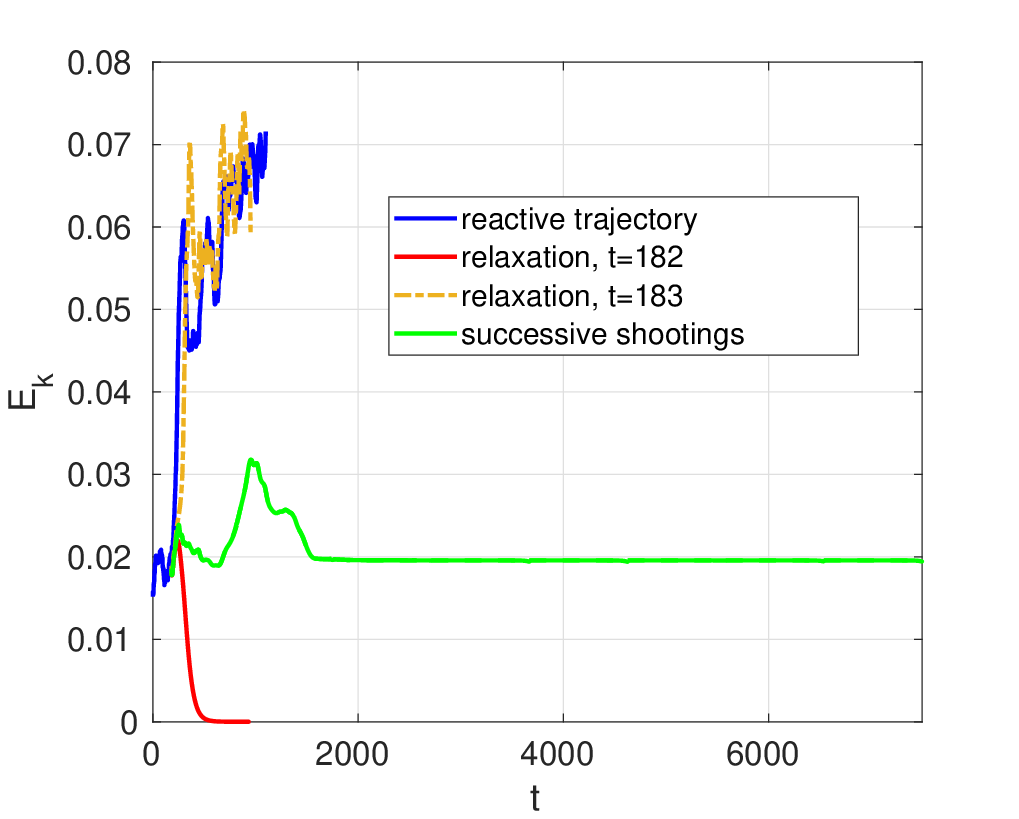}}
\rput(8.25,8){\textbf{(g)}}
\rput(5.75,9){\includegraphics[width=4cm,clip]{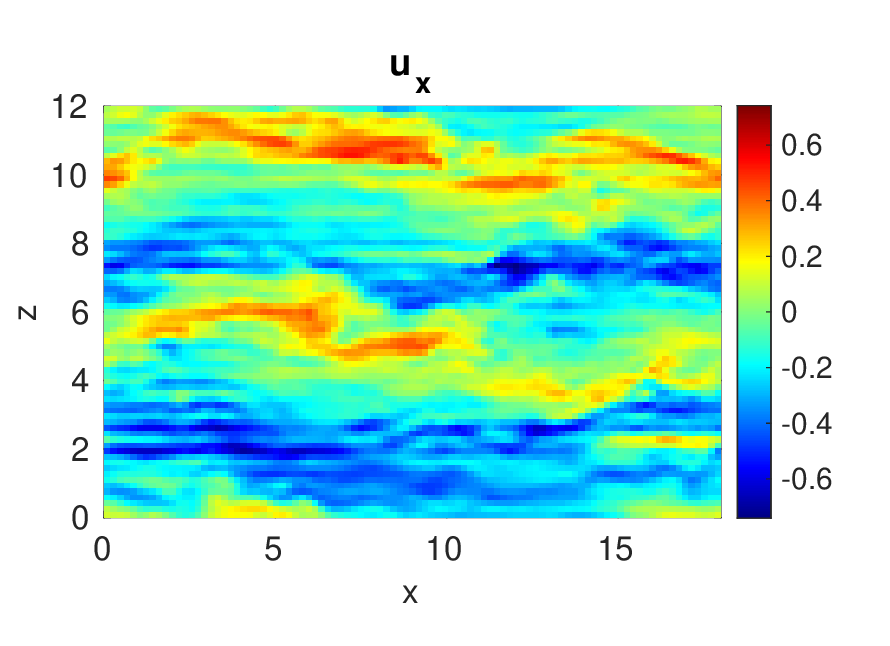}}
\rput(3.7,9.85){\textbf{(a)}}
\rput(5.75,6.5){\includegraphics[width=4cm,clip]{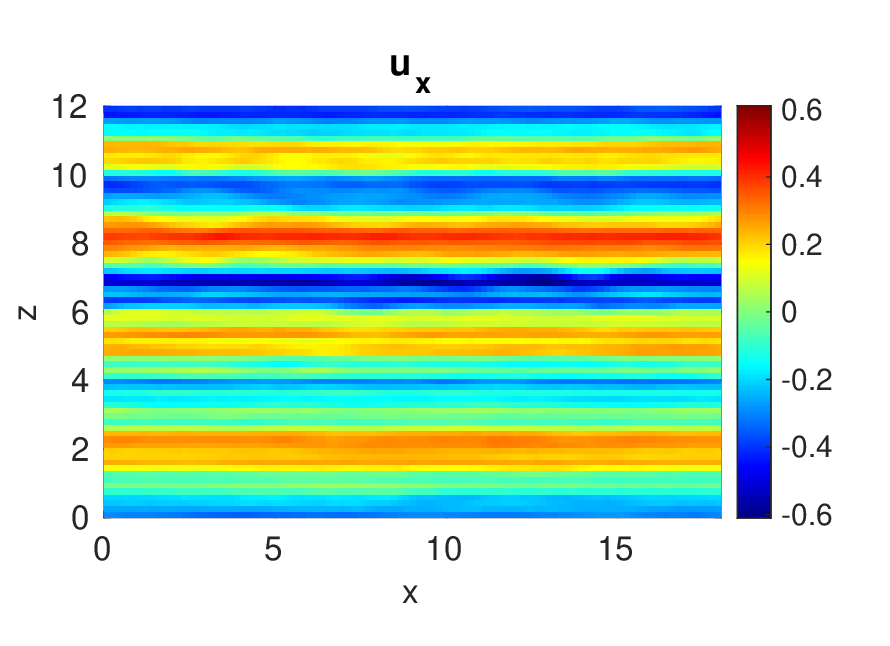}}
\rput(1.75,6.8){\includegraphics[width=4cm,clip]{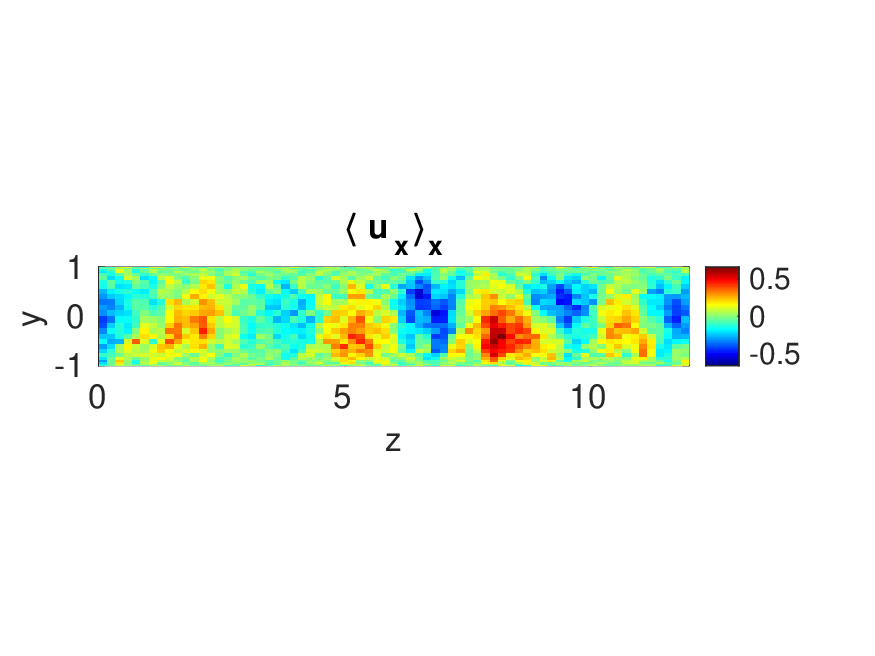}}
\rput(5.75,3){\includegraphics[width=4cm,clip]{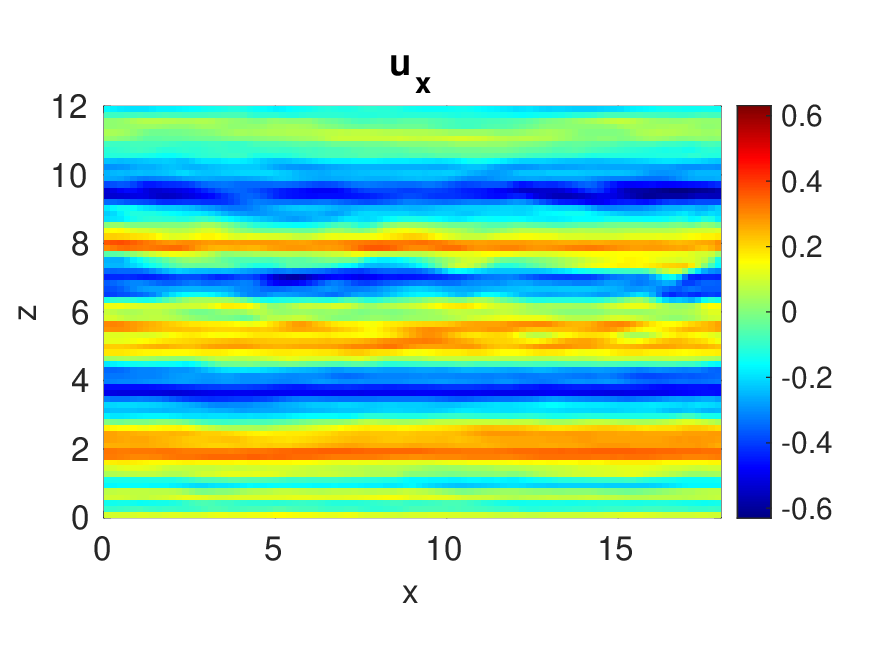}}
\rput(2,4.5){\includegraphics[width=4cm,clip]{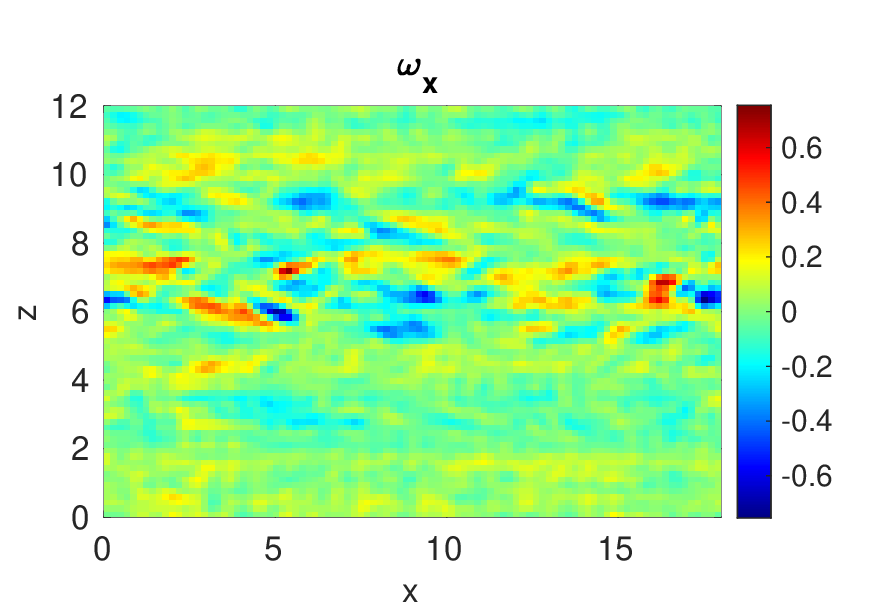}}
\rput(3.7,6.85){\textbf{(b)}}
\rput(0.05,5.3){\textbf{(c)}}
\rput(3.6,3){\textbf{(d)}}
\rput(10,2){\includegraphics[width=4cm,clip]{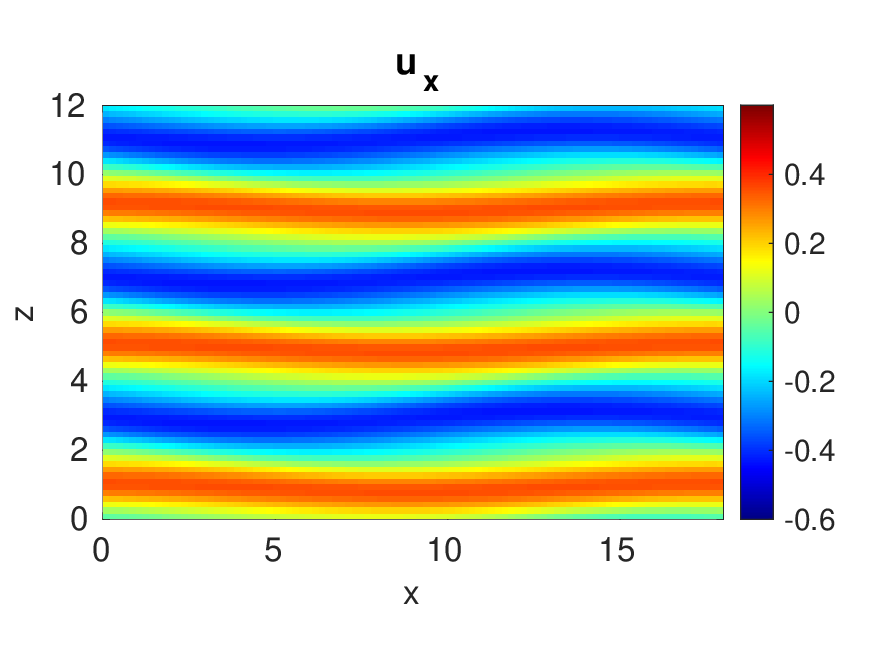}}
\rput(8,2.9){\textbf{(e)}}
\rput(15,2){\includegraphics[width=4cm,clip]{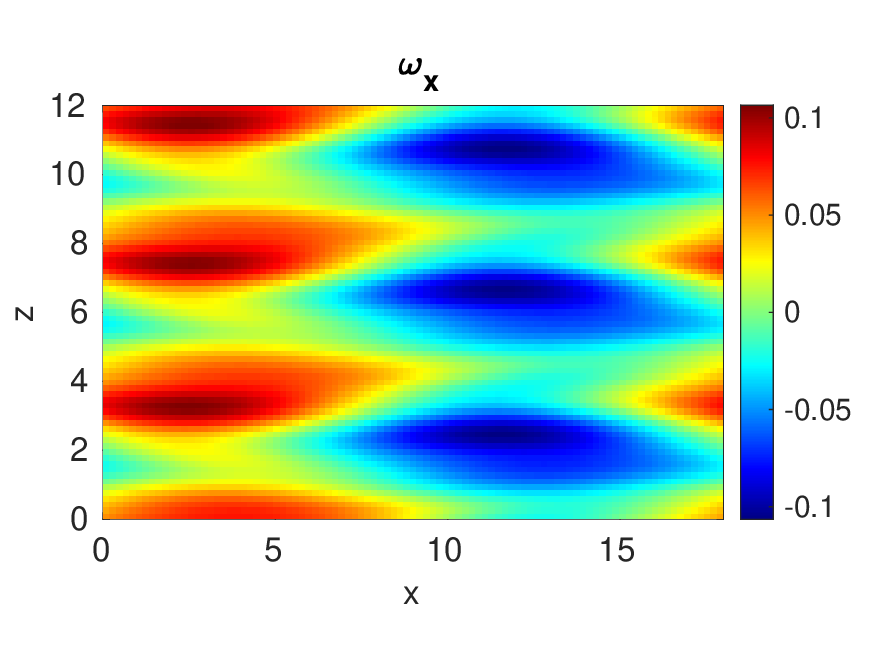}}
\rput(13,2.9){\textbf{(f)}}
\psline[linecolor=gray]{->}(9.95,8.2)(6.7,9.25)
\psline[linecolor=gray]{->}(9.25,5.5)(6.7,6)
\psline[linecolor=gray]{->}(9.35,5.5)(3,4.5)
\psline[linecolor=gray]{->}(9.35,5.5)(6.5,3)
\psline[linecolor=gray]{->}(14,5.55)(10.5,2.5)
\psline[linecolor=gray]{->}(14,5.55)(14.5,2.5)
\end{pspicture}
}
\caption{View of a reactive trajectory for a system of size $L_x\times L_z=18\times 12$ at Reynolds number $R=500$
under a divergence free forcing at $\beta=6.25\cdot 10^4$,
along with the relaxations and bisections started from this trajectory.
(a) Colour levels  of the streamwise velocity  in the plane $y=0$ when the flow has reached the turbulent state.
(b) Right: colour levels  of streamwise velocity in the plane $y=0$ at $t=50$ in the early stage of the reactive trajectory,
left: colour levels of the streamwise average of the streamwise velocity in a $y-z$ plane at the same time.
(c) Colour levels of the streamwise vorticity in the plane $y=0$ at time $t=182$
(last time step relaxing to laminar flow) from which bisections are started.
(d) Colour levels of the streamwise velocity in the plane $y=0$ at time $t=182$.
(e) Colour levels of the streamwise velocity in the plane $y=0$ when the bisections have reached a saddle.
(f) Colour levels of the streamwise vorticity in the plane $y=0$ when the bisections have reached a saddle.
(g) Time series of the kinetic energy during the reactive trajectory (blue), during two relaxations from states taken on the reactive trajectory,
at $t=182$ (last relaxation to laminar state, red), $t=183$ (first relaxation to turbulent state, orange)
and on the bisections started from these two successive velocity fields (green).}
\label{traj_18_12}
\end{figure*}

\begin{figure*}[!htbp]
\centerline{
\begin{pspicture}(17,11)
\rput(5.75,9){\includegraphics[width=4cm,clip]{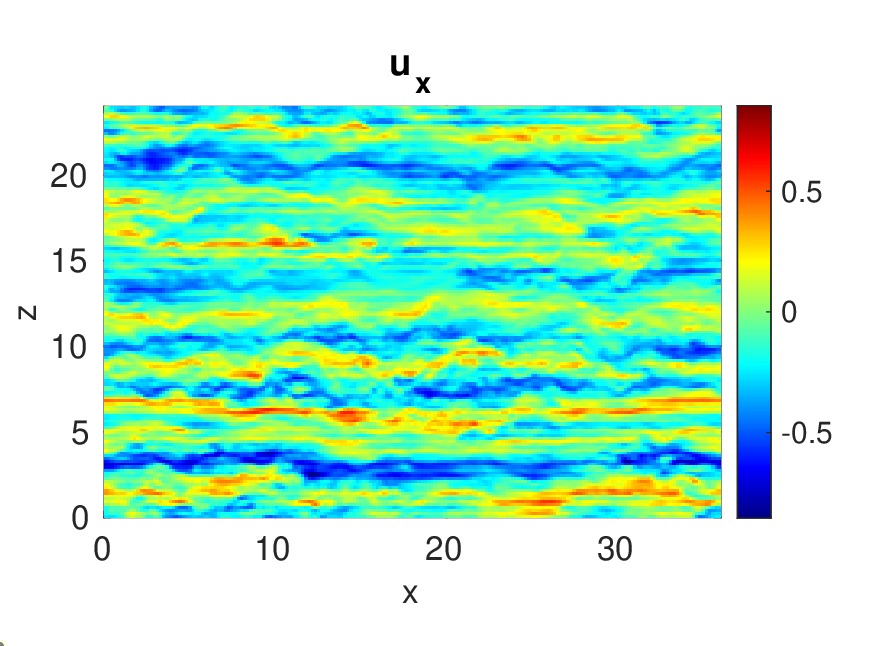}}
\rput(3.7,9.85){\textbf{(a)}}
\rput(5.75,6.5){\includegraphics[width=4cm,clip]{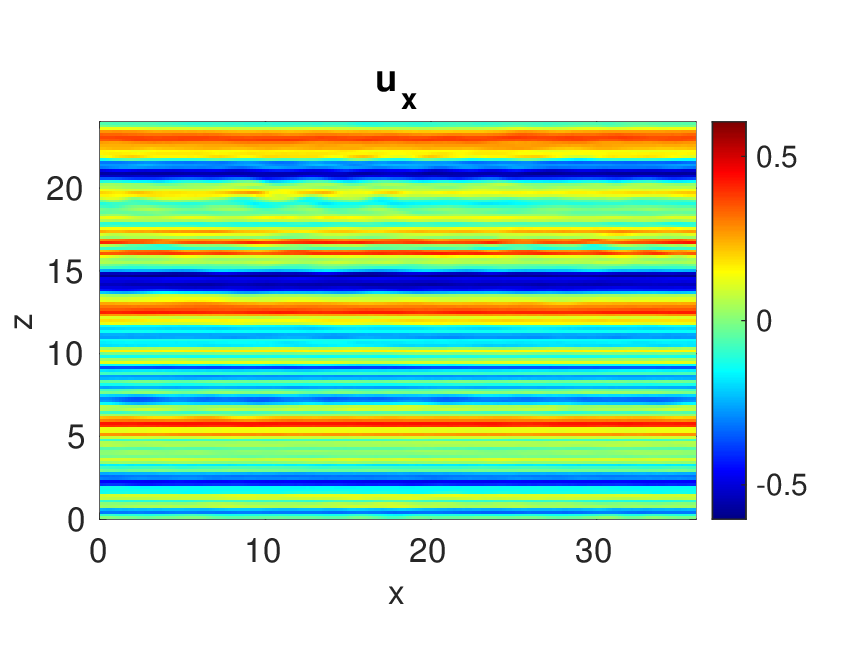}}
\rput(1.75,6.8){\includegraphics[width=4cm,clip]{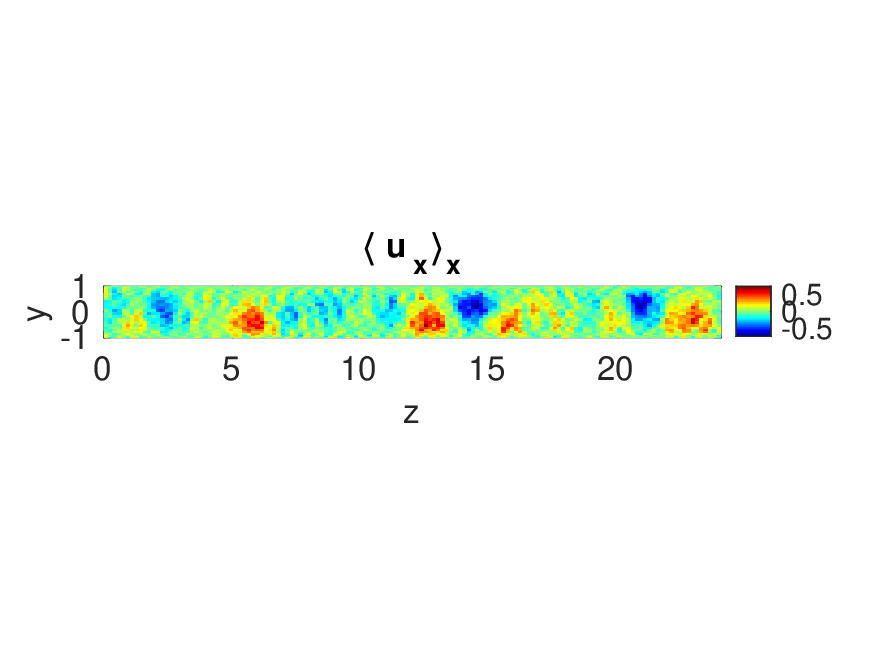}}
\rput(3.7,6.85){\textbf{(b)}}
\rput(5.75,3){\includegraphics[width=4cm,clip]{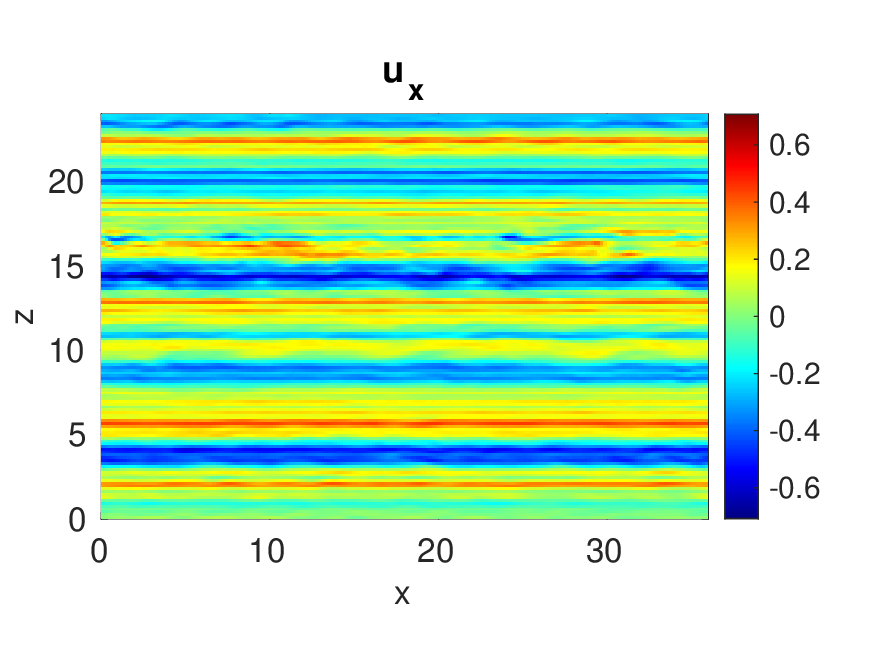}}
\rput(2,4.5){\includegraphics[width=4cm,clip]{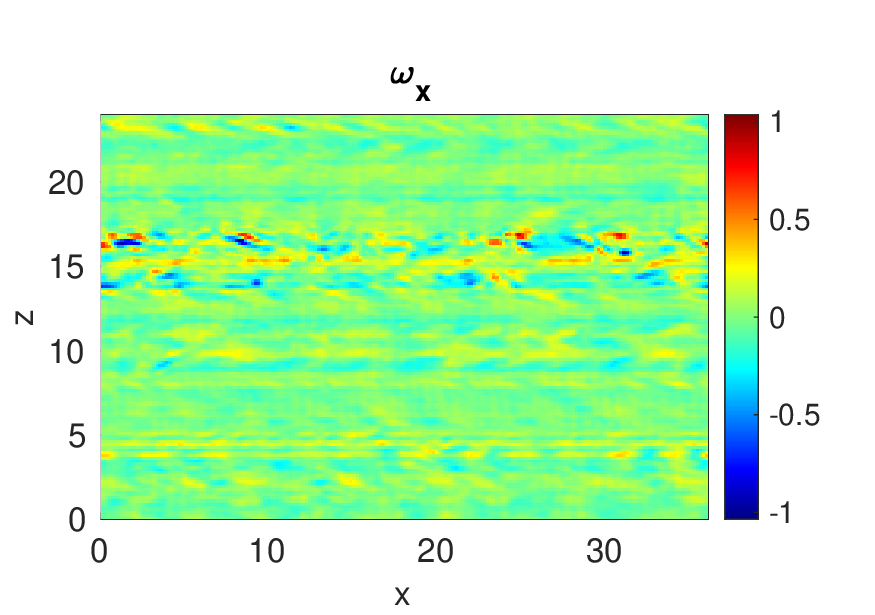}}
\rput(3.5,3){\textbf{(d)}}
\rput(0.05,5.25){\textbf{(c)}}
\rput(10,2){\includegraphics[width=4cm,clip]{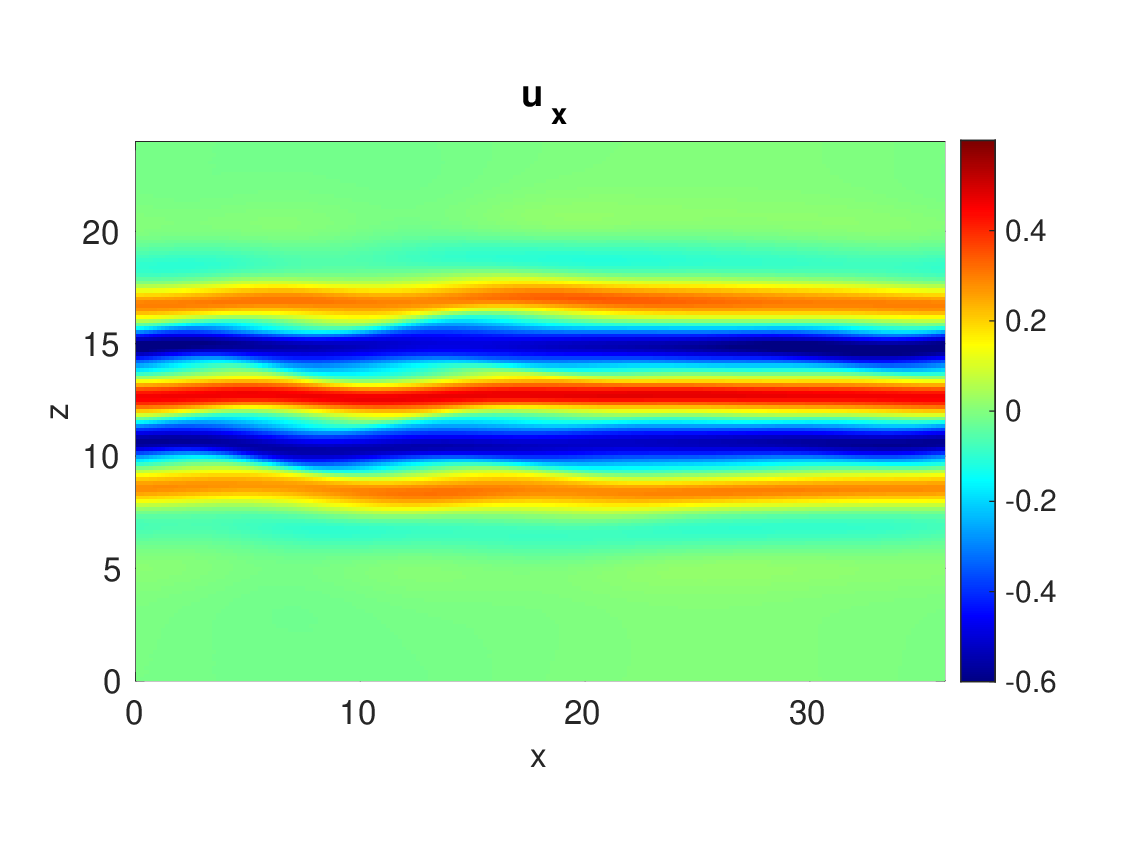}}
\rput(8,2.9){\textbf{(e)}}
\rput(15,2){\includegraphics[width=4cm,clip]{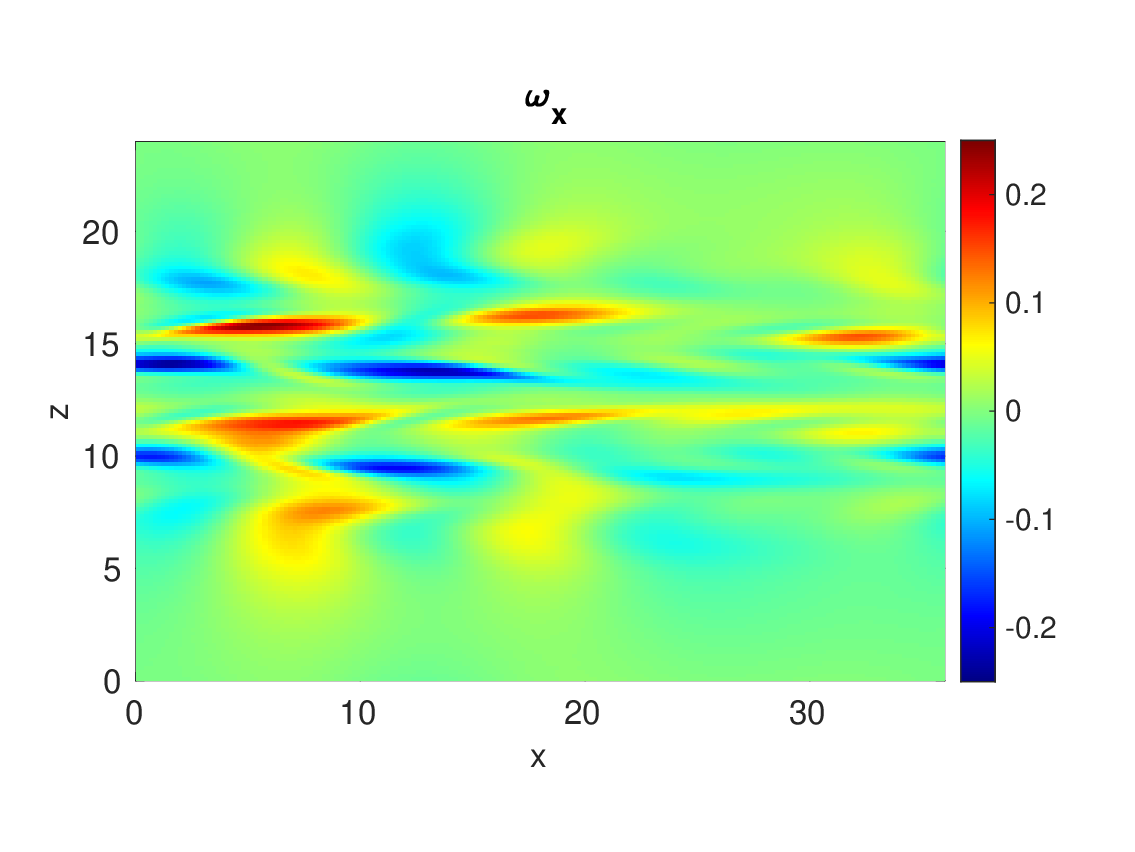}}
\rput(13,2.9){\textbf{(f)}}
\rput(11,6.5){\includegraphics[width=7cm]{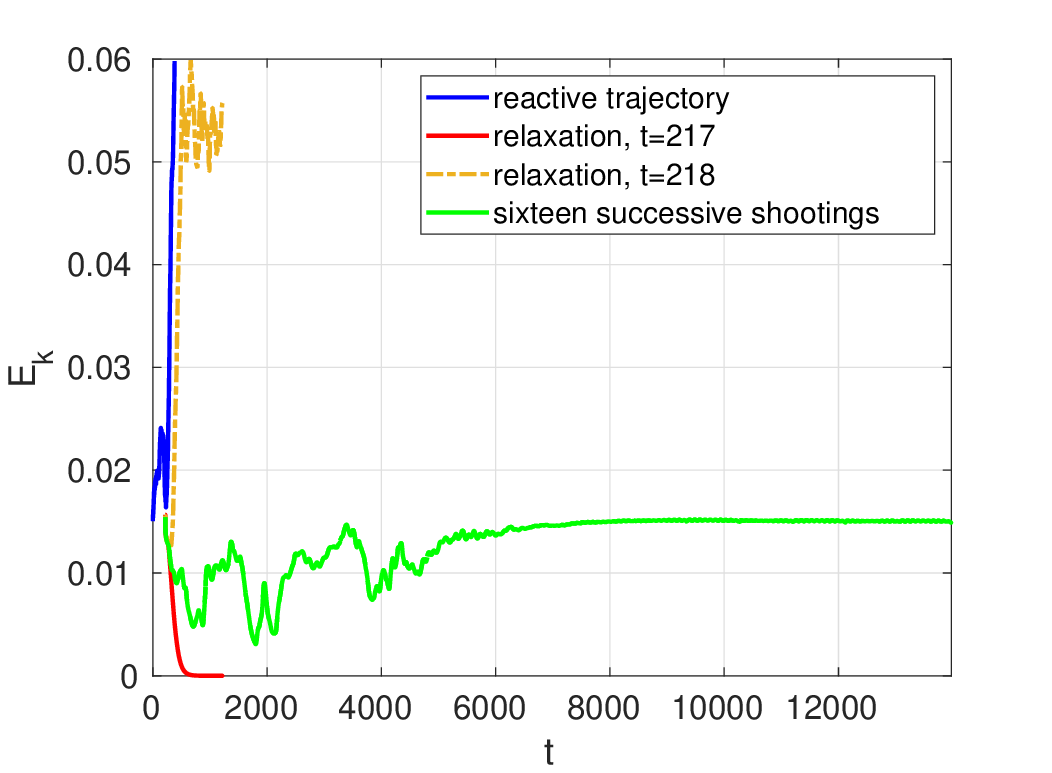}}
\rput(9.15,9){\textbf{(g)}}
\psline[linecolor=gray]{->}(8.7,8.4)(6.7,9.25)
\psline[linecolor=gray]{->}(8.62,5.7)(3,4.7)
\psline[linecolor=gray]{->}(8.62,5.7)(6.5,3.25)
\psline[linecolor=gray]{->}(8.54,5.85)(6.7,6)
\psline[linecolor=gray]{->}(13.8,5.6)(10.5,2.5)
\psline[linecolor=gray]{->}(13.8,5.6)(14.5,2.5)
\end{pspicture}
}
\caption{
View of a reactive trajectory for a system of size $L_x\times L_z=36\times 24$ at Reynolds number $R=500$
under a divergence free forcing at $\beta=3\cdot 10^4$,
and the bisections started from the trajectory.
(a) Colour levels  of streamwise velocity  in the plane $y=0$ when the flow has reached the turbulent state.
(b) Right: colour levels  of streamwise velocity in the plane $y=0$ at $t=50$ in the early stage of the reactive trajectory,
left: colour levels of the streamwise average of the streamwise velocity in a $y-z$ plane at the same time.
(c) Colour levels of the streamwise vorticity in the plane $y=0$ at time $t=217$ at which bisections are started.
(d) Colour levels of the streamwise velocity in the plane $y=0$ at the time $t=217$ at which bisections are started.
(e) Colour levels of the streamwise velocity in the plane $y=0$ when the bisections have reached a saddle.
(f) Colour levels of the streamwise vorticity in the plane $y=0$ when the bisections have reached a saddle.
(g) Time series of the kinetic energy during a reactive trajectory (blue), during two relaxation from states taken on the reactive trajectory,
at $t=217$ (last relaxation to laminar state, red), $t=218$ (first relaxation to turbulent state, orange)
and on the dichotomies started from these two successive velocity fields (green).}
\label{traj_36_24}
\end{figure*}
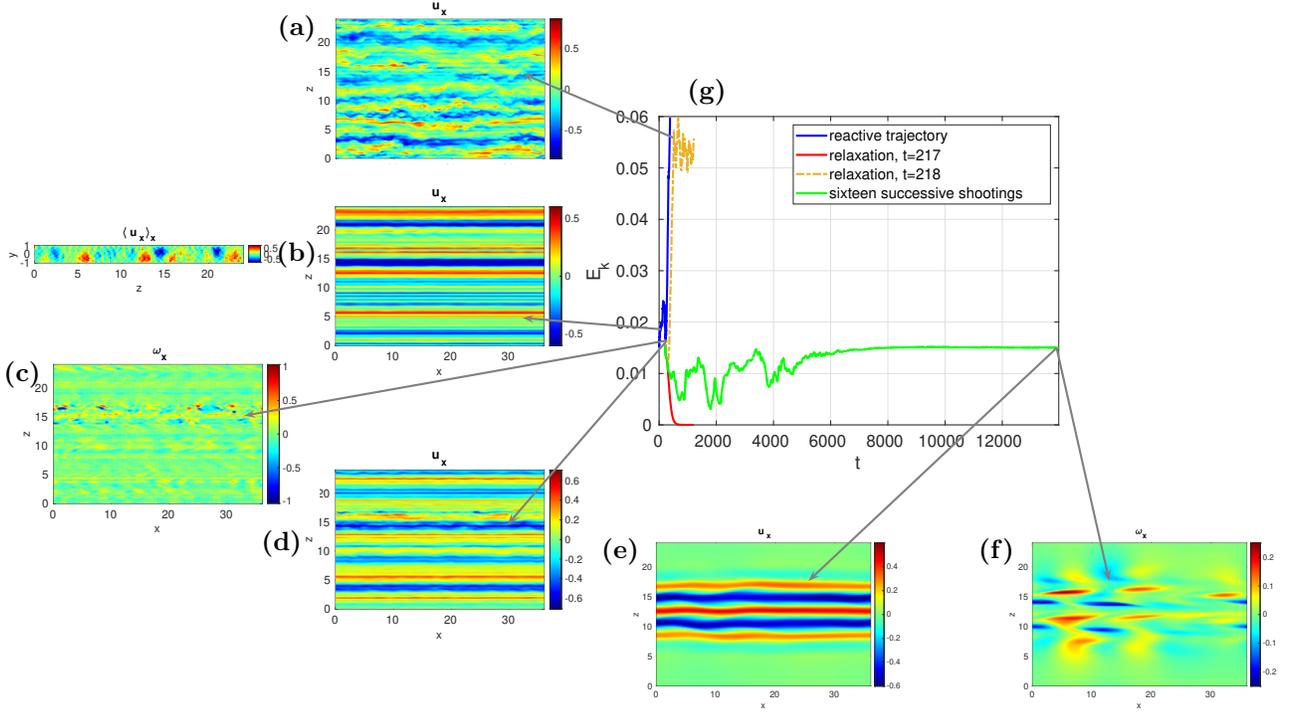

We will observe the transition paths on four examples encompassing the
three domain sizes considered in this article
(Fig.~\ref{traj_6_4_hres},~\ref{traj_6_4_wcx},~\ref{traj_18_12} and~\ref{traj_36_24}).
We list the property of the AMS computations, the parameter $\beta$ controlling the inverse of the variance of the forcing (and the energy injection rate),
 the estimated probability of transition and mean first passage time in table~\ref{tanpaths}.
Note that computations at $R=400$ in the MFU type domain and in a
domain of moderate size $L_x\times L_z=12\times 8$ at $R=500$ at standard resolution
were also performed, yielding similar results (not shown here).
Note also that regimes of parameters leading to much smaller probability of build up $\alpha$ and
much larger waiting times before build up will be presented later in this text (in this section and in \S~\ref{sstatp} and \S~\ref{seps}).
The same phenomenology was observed in those cases.
The parameter defining the reaction coordinate for these computations are presented in appendix~\ref{aparam}.
\begin{table*}[!htbp]
\centerline{
\begin{tabular}{|c|c|c|c|c|c|c|c|c|}
\hline $L_x \times L_z$& resolution& $N$ &$N_c$ &$\beta$&$\alpha$&$T$&forcing type& Figure \\ \hline
$6\times 4$&high &20&1&$5.5\cdot 10^{4}$& divergence free &0.35&$1.2\cdot 10^3$&  Fig.~\ref{traj_6_4_hres} \\ \hline
$6\times 4$&standard&50&1&$1.1\cdot 10^5$&on $u_x$&$2.0\cdot 10^{-3}\pm 4.0\cdot 10^{-4}$&$5.3\cdot 10^5\pm 1.8\cdot 10^5$& Fig.~\ref{traj_6_4_wcx}\\ \hline
$18 \times 12$&---&20&1&$6.25\cdot 10^4$& divergence free & $\alpha=0.32$& $T=6\cdot 10^3$& Fig.~\ref{traj_18_12} \\ \hline
$36\times 24$&---&20&4&$3\cdot 10^4$&divergence free &$\alpha=0.24$&$T=3.9\cdot 10^3$&  Fig.~\ref{traj_36_24} \\ \hline
\end{tabular}
}
\caption{List of the parameters used in AMS computations of transition paths in domains of several sizes $L_x\times L_z$
at Reynolds number $R=500$ presented in
figures~\ref{traj_6_4_hres},~\ref{traj_6_4_wcx},~\ref{traj_18_12} and~\ref{traj_36_24}, and resulting properties.
We indicate the resolution type (when relevant), the parameter $\beta$ controlling the inverse of the variance of the forcing (details on forcing spectra are given in appendix~\ref{stnoise}),
along with the type of forcing,
the number of clones used in AMS computation $N$, as well as the number of clones removed at each step of the algorithm $N_c$.
The estimated probability of build up $\alpha$ and mean first passage time before build up $T$ are given,
along with the figure presenting the path.
}
\label{tanpaths}
\end{table*}

Our visualisations will be centred on the time series of kinetic energy during reactive
trajectories and relaxation/bisection trajectories constructed from the reactive trajectories
(Fig.~\ref{traj_6_4_hres} (f), Fig.~\ref{traj_6_4_wcx} (e), Fig.~\ref{traj_18_12} (g), Fig.~\ref{traj_36_24} (g)).
In all four cases, the time series of $E_k(t)$ indicate that after crossing $\mathcal{C}$, the kinetic energy first grows gradually.
It does so by receiving energy in streamwise velocity tubes that show little trace of
streamwise modulation and have a typical spacing slightly shorter than that of velocity streaks
(Fig.~\ref{traj_6_4_hres} (c), Fig.~\ref{traj_6_4_wcx} (b), Fig.~\ref{traj_18_12} (b)), Fig.~\ref{traj_36_24} (b)).
The energy contained in the flow increases rapidly after $t\simeq 400$ in the MFU type system
at higher resolution (Fig.~\ref{traj_6_4_hres} (f)),
after $t\simeq 1000$ in the MFU type system forced along $\mathbf{e}_x$ (Fig.~\ref{traj_6_4_wcx} (e)),
after $t\simeq 150$ in the system of size $L_x\times L_z=18\times12$ (Fig.~\ref{traj_18_12} (g))
and after $t\simeq 200$ in the system of size $L_x\times L_z=36\times 24$ (Fig.~\ref{traj_36_24} (g)).
At that time, streamwise modulation of streamwise velocity, with a spanwise localisation in these larger domains,
can be can be observed (Fig.~\ref{traj_18_12} (d), Fig.~\ref{traj_36_24} (d)).
Concomitantly, streamwise vorticity with a near turbulent amplitude is also observed where $u_x$ is modulated
(Fig.~\ref{traj_6_4_hres} (b), Fig.~\ref{traj_18_12} (c), Fig.~\ref{traj_36_24} (c)).
The flow then has received enough energy to start the SSP on its own, even locally and finally evolve toward transitional turbulence globally.
We note that this two stage process is visible in the $\log(E_{k,x}),\log(E_{k,y-z})$ plane
for an example of reactive trajectory computed in a MFU type system (Fig.~\ref{glob_traj} (a)) at standard resolution
for divergence free forcing using $\beta=6\cdot 10^{5}$ (leading to $\alpha =2.2\cdot 10^{-25}\pm 0.8\cdot 10^{-25}$):
we first observe an increase of $E_{k,x}$, corresponding to the growth in intensity of velocity streaks,
at rather low $E_{k,y-z}$ (negligible streamwise vortices). The next stage of rapid growth of kinetic energy then corresponds
to the concomitant growth of $E_{k,x}$ and $E_{k,y-z}$.
The same reactive trajectory leaves a less clear trace in the plane formed by the energy injection $I$ and the dissipation $D$ (Fig.~\ref{glob_traj} (b)),
where both $I$ and $D$ grow in a correlated manner from $1$.
We note however that the dissipation $D$ fluctuates more than the energy injection $I$ and that we have $D>I$ on the reactive trajectory.
Being quadratic, $D$ is much more sensitive to fluctuations of velocity caused by forcing noise, which can average out
in the linear $I$ (Eq.~(\ref{total_bud})).
We find that $D-I\simeq 0.4\pm 0.1$ for $\beta=6\cdot 10^6$, corresponding to $(D-I)/R\simeq 8\cdot 10^{-4}$.
Since we have $\epsilon \ge 10^{-3}$ for our control parameters, the growth of total energy is
indeed driven by energy injection by the forcing, but part of that injected energy is spent in dissipation
of the smaller scales generated by the noise forcing.
We further checked this assertion by computing the average and standard deviation of $I-D$ on reactive trajectories
in the range $3\cdot 10^5 \le \beta \le 6\cdot 10^5$, both are affine when displayed in log-log scale
as a function of $\beta$ with a $-1$ slope, indicating that they scale like $\frac{1}{\beta}$.

In the MFU type domain, the comparison of trajectories obtained at higher resolution with divergent free forcing (Fig.~\ref{traj_6_4_hres}),
at standard resolution with divergent free forcing (no detailed example shown here)
and with forcing on the streamwise component at standard resolution (Fig.~\ref{traj_6_4_wcx}),
indicate that the scenario observed on the trajectory is the same in all three types of case: we find the same succession of events, the time series
of kinetic energy have the same shape.
This indicates that the mechanism generating build up during the computed reactive trajectories is independent
of the forcing family and that they can be obtained at standard resolution.
Note that reactive trajectories are random realisations of a given process.
The trajectories differ from a computation to the other in duration, in spatial organisation of the fluctuations,
because distinct realisations of the forcing noise are drawn.
This means that we cannot perform a strict one to one comparison of reactive trajectories are different resolutions
at this point.
In order to evaluate whether the computation is converged, we will perform additional comparisons in the next sections, between
distributions of values, averaged values or paths (\S~\ref{sstatp}), saddle organisation (\S~\ref{sbis})
as well as the scaling laws $\alpha$, $T$ and $\tau$ follow (\S~\ref{setau}).

In extended domains, at the beginning of the rapid increase of kinetic energy
(which will be identified as the crossing of separatrix between laminar and turbulent flow in \S~\ref{sbis}),
one can notice that streamwise modulation of streamwise velocity can be observed in a narrow band in $z$
($5\lesssim z\lesssim 10$ for $L_x\times L_z=18\times 12$, Fig.~\ref{traj_18_12} (d),
$13\lesssim z\lesssim 18$ for $L_x\times L_z=36\times 24$ Fig.~\ref{traj_36_24} (d)).
Streamwise vorticity of intensity comparable to what is found when the flow is turbulent appears in the same band
($L_x\times L_z=18\times 12$, Fig.~\ref{traj_18_12} (c), $L_x\times L_z=36\times 24$ Fig.~\ref{traj_36_24} (c)),
while its amplitude is almost ten times smaller outside of it.
The streamwise velocity tubes observed outside said band correspond to a response to the forcing,
but they tend to be narrower, less intense and showing less streamwise modulation than the streaks showing some streamwise modulation.
The subsequent growth of kinetic energy corresponds to the extension of the modulation in the $z$ direction, until
the whole flow
reaches the turbulent state, visualised in figure (Fig.~\ref{traj_6_4_hres} (a)),
(Fig.~\ref{traj_18_12} (a)), and (Fig.~\ref{traj_36_24} (a)), with colour levels of the streamwise velocity in the midplane $y=0$
and in (Fig.~\ref{traj_6_4_wcx} (a)) with isosurfaces of streamwise velocity in the simulation box.
The final turbulent states comprises of the expected velocity streaks, as visible in the visualisations, and streamwise vortices.

\subsubsection{Last state at the last step}\label{slast}

\begin{figure*}[!htbp]
\centerline{\includegraphics[width=6.5cm,clip]{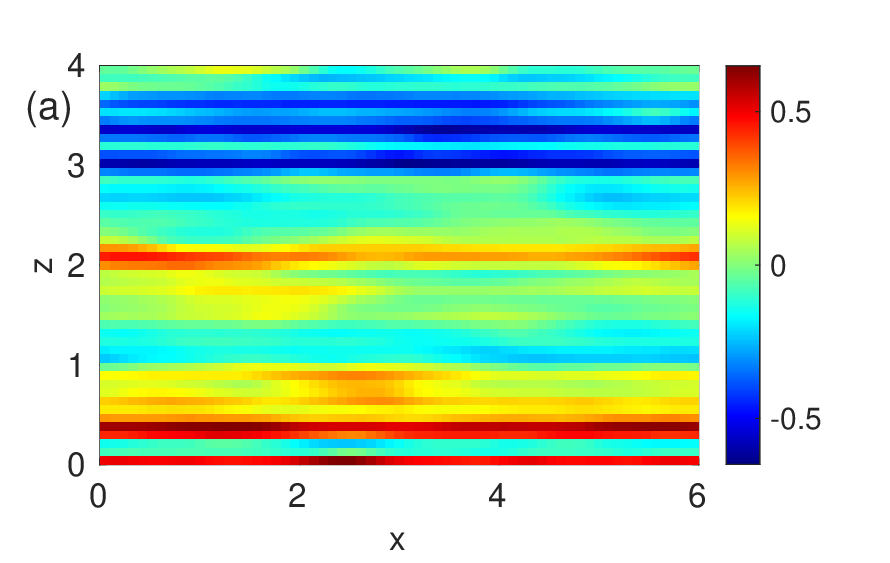}
\includegraphics[width=6.5cm,clip]{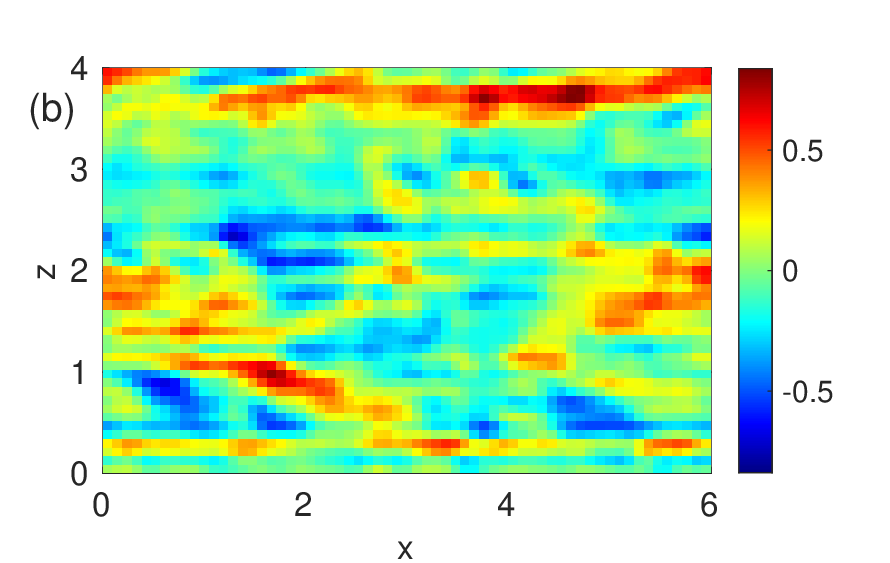}}
\centerline{
\includegraphics[width=6.5cm,clip]{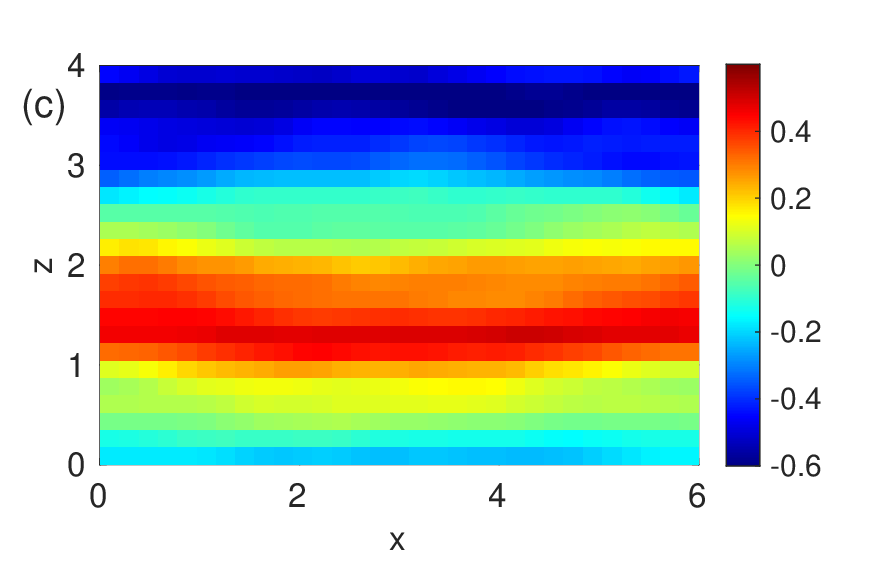}
\includegraphics[width=6.5cm,clip]{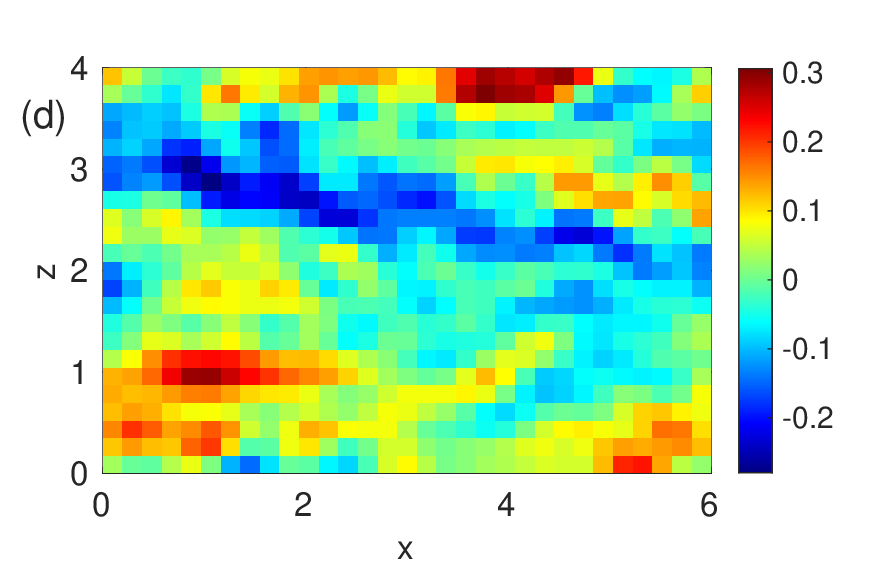}
}
\centerline{
\includegraphics[width=6.5cm,clip]{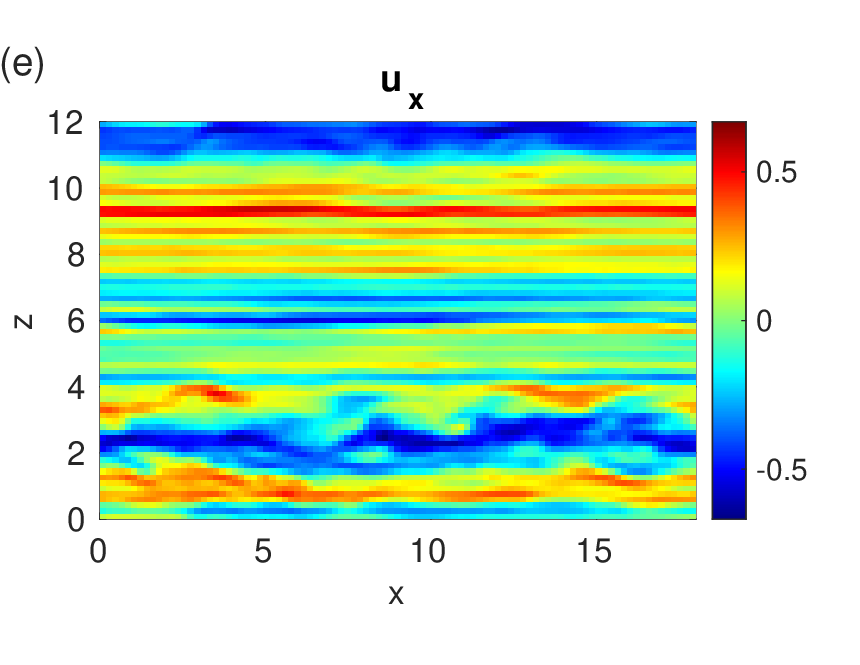}
\includegraphics[width=6.5cm,clip]{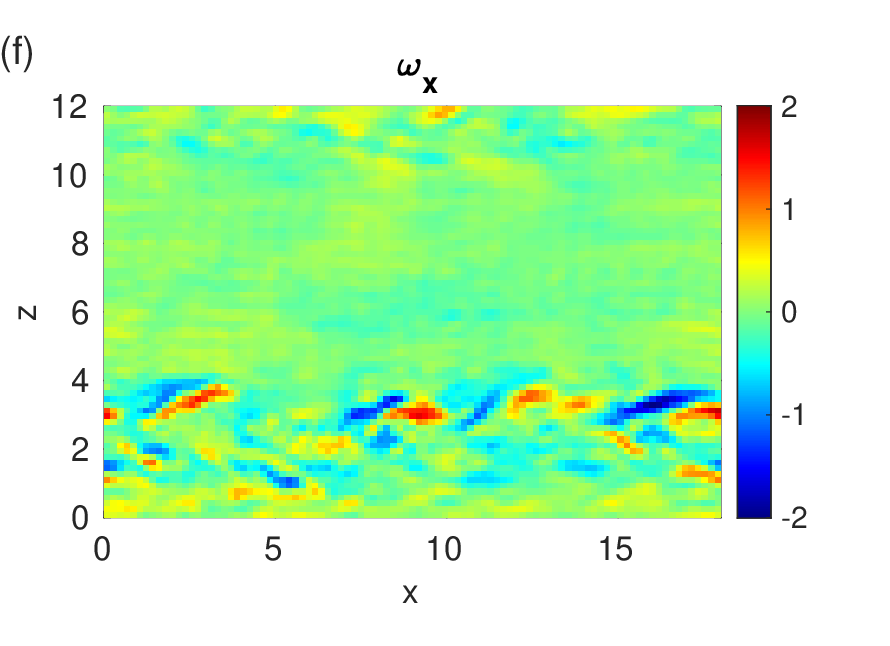}}
\centerline{\includegraphics[width=6.5cm,clip]{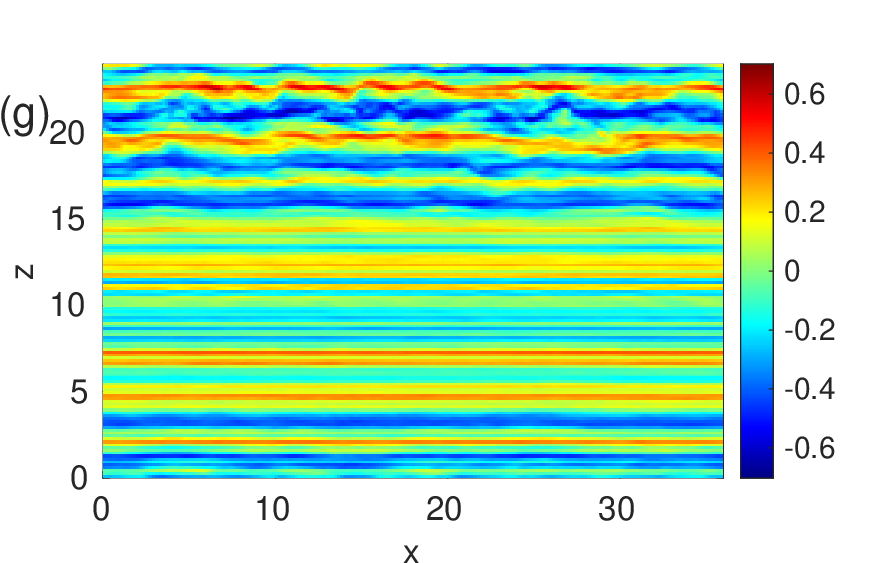}
\includegraphics[width=6.5cm,clip]{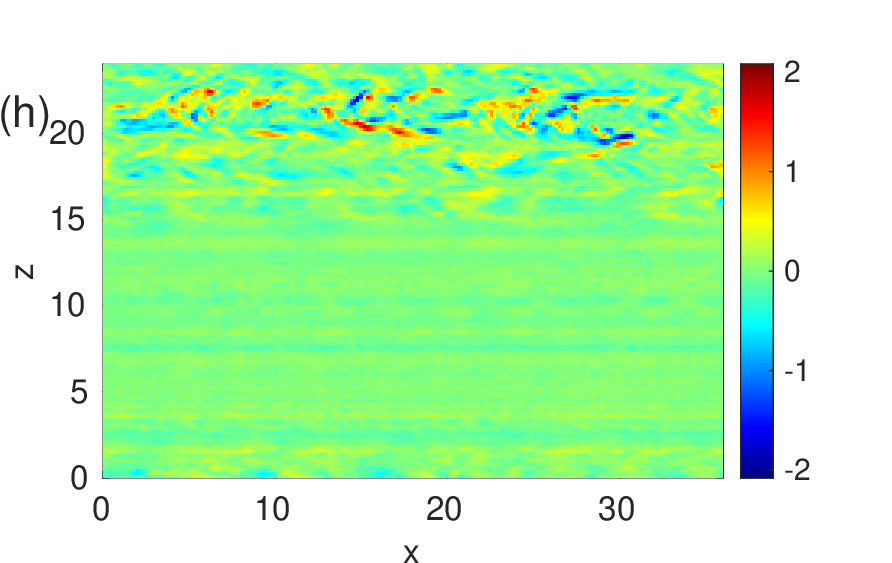}}
\caption{
From the AMS computation in the system of size $L_x\times L_z=6\times 4$ at $R=500$ and $\beta=5.5\cdot 10^5$ at high resolution,
(a) colour levels of the streamwise velocity in the plane $y=0$ on the last state at the last step for the AMS computation,
(b) colour levels of the streamwise vorticity in the plane $y=0$ on the last state at the last step for this AMS computation.
From the AMS computation in the system of size $L_x\times L_z=6\times 4$ at $R=500$ and $\beta=6\cdot 10^5$ at sandard resolution,
(c) colour levels of the streamwise velocity in the plane $y=0$ on the last state at the last step for the AMS computation,
(d) colour levels of the streamwise vorticity in the plane $y=0$ on the last state at the last step for this AMS computation.
From the AMS computation in the system of size $L_x\times L_z=18\times 12$ at $R=500$, and $\beta=8\cdot 10^4$, 
(e) colour levels of the streamwise velocity in the plane $y=0$ on the last state at the last step for the AMS computation,
(f) colour levels of the streamwise vorticity in the plane $y=0$ on the last state at the last step for this AMS computation.
From the AMS computation in the system of size $L_x\times L_z=36\times 24$ at $R=500$, at $\beta =3\cdot 10^4$,
(g) Colour levels of the streamwise velocity in the plane $y=0$ on the last state at the last step for this AMS computation,
(h) Colour levels of the streamwise vorticity in the plane $y=0$ on the last state at the last step for this AMS computation.
}
\label{figlast}
\end{figure*}

\begin{table*}[!htbp]
\centerline{
\begin{tabular}{|c|c|c|c|c|c|c|c|c|c|}
\hline $L_x\times L_z$&&resolution $N$& $N_c$&$\beta$&forcing type  &$\alpha$ &$T$&figure \\ \hline
$6\times 4$&high&400&80&$6\cdot 10^{5}$&divergence free&$2.2\cdot 10^{-25}\pm 0.8\cdot 10^{-25}$&  $6\cdot 10^{27}\pm 2\cdot 10^{27}$ & Fig.~\ref{figlast} (a,b) \\ \hline
$6\times 4$&standard&50&1&$1.1\cdot 10^5$ & on $u_x$&$2.0\cdot 10^{-3}\pm 4.0\cdot 10^{-4}$& $5.3\cdot 10^5\pm 1.8\cdot 10^5$ & Fig.~\ref{figlast} (c,d) \\ \hline
$18\times 12$&---&80&16&$8\cdot 10^4$& divergence free&$7.7\cdot 10^{-6}$& $1.8\cdot 10^7$ & Fig.~\ref{figlast} (e,f) \\ \hline
$36\times 24$&---&20&4&$3\cdot 10^4$& divergence free &$0.24$& $3.9\cdot 10^3$ & Fig.~\ref{figlast} (g,h) \\ \hline
\end{tabular}
}
\caption{
Table indicating the parameters used in AMS computations
where the last states at the last step presented in figure~\ref{figlast}
were computed.
This indicates the size of the domains $L_x\times L_z$ where the computations were performed
at Reynolds number $R=500$ and the resolution type (when relevant).
We give the number of clones $N$ used in AMS computations along with the number of clones removed at each step $N_c$,
the parameter $\beta$ controlling the inverse of the variance of the forcing and the energy injection rate,
as well as the forcing type.
The estimated properties (probability of transition $\alpha$ and mean first passage time $T$)
of the transition paths from which the states are extracted are provided, along with the label of the figure where
the states are shown.
}
\label{tablast}
\end{table*}
We add visualisations of the last state at the last step to our visualisations of the reactive trajectories,
following the definition of this state in section~\ref{sams}.
Those states were obtained from AMS computations using parameters given in table~\ref{tablast} for all three of our domain sizes,
both at standard and high resolution and for both forcing type in MFU type domains.
In figure~\ref{figlast}, we include the streamwise velocity in the $y=0$ plane of this  state
(Fig.~\ref{figlast} (a,c,e,g)) and the streamwise vorticity in the $y=0$ plane of this state
(Fig.~\ref{figlast} (b,d,f,h)), obtained form the AMS computations that yielded the reactive trajectories presented in section~\ref{Paths}.
This diagnostic indicates that at this turning point of AMS computations,
streamwise elongated streamwise velocity tubes are formed and display turbulent like intensity with $\max_{\text{space}} |u_x|\ge 0.5$,
although they present little trace of streamwise modulation in the smaller, MFU type, system (Fig.~\ref{figlast} (a,c)).
The corresponding streamwise vorticity (Fig.~\ref{figlast} (b,d)) thus has a maximum intensity of $|\omega_x|\lesssim 0.5$,
which is a fraction of what can be found when the flow is actually turbulent.
Again, we note that there is no striking qualitative difference in those states obtained two distinct types of forcing.
Streamwise velocity tubes are also formed in larger systems (Fig.~\ref{figlast} (e,g)).
However, we note that in a restricted range of $z$
(taking into account periodic boundary conditions, $0\lesssim z\lesssim 3$
and $10\lesssim z\lesssim 12$ for
$L_x\times L_z=18\times 12$ Fig.~\ref{figlast} (e), $15\lesssim z\lesssim 24$ for $L_x\times L_z=36\times 24$ Fig.~\ref{figlast} (g)),
wall turbulence like streamwise modulation of $u_x$ is present.
We note that streamwise vorticity of turbulent like intensity $\omega_x=\mathcal{O}(1)$ and with turbulent like scale
is now only found in a similar range of $z$ (taking into account periodic boundary conditions,
$10\lesssim z\lesssim 12$ and to a lesser extent $0\lesssim z\lesssim 3$ for
$L_x\times L_z=18\times 12$ Fig.~\ref{figlast} (f), $18\lesssim z\lesssim 24$ for $L_x\times L_z=36\times 24$ Fig.~\ref{figlast} (h)).
Vorticity is more intense in the larger system.
This further indicates that while the reactive trajectory corresponds to a response
to the forcing through the development of streamwise velocity tube,
the flow only reach a turning point when locally, the tubes become sufficiently intense
and display intense enough shear between one another to trigger
the formation of streamwise modulation and corresponding streamwise vortices.
Once this has happened, this streamwise modulation and said vortices will contaminate nearby regions in $z$ until they extend to the whole flow.

\subsubsection{Bisections}\label{sbis}

\begin{figure}[!htbp]
\centerline{
\begin{pspicture}(6.5,5)
\rput(3.25,2.4){\includegraphics[width=6.5cm]{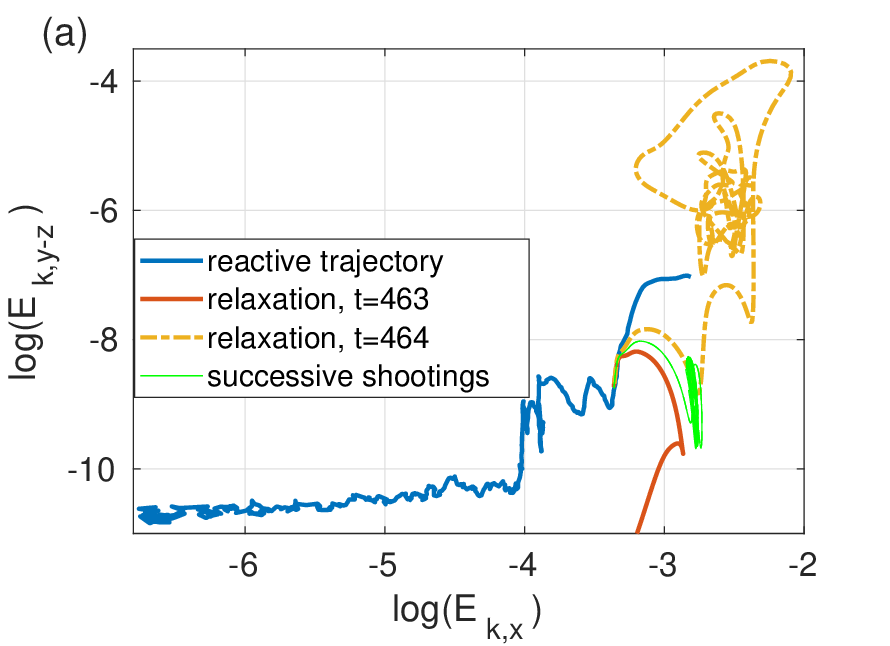}}
\rput(2,3.725){\includegraphics[width=1.84cm,clip]{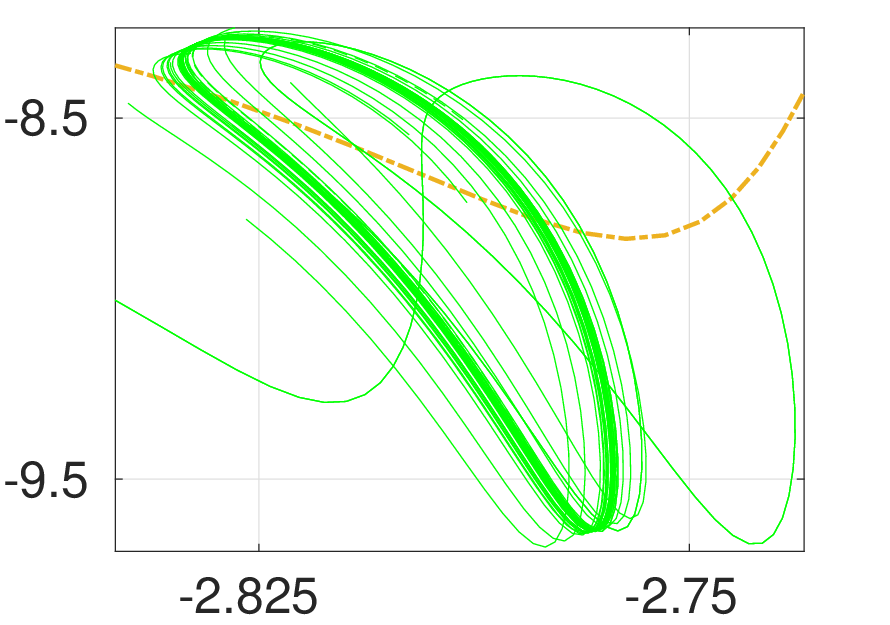}}
\psframe[linewidth=0.015](4.99,1.45)(5.18,2.17)
\psline{->}(5.085,2.17)(2.9,3.8)
\end{pspicture}
\begin{pspicture}(6.5,5)
\rput(3.25,2.4){\includegraphics[width=6.5cm]{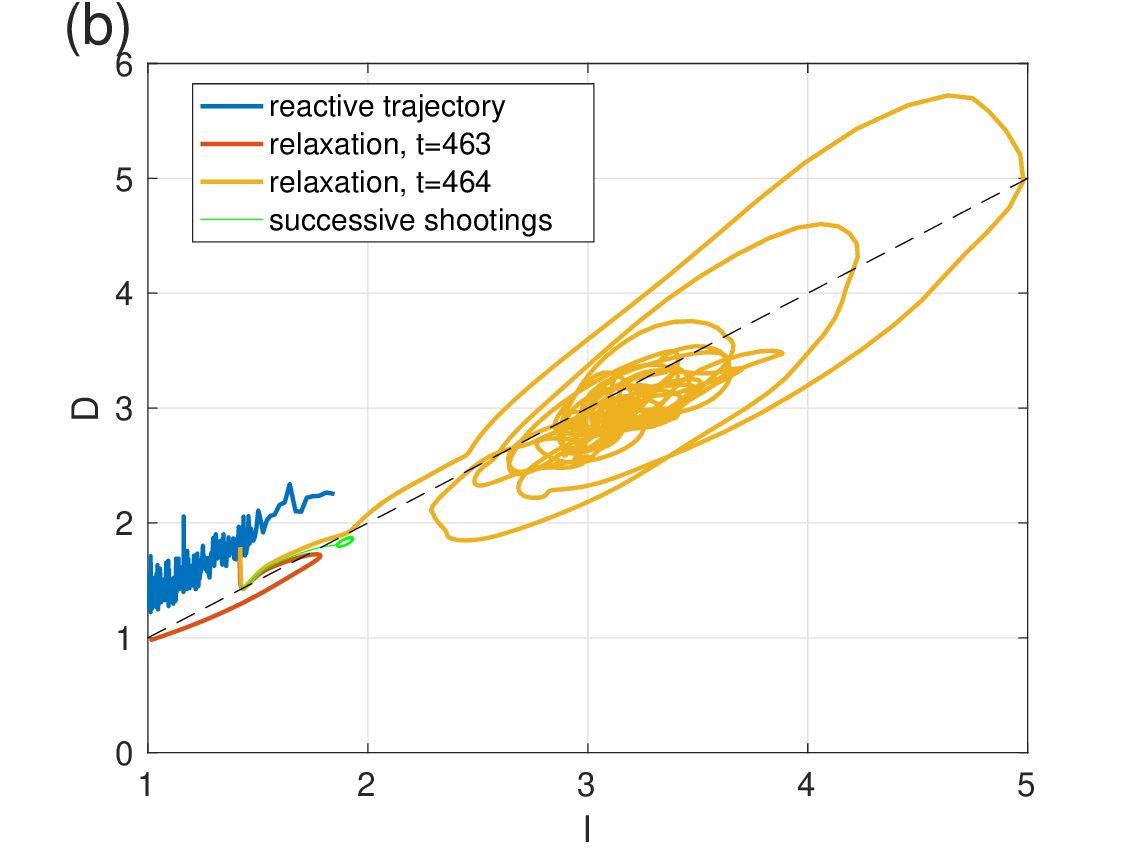}}
\rput(4.77,1.35){\includegraphics[width=2.15cm,clip]{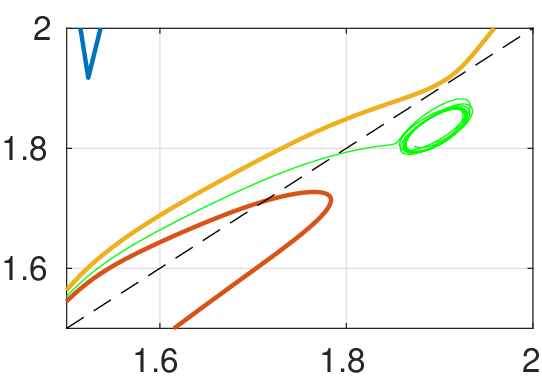}}
\psframe[linewidth=0.015](1.5,1.5)(2.12,1.865)
\psline{->}(2.12,1.68)(3.9,1.35)
\end{pspicture}
}
\caption{
(a) Plot of a reactive trajectory obtained by means of AMS in the MFU type domain at $\beta=600000$ (blue line),
relaxations toward laminar (red line) and turbulent flow (orange line) started from the reactive trajectory
as well as the result of the bisections (green line)
in the plane formed by the logarithm of kinetic energy on components $\log(E_{k_,x})$ and $\log(E_{k,y-z})$.
An inset zooms on the area where the dynamics winds around the unstable limit cycle.
(b) Plot of the same reactive trajectory, relaxation and result of bisections, in the plane $I,D$ where $I$
is the total energy injection and $D$ is the total dissipation (Eq.~(\ref{total_bud})), using the same colour code.
The dashed line $D=I$ is added. An inset zooms on the area where the dynamics winds around the unstable limit cycle.
}
\label{glob_traj}
\end{figure}
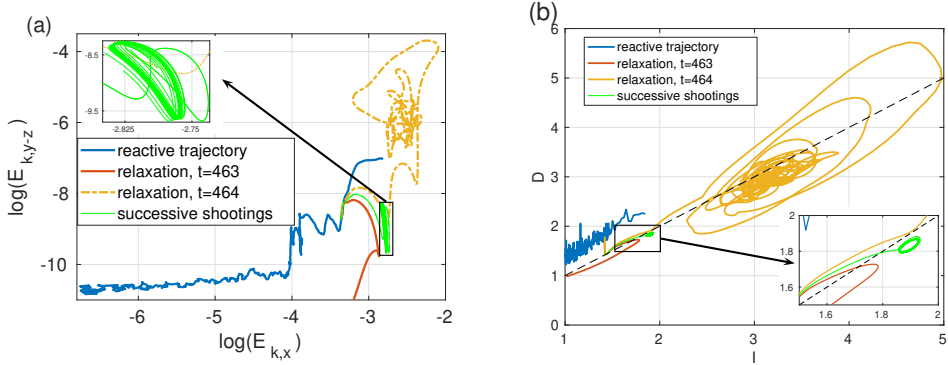

Following the procedure presented in section~\ref{metbis},
we perform relaxations and bisections using reactive trajectories computed by means of AMS.
We will examine the type of unstable attractor (unstable fixed point, unstable limit cycle, unstable travelling wave)
which is reached and describe its main features.
The result of these bisections will be compared to reactive trajectories and the last state at the last step examined in the former subsection.
We will check whether the system displays a standard instanton phenomenology,
which would be characterised by greater and greater correspondence between $\mathbf{u}_{\rm last}$
and the result of bisections: this would mean the trajectories would cross the separatrix very near the saddle.
The time series of these relaxations and the result of following bisections are added
to the plots:
\begin{itemize}
\item  of figure~\ref{traj_6_4_hres} (f) for a MFU type system at high resolution with divergence
free noise with $\beta=5.5\cdot 10^4$ and a sampling time of $\delta t=1$ on the trajectory, starting bisections at $t=439$,
\item of figure~\ref{traj_6_4_wcx} (e) for a MFU system at standard resolution with noise on component $\mathbf{e}_x$ with $\beta=1.1\cdot 10^5$,
and a sampling time of $\delta t=2$ on the trajectory, starting bisections at $t=1048$,
\item of figure~\ref{traj_18_12} (g) for a system of size $L_x\times L_z=18\times 12$ with divergence free noise with $\beta=6.25\cdot 10^4$
and a sampling time of $\delta t=1$ on the trajectory, starting bisections at $t=182$,
\item of figure~\ref{traj_36_24} (g) for a system of size $L_x\times L_z=36\times 24$ with divergence free noise with $\beta=3\cdot 10^4$
and a sampling time of $\delta t=1$ on the trajectory, starting bisections at $t=217$.
\end{itemize}
The corresponding values of $\alpha$ and $T$ are given in section~\ref{Paths}.
The first two relaxations as well as the result of bisections is also added to the plots in the $\log(E_{k,x}),\log(E_{k,y-z})$
plane (Fig.~\ref{glob_traj} (a)) and in the $(I,D)$ plane (Fig.~\ref{glob_traj} (b)).
Relaxation toward the laminar flow naturally corresponds to $\log(E_{k,x}),\log(E_{k,y-z})\rightarrow -\infty$, and $I,D\rightarrow 1$,
while relaxation toward turbulence leads to time fluctuations, naturally found near $I\simeq D\simeq 3$.
Note that during relaminarisation, we have $I>D$, energy injection is larger than dissipation leading to $E_{\rm tot}$
is actually growing (Eq.~(\ref{total_bud})).
Indeed, this energy is dominated by the streamwise flow which happen to be less and less depleted
as the flow goes from turbulent to laminar.
As the size of the system is increased (from $6\times 4$ to $36\times 24$),
longer and longer transients have to be passed until said saddle is reached.
During these transients, the smallest scales visible on snapshots of the reactive trajectories are smoothed out,
in particular in the wall normal and spanwise components on the velocity field.

\begin{figure}[!htbp]
\centerline{
\includegraphics[width=7cm]{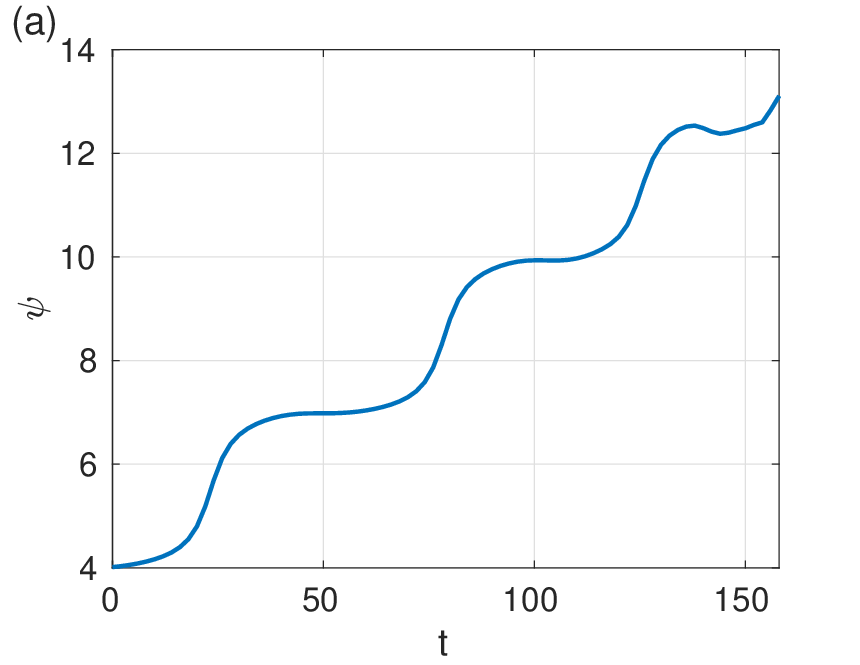}\includegraphics[width=7cm]{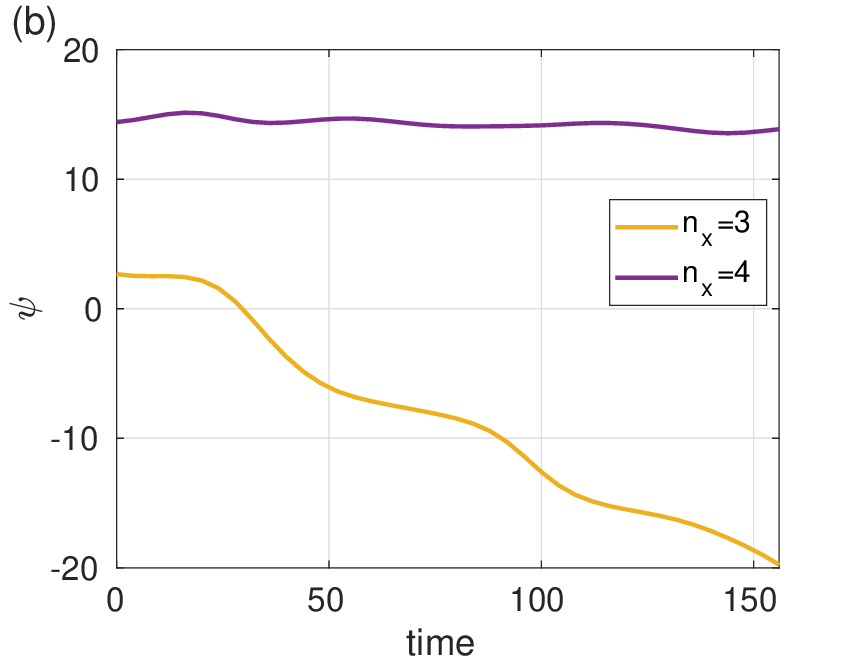}
}
\caption{
(a) Rescaled phase of the first $x$ Fourier mode of $u_z$ computed at $z=2$ and $y=0$ as a function of time computed for the limit cycle found
in MFU type domains (Fig.~\ref{traj_6_4_hres} (d,e), Fig.~\ref{traj_6_4_wcx} (c,d)).
(b) Rescaled phase of the third and fourth $x$ Fourier mode of $u_z$ computed at $z=12$ and $y=-0.75$
as a function of time computed for the travelling wave found
in the domain of size $L_x\times L_z=36\times 24$ (Fig.~\ref{traj_36_24} (e,f)).}
\label{phase_bis}
\end{figure}

For the smallest, MFU type system, we reach an unstable limit cycle that displays small oscillations of kinetic energy
(Fig.~\ref{traj_6_4_hres} (f), Fig.~\ref{traj_6_4_wcx} (e)).
This cycle can also be viewed in the $\log(E_{k,x})$, $\log(E_{k,y-z})$ plane (Fig.~\ref{glob_traj} (a)),
this indicates that these two components have a phase shift
($E_{k,x}$ is almost in phase opposition with $E_{k,z}$ and in phase quadrature with $E_{k,y}$).
This limit cycle consists in oscillations of the streamwise modulation
of a pair of streamwise velocity tubes (Fig.~\ref{traj_6_4_hres} (d), Fig.~\ref{traj_6_4_wcx} (c)).
As expected the low speed tubes are located in the upper part of the channel $y>0$ and the
high speed tubes is located in the lower part of the channel $y<0$.
The velocity in the tubes is rather intense, leading to $\max_{\text{space}} |u_x|=0.75\pm 0.05$.
The modulation in the two velocity tubes are best described as in phase quadrature with one another.
This streamwise modulation leads to a corresponding streamwise vorticity which has a rather low magnitude $|\omega_x|=\mathcal{O}(0.1)$.
Note that the spatial variation of streamwise vorticity is sharp, with an oblique $\omega_x=0$ stripe striking each vortex in the $y=0$ plane
(Fig.~\ref{traj_6_4_hres} (e), Fig.~\ref{traj_6_4_wcx} (d)).
This cycle is found both at lower and higher resolution and regardless of the forcing type
(Fig.~\ref{traj_6_4_hres} (f), Fig.~\ref{traj_6_4_wcx} (e), Fig.~\ref{glob_traj} (a)).
This indicates that the reactive trajectories cross the separatrix in the same region for both families of forcing,
and that the crossing location of the separatrix can be viewed as converged with spatial resolution.
The relative amplitude of oscillations of the streamwise component is small:
at high resolution the time average of kinetic energy is $\langle E_{k,x}\rangle_t=0.0620$,
while the rescaled root mean square is $\sqrt{2}\sqrt{\langle E_{k,x}^2\rangle_t-\langle E_{k,x}\rangle_t^2}=0.0023$.
The relative amplitude of oscillation of the wall normal and spanwise components is much larger,
we find that the time averages of kinetic energy on these components are
$\langle E_{k,y}\rangle_t=3.55\cdot 10^{-5}$ and $\langle E_{k,z}\rangle_t=1.44\cdot 10^{-4}$,
while the rescaled root mean square are $\sqrt{2}\sqrt{\langle E_{k,y}^2\rangle_t-\langle E_{k,y}\rangle_t^2}=1.22\cdot 10^{-5}$
and $\sqrt{2}\sqrt{\langle E_{k,z}^2\rangle_t-\langle E_{k,z}\rangle_t^2}=9.14\cdot 10^{-5}$.
The increase of resolution leads to minute changes to this limit cycle: we find a relative increase of 2\% on $\langle E_{k,x}\rangle_t$
and 6\% on $\langle E_{k,y}\rangle_t$ and $\langle E_{k,z}\rangle_t$,
with smaller changes on the amplitude of the cycles, as measured by much smaller change in the RMS kinetic energy.
The visualisation of $\omega_x$ at both resolution (Fig.~\ref{traj_6_4_hres} (e), Fig.~\ref{traj_6_4_wcx} (d)) confirms that the difference is minute.
We also observe the cycle in the total energy injection $I$, total dissipation $D$ plane (see Eq.~(\ref{total_bud})) in figure~\ref{glob_traj} (b).
Note that in line with the observation of the disappearance of very small scale fluctuations,
dissipation rapidly decrease from higher fluctuating value to a value close to $I$ during the relaxation.
Bisections lead to the limit cycle which lies slightly under to the $I=D$ line.
This is an effect of discretisation errors in the evaluation of $I$ and $D$ combined with the very small amplitude
of the limit cycle. As resolution is increased, this cycle moves toward the line.
Finally, we  measured the period of oscillations of kinetic energy by determining the frequency
of the largest amplitude peak discrete Fourier transforms of the concatenated times series of $E_k(t)$ obtained during successive shootings.
This gives  $\mathcal{T}=59.8\pm 0.6$ at low resolution (using a time series of duration $4537$) and $\mathcal{T}=58.2\pm 0.6$ at higher resolution
(using a time series of duration $5249$).
This again confirms the convergence of the numerical simulation.
Observation of the velocity fields indicate that every half period of kinetic energy,
the streamwise modulation of velocity fields nears zero, then grows with a $L_x/2$ phase shift.
This means that the actual time period of the velocity fields is twice that of the kinetic energy.
We observe this in figure~\ref{phase_bis} (a) using the rescaled phase of the first Fourier mode of $u_z$ computed
in the midplane at $z=2$ defined as
\begin{equation}
\psi(t)=L_x\left(\frac{\arg(\hat{u}_z(n_x=1,y=0,z=2,t))}{2\pi}+\frac12\right)\,.\label{rescphase}
\end{equation}
This rescaling matches with a $x$ position. It is incremented by $L_x$ each time $\arg(\hat{u}_z)$
goes from $2\pi$ to $0$ and decremented by $-L_x$ each time
$\arg(\hat{u}_z)$ goes from $0$ to $2\pi$.
The succession of plateau separated by a rapid jump every period of the kinetic energy
quantifies the fact that we periodically see the pattern in $\omega_x$ disappear and reappear at a distance $L_x/2$.
This edge state in the form of an unstable limit cycle has not been specifically studied in the literature to our knowledge, possibly because of
the smallness of the domain. However, a similar spatial organisation of the flow,
similar periods and average amplitudes have been observed in the $(I,D)$ plane for unstable limit cycles in comparably small
domains at smaller Reynolds numbers \cite{kreilos2012periodic}. Note however that the amplitude of the observed cycles was larger
than what we display here by an order of magnitude.
A very similar state has been observed at $R=400$ (not shown here).

As the system size is increased to larger values $12\times 8$ (not shown) and $L_x\times L_z=18\times 12$,
several changes occur in this saddle.
The saddle becomes time invariant, both for kinetic energy (Fig.~\ref{traj_18_12} (g))
and velocity and vorticity fields (Fig.~\ref{traj_18_12} (e,f)):
it is thus an actual saddle point.
It still presents streamwise modulated streamwise velocity tubes that occupy the whole domain,
however, these velocity tubes are less intense with $\max_{\text{space}} |u_x|=0.44\pm 0.01$.
This leads to a steady value of kinetic energy of $E_k=0.2$, a third of the time average found in the MFU type system.
Additionally, the
streamwise modulation in the adjacent streamwise velocity tubes is of varicose type.
The spatial organisation and average kinetic energy of this saddle is consistent
with saddles computed at $R=400$ in domains of comparable size \cite{schneider2008laminar}.
The spatial variation of the streamwise vorticity is less sharp than in the MFU type systems
and its amplitude is still moderate $|\omega_x|=\mathcal{O}(0.1)$ (Fig.~\ref{traj_18_12} (f)).
As the system size is further increased to $L_x\times L_z=36\times 24$,
yet another change appears in the saddle which is reached after bisections: the state becomes localised in $z$ (Fig.~\ref{traj_36_24} (e))
and modulated in $x$ to a smaller extent.
the most intense streamwise velocity tube has $u_x\ge 0$ in our computation,  with $\max_{\text{space}} |u_x|=0.78\pm 0.01$,
it is flanked by pairs of negative then positive streamwise velocity tube of comparable, though decreasing intensity.
The streamwise velocity then decays rapidly with $z$.
Moreover the modulation of streamwise velocity is localised in $x$,
leading to a streamwise localisation of streamwise vorticity (Fig.~\ref{traj_36_24} (f)).
The streamwise propagation of this modulation makes this saddle a travelling wave.
This travelling wave is consistent with those observed at smaller Reynolds numbers $R=400$,
in domains slightly larger in the spanwise direction and larger in the streamwise direction \cite{schneider2010localized}.

When the saddle state occupies the whole domain (at any given time for $L_x\times L_z=6\times 4$
and any time for $L_x\times L_z=12\times 8$  and $L_x\times L_z=18\times 12$), it has spatial symmetries:
it is invariant under the operation $y\rightarrow -y$, $z\rightarrow -z$ $x\rightarrow -x+\delta x$ with $\delta x\lesssim 1$.
The the domain of size $L_x\times L_z=6\times 4$, the time periodicity, the time periodicity can be viewed as a time symmetry.
Moreover, advancing in time by half a period and applying the afore mentioned spatial transformation leads to a space time symmetry.
Moreover, increasing the system size to $L_x\times L_z=12\times 8$ and to $L_x\times L_z=18\times 12$,
means that the saddle has respectively two  and three wavelengths of the same pattern in both $x$ and $z$ direction and
thus respectively introduces the symmetry groups $\mathbb{Z}/2\mathbb{Z}\times\mathbb{Z}/2\mathbb{Z}$
and $\mathbb{Z}/3\mathbb{Z}\times\mathbb{Z}/3\mathbb{Z}$.
The symmetry and the parity of velocity tubes numbers is broken in the larger domain $L_x\times L_z=36\times 24$, where the state is localised.
This broken symmetry is accompanied by a breaking of the steadiness in $x$ of the saddle.
Our trajectory led to a saddle which is a wave travelling in the $-\mathbf{e}_x$ direction.
We note that this system can also sustain a travelling wave obtained under the operation $y\rightarrow -y$,
$\mathbf{u}\rightarrow -\mathbf{u}$ \cite{schneider2010localized}.
This conditions the direction of propagation as the celerity of the wave is related to $\int_{y,z} u_x\,{\rm d}y{\rm d}z\ne 0$
because of the spanwise localisation.
More precisely, we find a carrier wave of wave number $n_x=4$ which is almost steady is space with an almost constant amplitude,
with a travelling modulation at lower wavenumbers.
We can observe this using equation~(\ref{rescphase}, but this time at $y=-0.75$ and $z=12$) giving the rescaled phase of Fourier modes $n_x=4$
(part of the carrier) and $n_x=3$ (modulation) in figure~\ref{phase_bis} (b),
this time at $z=12$ and $y=-0.75$ where $u_z$ is the most intense and most concentrated.
The rescaled phase of mode $n_x=4$ is near constant and confirms us the spatial steadiness of the modulation.
Meanwhile the monotonous decrease of the phase of mode $n_x=3$ confirms the travel of the pattern toward smaller $x$.
The length $\Delta \psi=-24$ by which it has decreased in the time window is consistent with the observed movement of the
pattern in the periodic domain.
Note that this propagation occurs at constant spatially averaged kinetic energy (Fig.~\ref{traj_36_24} (g)).

We further our comparison of the respective properties of the reactive trajectories: the corresponding last state at the last step on the one hand
and the saddles which are the closest to the crossing point of the separatrix on the other hand.
In particular, we perform this comparison in view of the properties of last states and saddles in
many of observed noise induced transitions.
This comparison highlights a striking difference.
In most known stochastically forced systems, where the saddle is clearly identified, both types of states are very similar:
in those cases, all the trajectories cross the separatrix in the neighbourhood of the saddle,
in particular when the noise variance is smaller and trajectories concentrate in phase space.
In the case of larger noise variance, trajectories tend to spread around the saddle,
instead of all avoiding it.
In our case, for all system sizes, the reactive trajectories cross the separatrix
with velocity fields (Fig.~\ref{traj_6_4_hres} (b,c), Fig.~\ref{traj_6_4_wcx} (b), Fig.~\ref{traj_18_12} (c,d), Fig.~\ref{traj_36_24} (c,d))
which remain fairly similar in scale and in velocity and vorticity amplitude.
These amplitudes and scales are all clearly distinct
from the nearest saddle on the separatrix
(Fig.~\ref{traj_6_4_hres} (d,e), Fig.~\ref{traj_6_4_wcx} (c,d), Fig.~\ref{traj_18_12} (e,f), Fig.~\ref{traj_36_24} (e,f)),
whose stable manifold they cross.
For all system sizes one common difference is that the amplitude of streamwise vortices is stronger
($\max_{\text{space}}|\omega_x|\lesssim 1$) on the trajectory as it crosses the separatrix  as
it is in the saddle. The spatial characteristic scale of said vortices is also smaller on the trajectory than it is on the saddle.
The same can be said of the modulation of the streamwise velocity tubes.
Other size dependent distinctions can be found. For the MFU type system,
the trajectory cross the separatrix at a kinetic energy $E_k\simeq 0.02$ lower than
that of the unstable limit cycle where $E_k\simeq 0.06$.
This value increases very little with $\beta$ in the tested range.
In the system of size $L_x\times L_z=12\times 8$ most reactive trajectories start localising in $z$ (not shown here).
All trajectories localise in $z$ in
the system of size $L_x\times L_z=18\times 12$, while the saddle is spread over the whole domain in those two systems.
In the largest system of size $L_x\times L_z=36\times 24$, streamwise vorticity is more localised in $z$ than it is the case for the traveling wave.
We will see in the next sections (\S~\ref{sstatp},~\ref{sea})
that this happens while all trajectories have the properties of concentrated trajectories for small enough forcing noise variance.

\subsection{Statistical description of trajectories}\label{sstatp}

We complement our description of trajectories with the discussion of the statistical
properties of large ensembles of trajectories obtained by means of AMS.
We will use diagnostics proposed in earlier
studies of multistability and rare events.
We will confirm that the cases presented above for the discussion of spatial organisation of the flow on paths are representative
and we will further compare our reactive trajectories to typical rare noise induced transitions.

\begin{figure}[!htbp]
\centerline{
\includegraphics[width=6.5cm]{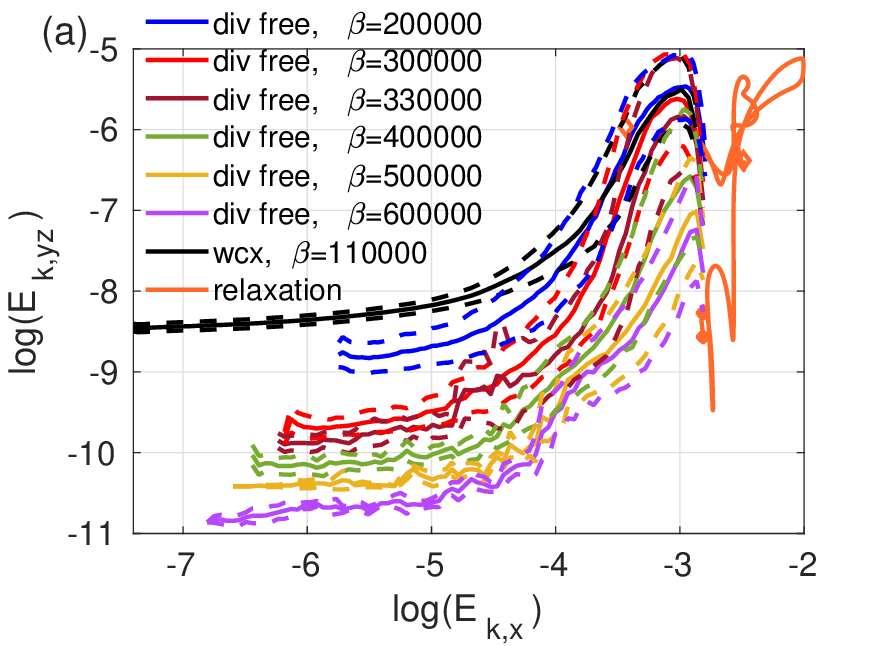}
\textbf{(b)}
\includegraphics[width=6.5cm]{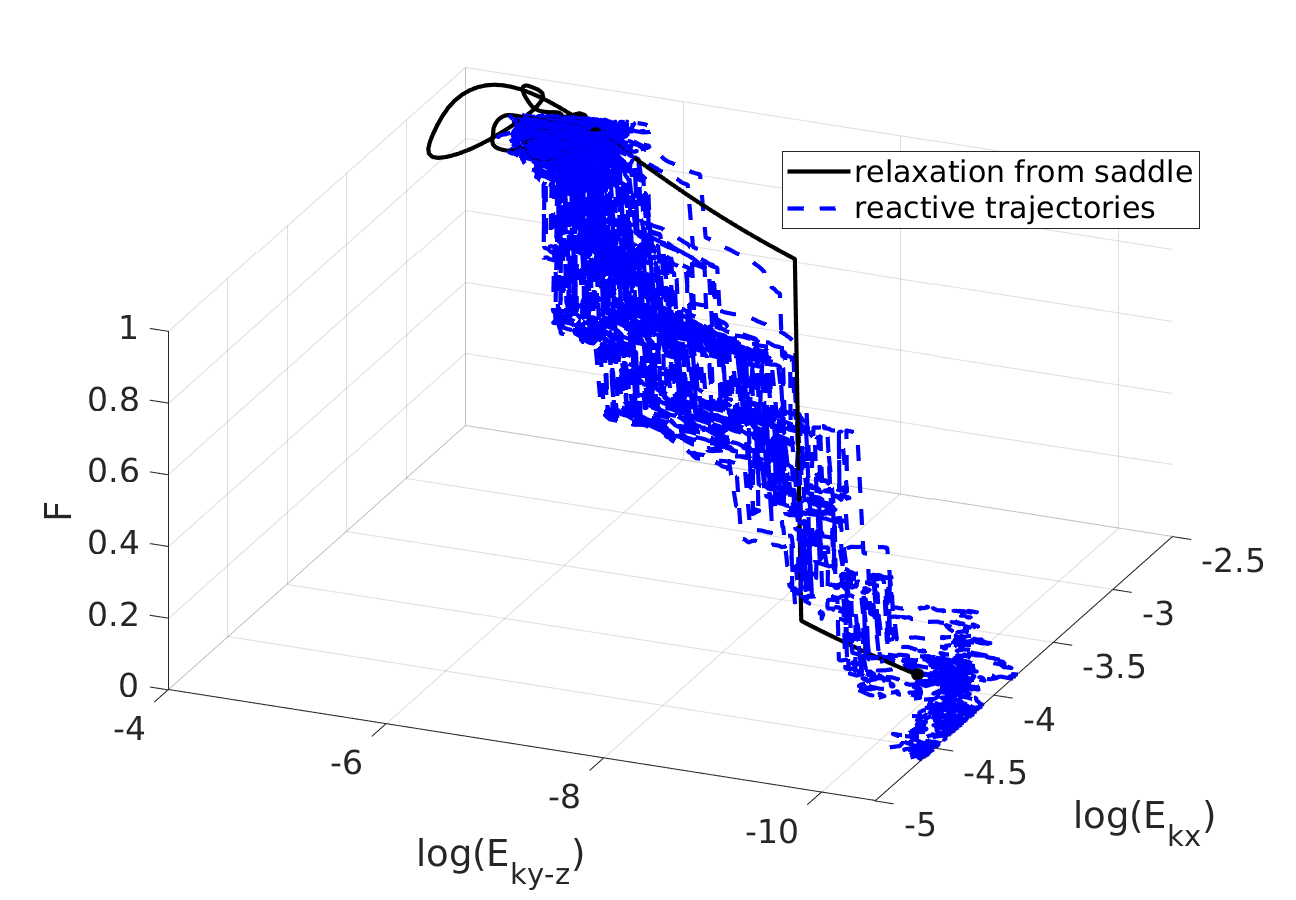}
}
\caption{
(a) Average plus and minus standard deviation of the logarithm of the kinetic energy
in $y-z$ components $\log(E_{k,y-z})$ conditioned on logarithm of the kinetic energy in $x$ component $\log(E_{k,x})$
obtained from reactive trajectories computed by means of AMS with a divergence free forcing and increasing $\beta$
as well as with forcing on the $x$ component (termed wcx in the caption).
The path in the $(\log(E_{k,x}),\log(E_{k,y-z}))$ plane of a relaxation from the saddle toward turbulence is added in orange, with a dot indicating its starting point.
(b) Three dimensional plot of 151 reactive trajectories obtained by means of AMS in the domain of size $L_x\times L_z=18\times 12$ at $\beta=80000$
(blue) as well as the relaxation from the fixed point (black)
in the space formed by the logarithm of kinetic energy contained in components
$\log(E_{k,x})$, $\log(E_{k,y-z})$ and turbulent fraction $F$.
}
\label{glob_traj2}
\end{figure}

We first present the average path and standard deviation around it in the plane formed by $\log(E_{k,y-z})$
and $\log(E_{k,x})$ (Fig.~\ref{glob_traj2} (a)),
in a system of size $L_x\times L_z=6\times 4$ at $R=500$.
The figure is generated using the same procedure as in \cite{rolland2022collapse} \S~3.1.2:
we compute the average and standard deviation of $\log(E_{k,y-z})$,
conditioned on the value of $\log(E_{k,x})$ on reactive trajectories.
The tube formed by the average plus/minus the standard deviation thus indicates us where $66\%$
of paths go (using the reasonable assumption of a Gaussian spread around the mean).
We compare the type of paths followed in the case of forcing on component $x$ (with $554$ paths)
to paths obtained under divergence free forcing at increasing $\beta$
($1600$ paths for $\beta=200000$, $687$ paths for $\beta=300000$, $738$ paths for
$\beta=330000$, $718$ paths for $\beta=400000$, $740$ paths for $\beta=5000000$ and 755 paths for $\beta=600000$).
These paths have an estimated probability of occurrence that range from $0.59\pm 0.01$ to $2.2\cdot 10^{-25}\pm 0.8\cdot 10^{-25}$
(which can be considered as rare).
Each set of paths starts from a slightly different neighbourhood.
Indeed, since $\mathcal{A}$ and $\mathcal{C}$ are defined using the average
and standard deviation of kinetic energy conditioned on no transition occurring
(\S~\ref{dreac}),
the smaller the energy injection rate is, the smaller are $E_{k-x}$ and $E_{k,y-z}$ at the beginning of the trajectories (Fig.~\ref{energy}).
However, each group of paths (for decreased $\beta$ or changed forcing type)
displays the same type of evolution, with clear convexity of the curve $\langle \log(E_{k,y-z})\rangle (\log(E_{k,x}))$.
This confirms in a statistical manner the observation that the two types of forcing eventually
lead to flows that follow the same type of paths as they build up toward turbulence,
in that they cross the separatrix at similar locations, with similar flow patterns.
Schematically, we
first observe a growth of $\log(E_{k,x})$ then a growth of $\log(E_{k,y-z})$.
This indicates that the velocity streaks would first receive energy from the forcing, mostly indirectly,  and are in an externally sustained process.
This part of the trajectory is similar to what is observed using deterministic targeted forcing \cite{rigas2021nonlinear}.
They do so until they are intense enough to naturally start the streaks instability and to go
from an interplay of streaks and vortices sustained by energy injection to the self sustaining process
of turbulence.
In all cases, the trajectories are more and more concentrated around the corresponding
average path as $\beta$ is increased and $\epsilon$ is decreased.
For the lowest values of $\epsilon$, we sampled these trajectories
from several independent AMS runs to ensure that this narrower variance was not an effect
of a smaller variability on the first part of the path  within an AMS run.

We perform a similar observation, this time using the curves of all the reactive trajectories
computed by means of AMS along with a relaxation from the fixed point computed using bisections started from a reactive trajectory in a system
of size $L_x\times L_z=18\times 12$ at $\beta=80000$.
The paths are represented in the three dimensional phase space formed by $\log(E_{k,x})$, $\log(E_{k,y-z})$
and the turbulent fraction $F$ (Fig.~\ref{glob_traj2} (b)).
This last quantity was added to discuss the effect of spatial localisation on the paths.
With a probability of observing the transition at this energy injection rate at $\alpha=\mathcal{O}\left(10^{-5}\right)$,
we observe a concentration of paths in phase space (this can even be observed at $\beta=6.25\cdot 10^{4}$ where $\alpha=\mathcal{O}\left(10^{-1}\right)$).
In that case we first have the growth of $E_{k,x}$ (from $\log(E_{k,x})\lesssim -4.5$ to $\log(E_{k,x})\gtrsim -4$)
followed by that of $E_{k,y-z}$ already observed in smaller systems,
while $F=0$.
This corresponds to the formation of a streamwise band where velocity streaks are intense enough,
modulated and in interplay with spanwise localised streamwise vortices.
Once this band is intense enough ($\log(E_{k,x}\gtrsim -4$, $\log(E_{k,y-z})\gtrsim -10$) the turbulent fraction $F$ become non-zero and turbulence expands in the $z$ direction.
This is indicated by the correlated increase of $\log(E_{k,x})$, $\log(E_{k,y-z})$ and $F$.								

We note two peculiar facts.
Firstly, while the paths are more and more concentrated around their average in the MFU sized system,
these averages are distinct from one value of $\epsilon$ to another and
the bands given by plus/minus the standard deviation do not overlap.
We do not observe the typical narrower and narrower concentration around an instanton.
We note however that these bands move in phase space as $\epsilon$ is decreased,
which means that said instanton could be reached asymptotically.
Secondly, we added to the graph of figure~\ref{glob_traj2} (a) the path followed in the  $\log(E_{k,y-z}), \log(E_{k,x})$
plane by the relaxation from the unstable limit cycle which is the nearest to the transition path
to turbulence, as computed in section~\ref{sbis}.
At that size, the relaxation starts from values of $E_{k,x}$ close to what
is found in turbulence and displays a burst of $E_{k,y-z}$ along with an overshoot
of $E_{k,x}$: the path followed in that plane is distinct from reactive trajectories.
																		
This can be viewed as a corollary of the first observation.
We similarly added the curve followed in the three dimensional space $\log(E_{k,x})$, $\log(E_{k,y-z})$, $F$
by the relaxation from the fixed point in the system of size $L_x\times L_z=18\times 12$ (Fig.~\ref{glob_traj2} (b)).
This path can be viewed as the relaxation path that would be followed by an instanton in this system.
We note that none of the group of transitions paths follows that relaxation path.
The relaxation starts from a low value of $E_{k,y-z}$ ($\log(E_{k,y-z})\simeq -9.7$) and
moderate value of $E_{k,x}$ ($\log(E_{k,x})\simeq -3.9$).
The zero value of $F$ is a consequence of the value of the threshold used to compute the turbulent fraction,
as the saddle is delocalised in the whole domain.
As the flow relaxes and follows the unstable direction, the departure to the edge state
grows globally in the domain, not locally to the point at $\log(E_{k,x})\simeq-3.6$, $E_{k,y-z}=-8.3$,
where values of $e$ (see Eq.~(\ref{defef})) are large enough everywhere to be above the
detection threshold and the turbulent fraction jumps to $1$.
The kinetic energy contained in both streaks and vortices then keep growing globally until the flow reaches transitional wall turbulence.
Following both relaxations thus confirms what had been visualised on the field taken from examples.																											

\begin{figure}[!htbp]
\centerline{\includegraphics[width=7cm]{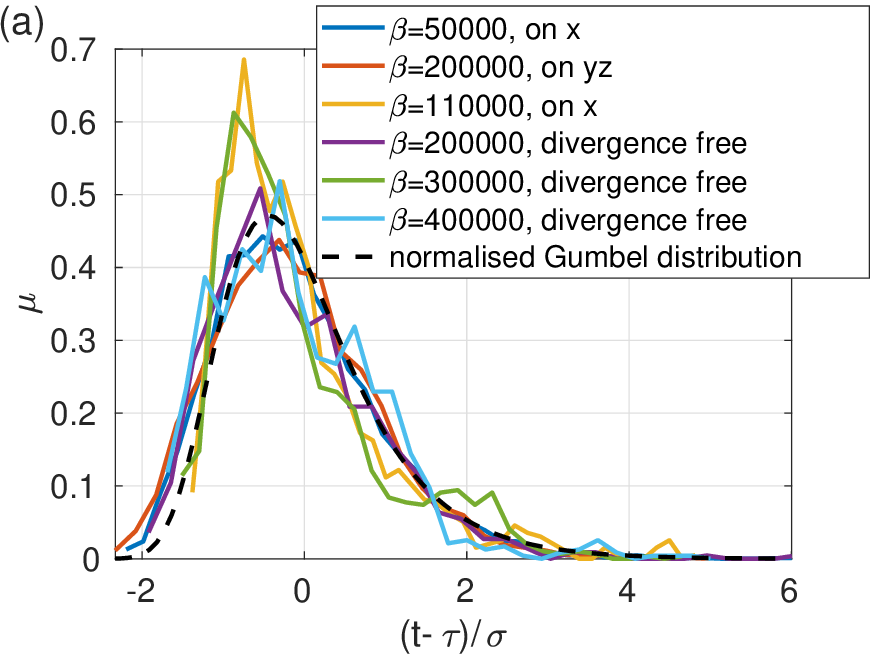}
\includegraphics[width=7cm]{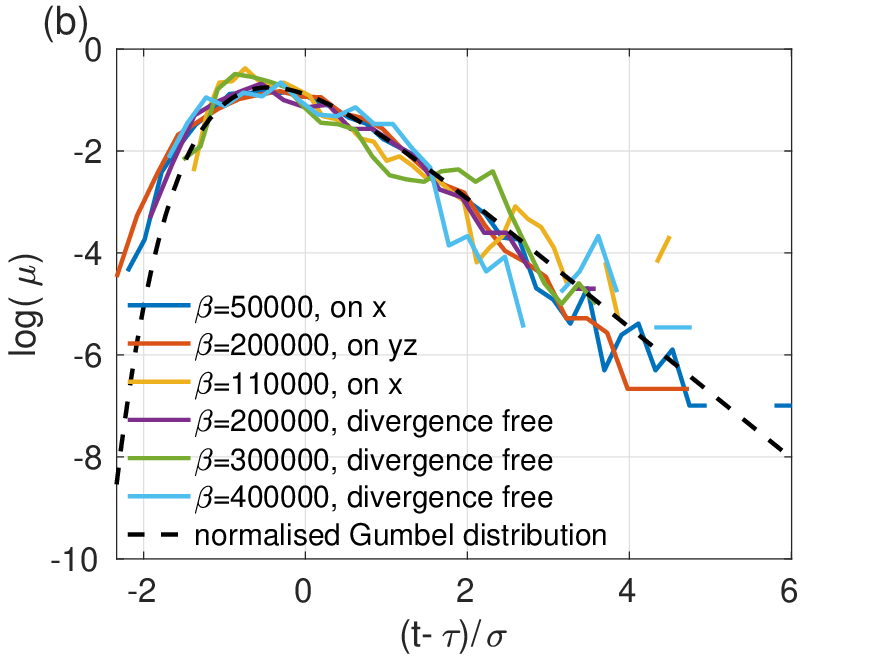}}
\caption{Distribution of duration of reactive trajectories rescaled by their sample average $\tau$ and sample standard deviation $\sigma$.
The trajectories are computed in a system of size $L_x\times L_z=6\times 4$ at Reynolds number $R=500$
under forcing on components or divergence free forcing for several energy injection rates.
A normalised Gumbel distribution (Eq.~(\ref{egum})) is added to the plots. (a) in linear scale. (b) in Logarithmic scale.
5320 trajectories were sampled for $\beta=50000$ with forcing on $x$ component.
3220 trajectories were sampled for $\beta=200000$ with forcing on $y$ and $z$ components.
1271 trajectories were sampled for $\beta=110000$ with forcing on $x$ component.
1600 trajectories were sampled for $\beta=200000$, 1399 were sampled for $\beta=300000$
and 1055 were sampled for $\beta=400000$ with divergence free forcing.}
\label{distrib}
\end{figure}

We then consider the distribution of durations of reactive trajectories,
in a system of size $L_x\times L_z=6\times 4$ at $R=500$.
They are sampled using AMS with forcing on components and divergence free forcing.
We display the distribution of durations rescaled by the sample average duration $\tau$
and the sample standard deviation $\sigma$ in figure~\ref{distrib} (a)
in linear scale and in  figure~\ref{distrib} (b) in logarithmic scale.
In both cases, the distributions are compared to a normalised Gumbel distribution defined using the Euler constant $\gamma=0.577\ldots$
\begin{equation}
\mu(s)=\frac{\pi}{\sqrt{6}}\exp\left(-\left( \frac{\pi}{\sqrt{6}} s+\gamma +\exp\left( \frac{\pi}{\sqrt{6}} s+\gamma \right) \right) \right)
\,.\label{egum}
\end{equation}
For one degree of freedom systems, in the limit of vanishing noise variance,
it has been demonstrated that this is the distribution of duration \cite{cerou2013length}.
It has been observed that the sampled distribution is close to a Gumbel
in more complex stochastic systems \cite{rolland2016computing,rolland_pre18}
when we have instanton phenomenology (with clear deviation when we do not,
for instance when the saddle becomes degenerate \cite{rolland2016computing}).
This distribution is also observed in systems not forced by noise \cite{rolland2022collapse,gome2022extreme}.
In our case, we observe a very good match between the sampled distribution and a Gumbel,
with small differences for $t\le \tau$ (which was also observed for instanton phenomenology in extended gradient systems \cite{rolland2016computing}).

Along with the concentration of trajectories and the shape of the distribution of duration of trajectories,
we saw that the computed reactive trajectories still displayed some properties of classical noise induced transitions.
We will test this comparison further in the next section with the dependence of transition on the energy injection rate.

\section{Effect of energy injection rate}\label{seps}

In this section, we analyse the results obtained from a series of AMS runs performed for increasing $\beta$ and decreasing $\epsilon$ in
the MFU type system using divergence free forcing and forcing on components.
We will follow quantities such as the probability of observing a transition $\alpha$,
the mean first passage time before observing a transition $T$ as well as the average
duration of reactive trajectories $\tau$ as a function of $\beta$ and $\epsilon$.
We will investigate whether we find exponential scaling for the first two,
in our ongoing comparison of the transition with typical behaviour.

\subsection{Probability of crossing and mean waiting time}\label{sea}

\begin{figure}[!htbp]
\centerline{
\includegraphics[width=7cm]{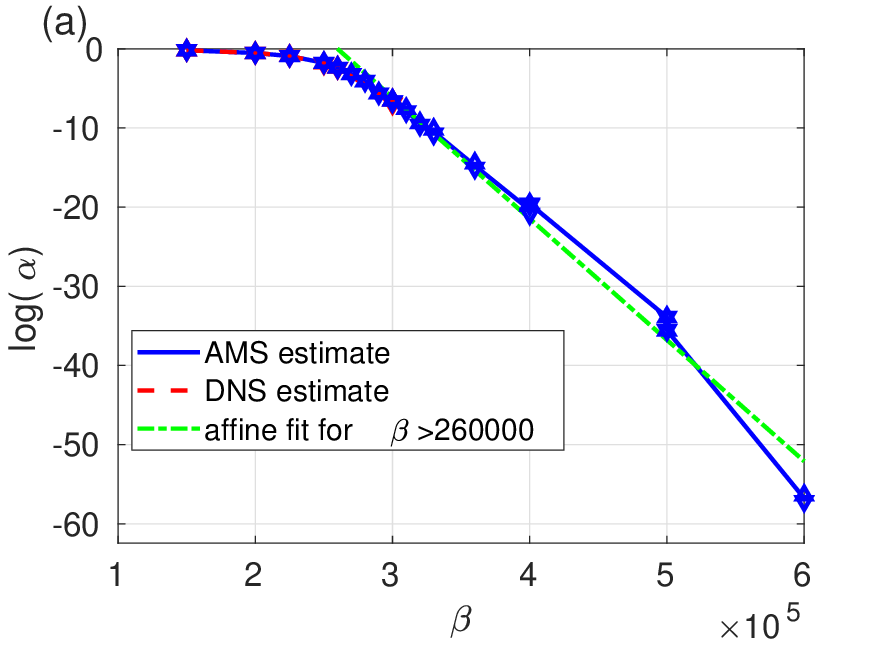}
\includegraphics[width=7cm]{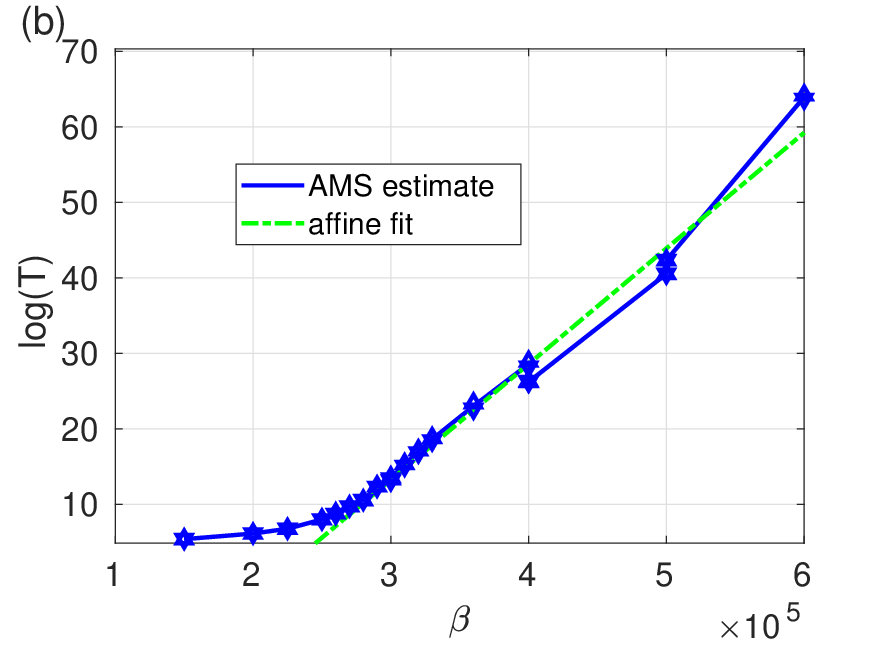}
\includegraphics[width=7cm]{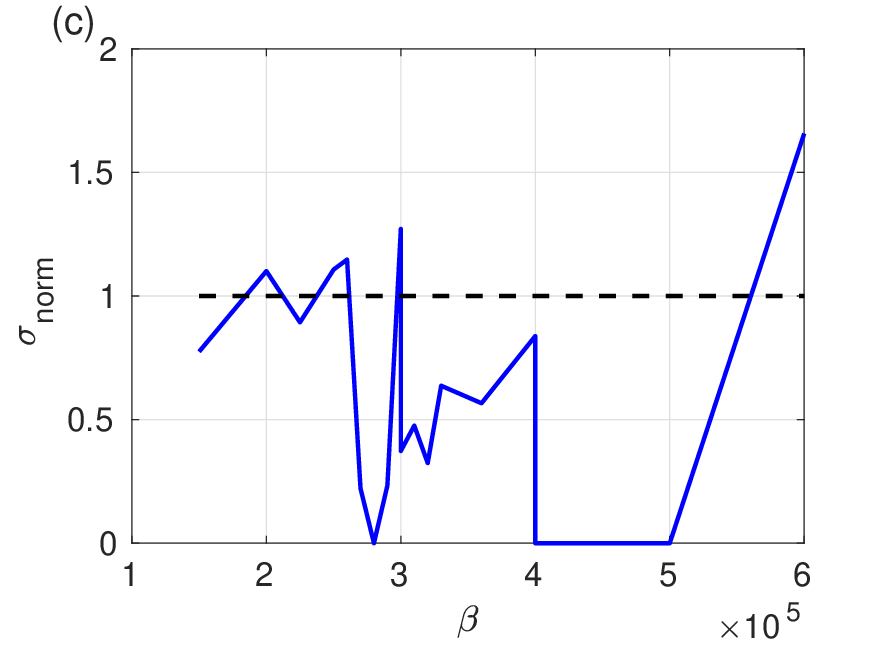}
}
\caption{From computations in  system of size $L_x\times L_z=6\times 4$ at Reynolds number $R=500$ under divergence free forcing,
by means of AMS and up to  $\beta =3\cdot 10^{5}$ by means of DNS.
(a) : Probability to cross from the defined neighbourhood of the laminar baseflow to the turbulent flow as a
function of $\beta$, an affine fit is performed for $\beta \ge 2.7\cdot 10^5$ and is added to the plot.
(b) : Mean first passage time before transition
to the turbulent flow as a function of $\beta$, an affine fit is performed for $\beta \ge 2.7\cdot 10^5$ and is added to the plot.
(c) : normalised standard deviation of the estimate of $\alpha$ as a function of $\beta$, the black dashed line indicates
the ideal value $1$.
}
\label{beta_dvf_6_4}
\end{figure}

\begin{figure}[!htbp]
\centerline{\includegraphics[width=7cm]{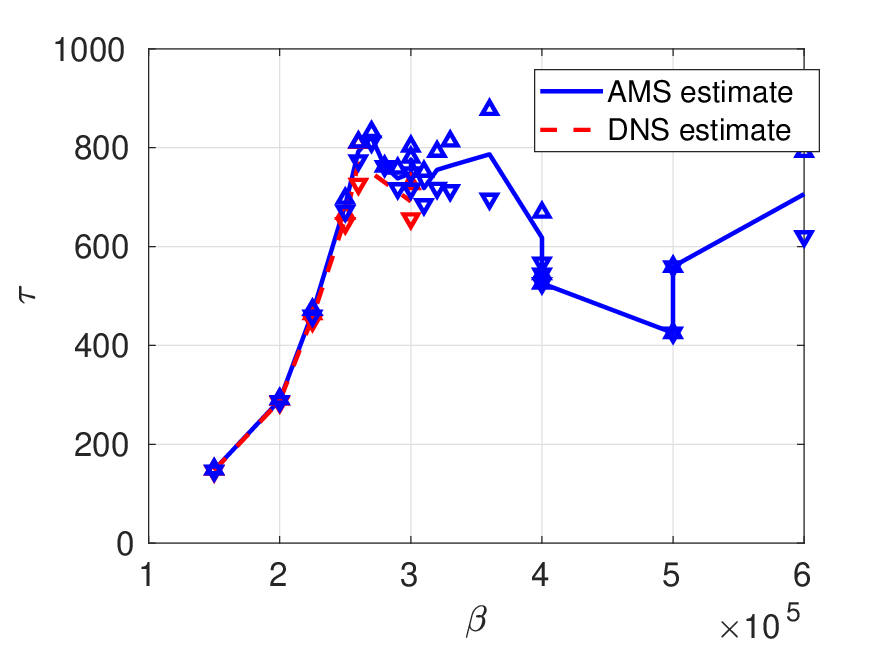}}
\caption{From the same AMS computations as in figure~\ref{beta_dvf_6_4} average duration of reactive trajectories as a function of $\beta$.}
\label{beta_dvf_6_4t}
\end{figure}

In this section, we present the probability of crossing from a neighbourhood of the laminar flow
to turbulent flow under a divergence free forcing as well as the corresponding mean first passage time.
Both of them will be presented as a function
of $\beta$ with a fixed spectrum shape.
The mean first passage times are computed by means of AMS, following the procedure described in section~\ref{detams}.
The probability of crossing is computed by means of AMS for all $\beta$ of interest
as well as by means of DNS (following the procedure of \S~\ref{detdns}) up to $\beta=3\cdot 10^5$.
These computations were performed  in the MFU type  system of size $L_x\times L_z=6\times 4$ at Reynolds number $R=500$.
The estimates are performed by averaging over the values of several realisations, while the error bars are constructed from the $66$\% interval
of confidence obtained from the sample standard deviations. Up to $\beta=2.8\cdot 10^5$,
$N=200$ clones are used in AMS computations, $N=100$ clones are
used for $\beta=2.9\cdot 10^5$ and $\beta=3\cdot 10^5$ and $N=50$ clones are used for computations at $3.1\cdot 10^5 \le \beta \le 4\cdot 10^5$.
In these simulations, one clone was suppressed at each iteration. We performed further computations at $\beta=5\cdot 10^5$ using $N=800$ clones and
removing $N_c=160$ clones per iteration and at $\beta=6\cdot 10^5$ using $N=400$ clones and removing $N_c=80$ clones per iteration.
At least 10 realisations of AMS computations per control parameters are performed for $\beta < 4\cdot 10^5$. This number goes up to more than 30
for $2.9\cdot 10^5 \le \beta \le 3.3\cdot 10^5$, while it is of the order of several units for $\beta \ge 4\cdot 10^5$.

We present the probability to observe a build up computed by means of AMS and DNS in figure~\ref{beta_dvf_6_4} (a).
The corresponding mean first passage time before crossing to turbulence, computed solely by means of AMS,
is displayed in figure~\ref{beta_dvf_6_4} (b).
Both are presented in logarithmic scale.
These two quantities indicate that the noise induced transition becomes rapidly very rare
as $\beta$ is increased: the estimate of $\log(\alpha)$ decreases by more than five decades, while that
of $\log(T)$ increases by five decades in the range of $\beta$ considered here.
This corresponds to variations by more than twenty orders of magnitude in linear scale.
Most of this rapid variation takes place between $\beta=2.7\cdot 10^5$ and $\beta=6\cdot 10^5$.
In that range both $\log(\alpha)$ and $\log(T)$ are very well approximated by affine functions.
The fit are added to figure~\ref{beta_dvf_6_4} (a,b) and respectively
give the slopes $(-1.53\pm 0.01)\cdot 10^{-4}$ and $(1.53\pm 0.01)\cdot 10^{-4}$.
This confirms that the probability $\alpha$ decays exponentially and that the mean first passage time $T$ increases
exponentially and that their rate are opposite of one another.
This scaling of $T$ and $\alpha$ in $\beta$ is the same as the one observed in most systems
where rare bistability is present.
In that framework, as we had stated in section~\ref{intstat}, the decay rate of the probability $\alpha$
and the growth rate of the mean first passage time $T$ is formally given
by the difference of quasipotential between the visited saddle and the first multistable state \cite{bouchet2016generalisation}.
In our case, this would correspond to the edge state and the laminar baseflow.
Note however that the computation of the quasipotential and the estimation of its value at
specific flow states is beyond the scope of this work.

We perform two tests to validate the computation of $\alpha$. Up to $\beta=3\cdot 10^5$,
computations by means of DNS are conceivable since $\alpha\ge 10^{-3}$.
The results are added to figure~\ref{beta_dvf_6_4} (a)
and show an overlap between the AMS and DNS intervals of confidence.
This indicates that the values computed by means of AMS can be trusted in that range.
This indicates that the exponential decrease of $\alpha$ is valid in that range.
We complement this test by the computation of $\sigma_{\rm norm}$,
the normalised variance of the estimate of $\alpha$ (see Eq.~(\ref{snorm}), \S~\ref{aconv}),
which is presented in figure~\ref{beta_dvf_6_4} (c).
In the whole range of interest,
we can see that this quantity is around one for the whole range of $\beta$ considered here.
This confirms that the estimates performed by means of AMS are not tainted by an uncontrolled error, and that
the exponential decrease of $\alpha$ is valid in the whole range of interest.
Further tests of validity of computations using this indicator are presented in appendix~\ref{aconv}.

\subsection{Duration of trajectories}\label{setau}

After examining whether the probability of crossing and the mean first passage time
before crossing displayed typical exponential behaviour with $\beta$,
we will test another statistical property of reactive trajectories: their average duration.
We will investigate whether we find the typical slow growth with $\beta$, and we will separate the case of the divergence free forcing and the case
of the forcing on components, where we will further test the effect of the energy injection rate $\epsilon$.

\begin{figure}[!htbp]
\centerline{\includegraphics[width=6.5cm]{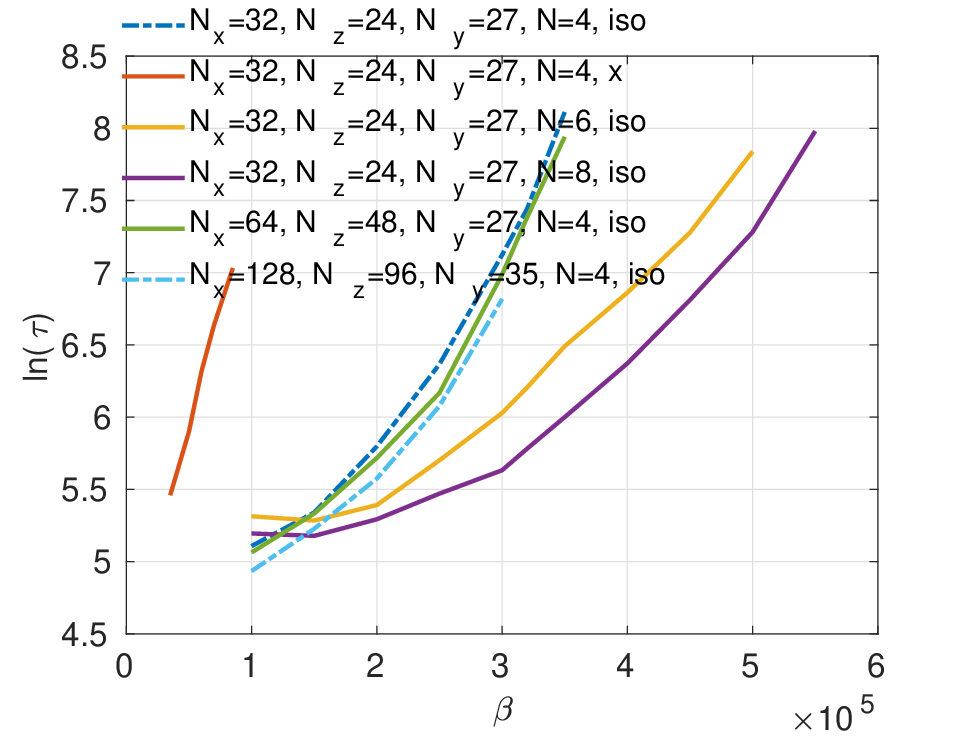}}
\caption{Logarithm of the duration of reactive trajectories as a function of $\beta$
for various forcing shapes (spectrum, resolution, forced components) computed by means of AMS in the MFU size system.
}
\label{tauepsilon}
\end{figure}

\subsubsection{Forcing on components}

The average duration $\tau$ of build up trajectories is computed as a function of $\beta$
for several shape of the spectrum of the forcing on components ($\sigma_{y,z}=0$ or not,
several values of $N_r$ at several resolutions). We first observe the logarithm of  $\tau$ as a function of $\beta$,
for all these cases in figure~\ref{tauepsilon}, (a).
We note that, as is commonly reported, $\tau$ increases with $\beta$ at fixed spectrum shape.
Moreover, the growth of the logarithm of  $\tau$ is faster than linear
with the parameter $\beta$ controlling the energy injection rate independently of the structure of the forcing.
We can observe this plot differently and consider $\tau$ at fixed $\beta$ as the spectrum shape is changed.
In particular, we consider the effect of a decreased energy injection rate as less modes are forced (as indicated in Eq.~\ref{defepsilon}).
Among forcing on all three components, we note that as we decrease $N_r$ from $8$ to $4$,
thus reducing the range of low wavenumbers where the spectrum is white and consequently reducing $\epsilon$,
$\tau$ is larger and larger. At fixed $N_r=4$, as we go from forcing on three components to forcing
on the one streamwise component, $\tau$ is even larger.
From these two observations, we can conclude that our estimates of $\tau$ all increase as the energy injection rate $\epsilon$ is decreased:
indeed, equation~(\ref{defepsilon}) indicates that both increasing $\beta$ and decreasing $N_r$ reduces the energy injection rate.
From these AMS computations, we also estimated the probability of crossing and the mean first passage time before crossing (not shown here).
Similarly to what has been observed in the former section, we have respectively a rapid decrease of $\alpha$ and a rapid increase
of $T$ as $\beta$ is increased.
Moreover, we also note that the effect of the spectrum shape is similar to what has been observed with $\tau$ : at fixed $\beta$,
injecting less energy leads to a decrease of $\alpha$ and an increase of $T$.
We have observed as a function of $\epsilon$ the average duration of trajectories $\tau$, the probability of observing a trajectory $\alpha$
and the mean first passage time before a trajectory occurs (not shown here).
While this rescaling tend to regroup the curves, the precise collapse is debatable.
The matter is not trivial:
 we expect $\tau$ to generally grow with $\frac{1}{\epsilon}$:
more energy injected per unit time should mean faster travel toward turbulence.
However, there is also no reason why the shape of the spectrum of the forcing noise should have no subtle effect,
we have not examined the energy transfer, direct and indirect, in detail.

From all this we are drawn to address two matters
\begin{enumerate}[i)]
\item
Firstly, we find that the energy injection rate $\epsilon$ has the same type of effect on the average kinetic energy,
on the type of path generated (in term of stages) and on quantitative properties of the paths like their average duration $\tau$,
with moderate effect of the forcing structure within this forcing family.
We can further back the assertion that the computed reactive trajectories are selected by the non linear dynamics and not by the forcing spectrum.
Behind these tests, we aim to investigate the question of universality in these paths:
to what extent are they a product of the Navier--Stokes equations alone and
the energy injection rate as a sole parameter,
and to what extent do they depend on the forcing spectrum.
\item Secondly, the growth of $\tau$ with the parameter controlling the noise variance (or energy injection rate) is apparently exponential:
its logarithm is convex.
As such it is quite faster then what had been previously observed in simpler systems \cite{rolland2015statistical,rolland2016computing}.
Indeed, it has been demonstrated in one degree of freedom systems having a simple saddle between both metastable states \cite{cerou2013length}
and verified numerically in such systems having more DoF \cite{rolland2015statistical,rolland2016computing}
that the duration of reactive trajectories grew logarithmically with the variance of the noise forcing.
When the saddle between both states become extremely flat, the reactive trajectories undergoes a random walk on the plateau between both states.
In that case, a faster linear increase of the duration of reactive trajectories is measured \cite{rolland2016computing}.
However, this leads to a Gaussian distribution of trajectory durations, unlike the Gumbel observed here (Fig.~\ref{distrib}).
An exponential increase can be seen in simple models if there is a local (weakly) metastable minimum standing between the two main metastable states.
In that case, the exponential increase is caused by the waiting time before escape of the secondary metastable state \cite{rolland2015statistical}.
In the case of build up of turbulence, there is no known intermediate metastable state.
However the complexity of phase space could play a similar role.
\end{enumerate}
\subsubsection{Divergence free forcing}

We present the average duration of reactive trajectories obtained under divergence free forcing as a function of $\beta$
in figure~\ref{beta_dvf_6_4t}.
In contrast with the behaviour of reactive trajectories obtained with forcing on components,
the average duration of reactive trajectories grows rather slowly.
Indeed, we can observe an increase of $\tau$ with $\beta$ up to $\beta\lesssim 260000$,
followed by a plateau for $260000 \lesssim\beta\lesssim 3300000$.
Above that value of $\beta\gtrsim 330000$, the average duration of reactive trajectories estimated by means of AMS is smaller.
As such, it is not straightforward to precisely fit a known functional dependence of $\tau(\beta)$ on that data,
even if the slow logarithmic growth would be the closest to what we estimated.
There are apparent distinctions with that behaviour, such as the local maximum.
Such maxima are not commonly reported in simpler noise driven transitions.
An effect of the large error bars cannot be ruled out,
especially since no change of tendency is observed in $\log(\alpha)$, $\log(T)$ nor in the trajectories.

\section{Conclusion}\label{sconcl}

\subsection{Summary}

In this text, we studied the trajectories going from laminar flow to transitional turbulence in plane Couette flow under a stochastic forcing,
in systems of increasing sizes.
The energy injection rate was decreased so that the mean waiting time before observing said trajectory became too large to be estimated by means of
DNS computations.
A rare events simulation method, Adaptive Multilevel Splitting, was therefore used for the systematic computations of trajectories
along with the probability that they occur and the mean first passage time before they occur.
In order to determine whether trajectories crossed a separatrix, and in that case where they crossed the separatrix, several diagnostics were used.
Firstly, the last state at the last branching stage was systematically computed.
Moreover, dynamics without forcing, that relax either to the laminar or to the turbulent flow,
were performed from the successive states on the trajectory.
Finally, using the last state relaxing to the laminar flow and the first state relaxing to the turbulent flow, bisections were
 performed until they converged to an unstable fixed point or an unstable limit cycle.
These computations were performed in domains of increasing sizes, from a MFU type flow  of size $L_x\times L_z=6\times 4$
to a system of size $L_x\times L_z=36\times 24$, to investigate effects of localisation.
In the MFU type domains, the effect of decreasing energy injection rate was investigated to study how these transitions became rare.
The numerical computations were validated: doubling of modes in each direction indicated
that numerical convergence with respect to space and time discretisation was obtained,
while comparison of AMS computations with DNS and tests of the variance of the estimator indicated that reactive trajectories were correctly computed.
The flow displayed on the paths was described, in view of the literature on transition to turbulence.
The avoidance of the saddles by the computed reactive trajectories was particularly noted.
Finally, the rarity of transitions was highlighted
through a parametric study in energy injection rate in the MFU type domain.

\subsection{Mechanical description of the paths}

Observation of the trajectories indicated that no matter the domain size and the structure of forcing,
the flow responded to the added noise by generating streamwise invariant streamwise velocity tubes, along with weak streamwise vorticity.
On the trajectory, these tubes gradually accumulate energy obtained from the forcing,
in that way the non trivial flow is sustained by the statistically steady energy injection.
The tubes finally become intense enough for the self sustaining process of turbulence to start,
after the trajectory has crossed the separatrix between laminar and turbulent flow.
It is notable that plane Couette flow was efficiently able to generate these tubes, even if the
wavenumber forcing range was distinct from the natural wavenumber range of transitional turbulence.
In that respect, these trajectories are distinct from the one where the wavenumber of velocity streaks and streamwise vortices
are directly non-linearly forced \cite{rigas2021nonlinear}.
This indicates that the flow is able to redistribute the energy from one range of wavenumbers to another, even in the regime
where the energy injection rate is small.
Moreover, this hints that the resulting coherent structures (in the form of streamwise velocity tubes)
are naturally selected by plane Couette flow, not imposed by the forcing.
In that respect, this would mean that we observed a non linear receptivity of the flow in the first part of the trajectory.
Because that receptivity was independent from the several types of forcing we applied,
it could indicate some universality in the response of the flow to the forcing.
That search for universal features in the transition, such as a mediator flow state arising solely from the Navier--Stokes equations in plane Couette flow,
and the corresponding mechanical processes, is then in line with the computation of the properties of the separatrix such as the saddle on it
or the point on it which is the closest to the laminar flow, as measured by kinetic energy.
The build up is distinct from these last two flow states because minimal seeds are precisely defined non linear flows and would not naturally emerge from
finite variance noise and, as we will further discuss, edge states are avoided by the computed build up trajectories.
What we observe here is closer to (non)-linear response,
with the distinction that we have not optimised the forcing,
and we are not observing the optimal response to any regular enough forcing.

Similarly to paths to turbulence identified using other approaches (forcing, bisections, optimisations),
the reactive trajectories computed here display a spatial localisation
of the triggered velocity streaks as the domain size is increased.
This localisation is visible in the last state at the last stage.
The main difference is that in those flow states the actual self sustaining process of turbulence is
triggered in a narrow region of the domain.
We have been able to observe spanwise localisation in domains of size $L_x\times L_z=18\times 12$
and $L_x\times L_z=36\times 24$.
We expect that with further increased domain size, such as $L_x\times L_z=68\times 34$, we will also
observe streamwise localisation of triggered turbulence, as it is the case in edge states and minimal seeds.
We have not performed these computation for this text due to the increased numerical cost.
As explained in appendix~\ref{aconv}, we expect that AMS computations in such a larger domain would function
as well as those presented in this text.

\subsection{Statistical description of the paths}

After regrouping our observations on the spatial organisation of the coherent flow structures
encountered during the build up, we now discuss the build up in a statistical manner.
This means considering trajectories as a group, histograms and averages, their parametric dependence,
in view of what is known on bistability in other noise driven system.

On several aspects, there is a similarity between the trajectories of build up of turbulence under forcing
and that of collapse of turbulence. The paths followed in the plane $\log(E_{k,x}) - \log(E_{k,y-z})$ both show a strong convexity.
The succession of events : first the generation of streamwise velocity tubes followed by the appearance of streamwise vorticity
(with spanwise localisation when domain size is increased)
can qualitatively seem as a rewind of those observed in the collapse of turbulence \cite{rolland2022collapse,liu2021decay,gome2022extreme}.
Elements of this scenario had been noted in earlier studies without being
at the centre of attention yet \cite{philip2011temporal,chantry2014studying}.
However, the build-up and collapse trajectories are \emph{by no mean} time reversed paths of one another.
Firstly the aforementioned  similarity is not exact.
Secondly, they take place under  clearly distinct conditions: the build up is forced,
while collapses are purely deterministic.
A corollary is that build up paths cross a separatrix, with the help of the forcing,
while the collapse paths do not: they avoid it.
This induces further differences, such as the duration of each phase:
the generation of the streamwise vorticity tubes shows a polynomial time dependence,
while the decay of said tube is exponential in time \cite{rolland2022collapse}.
The distinction between the forward and backward types of paths
is also different from what is found in noise driven non-gradient system.
In those system, there is a clearly distinction between the backward and forward path.
However, they both comprise of a fluctuation path toward the same saddle, followed by a relaxation path.
Meanwhile, the build up has a fluctuation path toward the separatrix then relaxes, while the collapse can
be viewed as a pure relaxation path.

\begin{figure*}[!htbp]
\centerline{
\begin{pspicture}(7,6)
\rput(0.25,5.75){\textbf{(a)}}
\pscircle[fillstyle=solid,fillcolor=pink](1,1){0.75}
\rput(1,1){$\mathcal{A}$}
\pscircle[fillstyle=solid,fillcolor=green](5,5){0.75}
\rput(5,5){$\mathcal{B}$}
\pscurve[linecolor=OliveGreen](0.5,3.5)(4,2.8)(6,0)
\pscircle[linecolor=OliveGreen,fillstyle=solid,fillcolor=OliveGreen](0.5,3.5){0.03}
\pscurve[linecolor=yellow](1.4,1.4)(3,2.25)(4,2.8)(4.9,4)(5,4.5)
\pscurve[linecolor=yellow,linestyle=dashed](1.4,1.6)(3.75,2.92)(4.83,4.5)
\pscurve[linecolor=yellow,linestyle=dashed](1.4,1.2)(4.25,2.68)(5.17,4.5)
\pscurve[linecolor=cyan](1.4,1.4)(4,2.8)(5,4.5)
\psline[linecolor=red]{->}(4,2.8)(4.25,3.1)
\psline[linecolor=red]{->}(4,2.8)(3.75,2.5)
\psline[linecolor=red]{<-}(4,2.8)(3.7,3.0)
\psline[linecolor=red]{<-}(4,2.8)(4.3,2.55)
\rput(5.4,2.8){\textcolor{red}{transition state}}
\pscurve[linecolor=blue](1,1.5)(1.5,4)(4.5,5)
\pscurve[linecolor=blue,linestyle=dashed](0.85,1.5)(1.35,4)(4.5,5.15)
\pscurve[linecolor=blue,linestyle=dashed](1.15,1.5)(1.65,4)(4.5,4.85)
\rput(1.5,5){\textcolor{blue}{Observed path}}
\rput(1.5,4.75){\textcolor{blue}{smallish $\epsilon$}}
\rput(5.75,4){\textcolor{cyan}{Instanton}}
\rput(5.75,3.75){\textcolor{cyan}{$\epsilon\rightarrow 0$}}
\rput(3,1){\textcolor{yellow}{Expected path}}
\rput(3,0.75){\textcolor{yellow}{smallish $\epsilon$}}
\rput(6.65,0.4){\textcolor{OliveGreen}{Separatrix}}
\end{pspicture}
\hspace{1.5cm}
\begin{pspicture}(7,6)
\rput(0.25,5.75){\textbf{(b)}}
\pscircle[fillstyle=solid,fillcolor=pink](1,1){0.75}
\rput(1,1){$\mathcal{A}$}
\pscircle[fillstyle=solid,fillcolor=green](5,5){0.75}
\rput(5,5){$\mathcal{B}$}
\pscurve[linecolor=OliveGreen](0.5,3.5)(1.2,3.4)(2.7,3.05)(4,2.8)(6,0)
\pscurve[linecolor=cyan](1.1,1.4)(1.3,3.3)(1.45,3.27)(1.9,3.15)(2.7,3.0)(3.7,2.85)(3.9,2.82)(4,2.8)(5,4.5)
\pscurve[linecolor=blue](1,1.5)(1.2,3.4)(1.5,4)(3,4.75)(4.5,5)
\pscurve[linecolor=blue,linestyle=dashed](0.85,1.5)(1.35,4)(4.5,5.15)
\pscurve[linecolor=blue,linestyle=dashed](1.15,1.5)(1.65,4)(4.5,4.85)
\pscircle[linecolor=OliveGreen,fillstyle=solid,fillcolor=OliveGreen](0.5,3.5){0.03}
\pscircle[linecolor=cyan,fillstyle=solid,fillcolor=cyan](1.2,3.4){0.06}
\rput(0.2,3.15){\textcolor{cyan}{mediator}}
\psline[linecolor=red]{->}(4,2.8)(4.25,3.1)
\psline[linecolor=red]{->}(4,2.8)(3.75,2.5)
\psline[linecolor=red]{<-}(4,2.8)(3.7,3.0)
\psline[linecolor=red]{<-}(4,2.8)(4.3,2.55)
\rput(1.5,5){\textcolor{blue}{Observed path}}
\rput(1.5,4.75){\textcolor{blue}{smallish $\epsilon$}}
\rput(5.75,4){\textcolor{cyan}{Instanton}}
\rput(5.75,3.75){\textcolor{cyan}{$\epsilon\rightarrow 0$}}
\rput(6.65,0.4){\textcolor{OliveGreen}{Separatrix}}
\end{pspicture}
}
\caption{(a) Sketch of the two metastable states $\mathcal{A}$ and $\mathcal{B}$ in phase space,
separated by a separatrix (green curve), on which we have the transition state:
the most accessible saddle point (red arrows indicating stable and unstable directions).
In the limit of vanishing energy injection rate $\epsilon\rightarrow 0$, the reactive trajectory should follow an instanton (in cyan):
on this first sketch, we represent an instanton that nears the separatrix only as it reaches the saddle along mostly unstable directions.
For small but finite energy injection rate $\epsilon$, we would expect that trajectories concentrate around the instanton (in yellow).
Instead, we systematically observe trajectories that concentrate around a path that crosses the separatrix away
from the instanton and the most accessible fixed point (in blue).
For these two types of paths: the full line indicates the average trajectory,
while the dashed lines indicate where the zone that contains the majority of paths.
(b) Modification of sketch (a) in the case where the instanton approaches the separatrix near a mediator state,
then goes along the separatrix approaches the nearest fixed point and finally travels towards $\mathcal{B}$, with a scenario similar to the one described by \cite{wan2015model}.}
\label{skconcl}
\end{figure*}
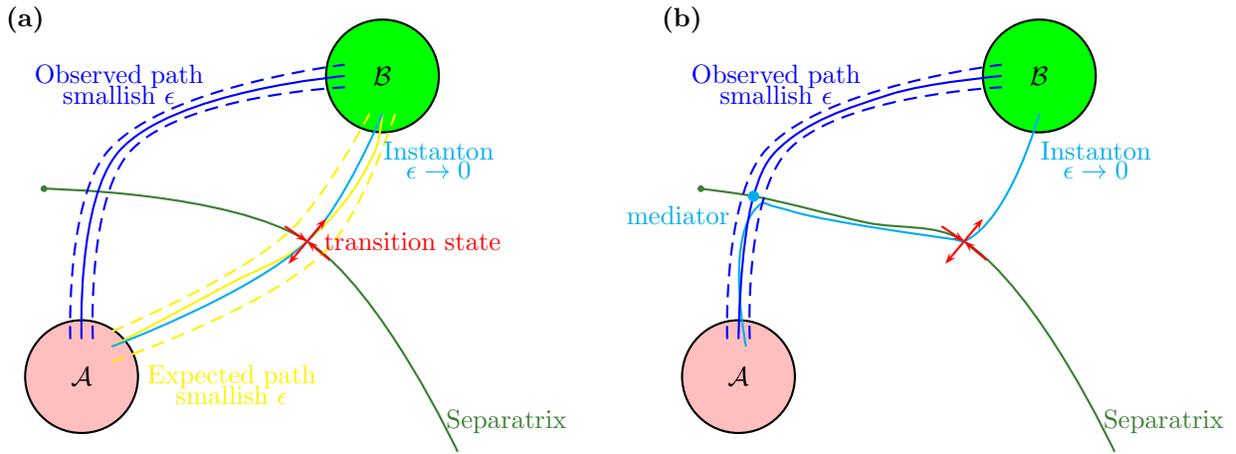

The relaxations that preceded bisections indicate that the reactive trajectories do cross the separatrix between laminar and turbulent flow.
They do so with a kinetic energy which is several orders of magnitude larger than that of
minimal seeds, typically of the order of $\mathcal{O}\left(10^{-6}\right)$.
It is doubtful that this difference could be solely explained by the fact that the typical slow
power law decrease of such energies starts at larger Reynolds numbers.
The ensuing bisection indicates that the separatrix is crossed in the stable manifold of the edge state, which also
has a relatively large kinetic energy. 
For all sizes and Reynolds numbers considered, this edge is fairly distinct from the flow
configuration observed on the last state at the last stage and in the two flow states on the reactive trajectories
just before and just after the crossing of the separatrix.
The edge has larger spatial scales and less intense, more coherent structures in wall normal and spanwise components than what is observed
in the reactive trajectories.
The edges are less localised than the trajectories in larger domains.
In the MFU type domain, The unstable limit cycle has more intense velocity streaks than what is seen on trajectories.
This avoidance of the saddle by the reactive trajectories is sketched in figure~\ref{skconcl} (b),
where this atypical case is shown in comparison with the typical case of figure~\ref{skconcl} (a).

The comparison between the reactive trajectory and the nearest fixed point on the crossed separatrix is very often the focus of attention.
This is because, as stated earlier, in systems which do not have a degenerate saddle \cite{bouchet2012non},
in the limit of energy injection and/or forcing noise variance going to zero,
the reactive trajectories are structured by instantons, which cross the separatrix right
at a saddle point (see the sketch in figure~\ref{skconcl} (a)).
In said regime, one finds a structuration of the bistability: an exponential dependence of the mean first passage time and crossing probability in
the noise variance, a concentration of trajectories in phase space and a specific distribution of reactive trajectories duration,
to mention the most important features.
Surprisingly enough, all these properties were found in our reactive trajectories.

This raises the question of why do we have this apparent contradiction.
There can be several causes for such a situation:
\begin{enumerate}[i)]
\item The paths sampled by means of AMS could be incorrect.
This can happen when the reaction coordinate is not adapted.
One then observes the apparent bias phenomenon: most sample reactive trajectories
do not follow the most likely path \cite{brehier2016unbiasedness,lucente2022coupling}.
However, that type of behaviour would show a clear distinction between the trajectory durations computed by means of AMS and those computed by means
of DNS as well as lead to a very large relative sample variance of the probability of crossing, which is not the case here.
Moreover, when this phenomenon manifests itself,
the reactive trajectories tend to select a path that still visit the neighbourhood of a saddle \cite{rolland2015statistical}.
\item One can observe correctly computed reactive trajectories which are outside of the instanton regime if the
relevant saddle is very flat \cite{rolland2016computing}.
In that situation, one does observes an exponential growth of mean first passage time and an exponential
decay of the crossing probability, which is our case.
Meanwhile, one observes a Gaussian distribution of trajectory duration and a linear increase of their average duration and variance,
which is not our case.
Moreover, this regime still visits the neighbourhood of a saddle, which is not our case.
\item When the saddle in the system can become degenerate, one can observe sub-instantons,
that do not follow the expected path \cite{bouchet2012non}.
However, in that case, the probability of crossing and the mean first passage time have a very distinct scaling
in exponential of the square root of the noise variance, which is not our case.
\item\label{sc4} Finally, a peculiar situation has been reported following the computation of the instanton going from a non linear traveling wave
to the laminar flow in two dimensional plane Poiseuille flow \cite{wan2015model}.
In that case, sketched in figure~\ref{skconcl} (b), the instanton reaches the neighbourhood of the separatrix near a flow configuration
termed a mediator state,
travels along the separatrix toward the unstable fixed point then relaxes toward its destination.
The peculiarity of that situation is that once the flow has reached the mediator point,
a small numerical error in instanton calculation, or a small amount of noise in a stochastic system,
can push the dynamics on the other side of the separatrix : the dynamics then travels near deterministically toward the final state $\mathcal{B}$.
Let us term $\mathcal{I}_{\mathcal{A}\rightarrow\mathcal{B}}$ the full action of the instanton going from $\mathcal{A}$
to $\mathcal{B}$, $\mathcal{I}_{\mathcal{A}\rightarrow m}$, the action of a trajectory following the instanton
from $\mathcal{A}$ to the mediator state, $\mathcal{I}_{m\rightarrow s}$ the action of a trajectory following the instanton
from the mediator state to the saddle and $\mathcal{I}_{m\rightsquigarrow \mathcal{B}}$
the action of a trajectory starting on the mediator and being slightly perturbed to cross
the separatrix and then travel deterministically toward $\mathcal{B}$.
The action of deterministic paths is always zero: this means that in the limit of noise variance going to zero,
the action of any path starting on the $\mathcal{B}$ side of the separatrix to $\mathcal{B}$ that has any chance of being observed is zero,
because it will follow a deterministic path.
This means that $\mathcal{I}_{\mathcal{A}\rightarrow\mathcal{B}}=\mathcal{I}_{\mathcal{A}\rightarrow m}+\mathcal{I}_{m\rightarrow s}\simeq \mathcal{I}_{\mathcal{A}\rightarrow m}$:
the action of the instanton is that of traveling toward the mediator point then to the saddle and then relaxing.
Moreover the part of the path along the separatrix has a negligible action $\mathcal{I}_{m\rightarrow s}\ll \mathcal{I}_{\mathcal{A}\rightarrow m}$
because it is a near deterministic path in the stable manifold of the saddle.
Similarly, the action of a path traveling toward the mediator along the instanton
then receiving a small amount of noise to cross the separatrix there
(as sketched in blue  sketched in figure~\ref{skconcl} (b))
is $\mathcal{I}_{\mathcal{A}\rightsquigarrow\mathcal{B}}=\mathcal{I}_{\mathcal{A}\rightarrow m}+\mathcal{I}_{m\rightsquigarrow \mathcal{B}}\simeq \mathcal{I}_{\mathcal{A}\rightarrow m}$.
The action $\mathcal{I}_{\mathcal{A}\rightarrow \mathcal{B}}$ and $\mathcal{I}_{\mathcal{A}\rightsquigarrow\mathcal{B}}$ are very similar and the corresponding probabilities
and mean first passage time respectively go to zero and to infinity exponentially
as very similar rates.
At finite though small noise variance, the probability of observing a trajectory going along the instanton all the way becomes small compared to
observing a trajectory going toward the mediator then crossing toward $\mathcal{B}$ there,
because the instanton requires the trajectory not to be perturbed once the mediator has been reached.
Therefore, computations by means of numerical simulation of the stochastic process (like DNS or AMS)
would give the trajectory crossing near the mediator. Moreover, trajectories would still concentrate up to the mediator.
However, the estimate of the probability and the mean first passage time would go like $\exp(\pm \beta \mathcal{I}_{\mathcal{A}\rightarrow m})$ and thus be exponential
and very close to the one predicted by the Freidlin Wentzel principle of large deviations.
In our case, the exponential dependence of $\alpha$ and $T$, the crossing of the separatrix away from the saddle and the concentration of
paths are consistent with this scenario.
We note however that the computed path do not concentrate on top of one another in planes computed using kinetic energy.
This could be an effect of the finite noise variance in these quadratic spatially averaged quantities. Indeed,
the mechanical scenario is not changed from one value of $\beta$ to another.
\end{enumerate}

As far as we can deduce from the data presented here, scenario~\ref{sc4} is the most similar to build up in plane Couette flow.
If it is the case, this would mean that we would observe an actual instanton all along the trajectory, not just until the mediator or turning point
only, for exceedingly small values of the noise.
Moreover, if we follow this line of reasoning, we cannot rule out the possibility that there could be a mediator even closer to the laminar state,
comparable to a minimal seed, that would only be reached at minuscule values of noise variance.
It could be even beyond the reach of rare events simulations methods.
The case of build up in plane Couette flow thus helped us identifying a different regime of convergence toward instantons,
already pointed out in two dimensional plane Poiseuille flow \cite{wan2015model}
and can be linked to observations made in other systems where crossing away from the saddle
and very flat dependence of the action on the crossing point in model of oceanic flows \cite{borner2023saddle}.
This behaviour could thus manifest itself in more complex bistable geophysical flows.
A similar behaviour was observed in a model representing a first order phase transition \cite{grafke2017non},
indicating that this type of analysis is also relevant for statistical physics.

\section*{Acknowledgment}
The author thanks Institut PPrime, ENSMA and their computational facilities, where this work was started, for its hospitality.
The author thanks the help of the computational resources of the PSMN platform of ENS de Lyon, as well as the hospitality of the laboratoire
de Physique de l'ENS de Lyon,
where part of the computation of this work have been performed.
Additional computational resources were granted by \emph{M\'esocentre de Calcul Scientifique Intensif} of the University of Lille and
this work was granted access to the HPC resources of IDRIS under the allocation 2022-A0122A01741 made by GENCI.
The author also thanks the support from CICADA \emph{(centre de calcul interactifs)} (now \emph{m\'esocentre Azzurra})
of University Nice Sophia-Antipolis where some early computations have been performed.
The author finally thanks F. Bouchet, R. B\"orner, J.-C. Vassilicos and J.-P. Laval for inspiring discussions,
as well as E. Vanden--Eijnden for pointing interesting references.
\appendix

\section{Structure of the forcing noise}\label{stnoise}

In section~\ref{sfns}, we have introduced a forcing $\mathbf{f}$ in the Navier--Stokes questions for the departure
to laminar plane Couette flow (Eq.~(\ref{nsf})).
We had stated that this forcing was prescribed using the spectrum of its correlation function (Eq.~(\ref{spcor})),
governed by the shape factor $\Gamma_{l,n_x,n_z}$ on component $l$ and in plane wavenumbers $n_x$ and $n_z$.
We had mentioned the two families of forcing used to generate the build up trajectories, illustrated with a visualisation of $\Gamma_{l,n_x,n_z}$
(Fig.~\ref{fact} (a,b)).
In this appendix, we will give a quantitative description of the procedure followed to generate these two forcings:
\begin{itemize}
\item Forcing on components. In that case, we prescribe the shape factor
of the Fourier transform of the correlation function: we rewrite $\Gamma_{l,n_x,n_z}$ using tensors  $\sigma_l\gamma_{n_x}\gamma_{n_z}$
on Fourier component $n_x,n_z$.
The tensor $\sigma_l$ scales the correlation function on each component: it can also be used
to select a forcing on one component rather than the other, using $\sigma_l=0$.
Meanwhile, the tensors $\gamma$ prescribe the shape of the spectrum of the correlation function.
They all have in common the property that $\gamma_{n_x}=0$ if $n_x>\left\lfloor\frac{N_x}{2} \right\rfloor$,
$\gamma_{n_z}=0$ if $n_z>\left\lfloor \frac{N_z}{2}\right\rfloor$,
$\gamma_{n_x}=0$ if $n_x<\left\lfloor\frac{L_x}{L_c}\right\rfloor$, $\gamma_{n_z}=0$ if $n_z<\left\lfloor\frac{L_x}{L_c}\right\rfloor$.
This first restriction is used to avoid forcing aliased modes.
We always used $L_c=1.0$ for the largest scale of cut-off of the forcing.
This second restriction ensures that the low wavelength of transitional wall turbulence are not directly forced.
Using a cut-off wavenumber $N_r$,  we set $\gamma_{n_x}=1$ for $\left\lfloor\frac{L_x}{L_c}\right\rfloor\le n_x\le N_r$
and $\gamma_{n_z}=1$ for $\left\lfloor\frac{L_z}{L_c}\right\rfloor\le n_z\le N_r$: we have a white spectrum at small wavenumbers.
We then set $\gamma_{n_x}=\frac{N_r}{n_x}$ for
$N_r\le n_x\le \left\lfloor\frac{N_x}{2} \right\rfloor$ and $\gamma_{n_z}=\frac{N_r}{n_z}$
for $N_r\le n_z\le \left\lfloor \frac{N_z}{2}\right\rfloor$:
we have a red spectrum at larger wavenumbers.
The forcing on component $l$ at wavenumber $n_x$, $n_z$ is then a Gaussian random number
of variance $\frac{1}{\beta}$ times $(\sigma_l\gamma_{n_x}\gamma_{n_z})^2$.
This means that $\Gamma_{l,n_x,n_z}=(\sigma_l\gamma_{n_x}\gamma_{n_z})^2$.
We considered such a forcing on the three components (termed ``iso'', with $\sigma_x=\sigma_y=\sigma_z$)
or only on the streamwise component (termed ``x'' with $\sigma_x=1$, $\sigma_y=\sigma_z=0$).
We only used it in MFU type systems of size $L_x\times L_z=6\times 4$ for the standard resolution.
We have used several values of $N_r=4$, $6$ and $8$ (which are specified when the results are presented.

\item  Divergence free forcing obtained from a curl.
For this matter, we use a potential vector $A_{n_x,n_y}$, such that forcing on component $x$ at wavenumber $n_x$, $n_z$ is
$-\frac{1}{\sqrt{\beta}}\frac{2\pi n_z}{L_z}A_{n_x,n_z}$, there is no forcing on component $y$ and the forcing on component $z$ is
 $\frac{1}{\beta}\frac{2\pi n_x}{L_x}A_{n_x,n_z}$, with $k_x=\frac{2\pi n_x}{L_x}$ and $k_z=\frac{2\pi n_z}{L_z}$.
We then have $A_{n_x,n_z}=\gamma_{n_x}\gamma_{n_z}\xi$, with $\xi$ a Gaussian random number of variance $1$.
These factors are non zero in a set range of wavenumber: we set that
$\gamma_{n_x}=0$ if $n_x>\left\lfloor \frac{N_x}{2} \right\rfloor$,  $\gamma_{n_z}=0$ if $n_z>\left\lfloor \frac{N_z}{2} \right\rfloor$,
$\gamma_{n_x}=0$ if $n_x<\left\lfloor\frac{L_x}{L_c}\right\rfloor$, $\gamma_{n_z}=0$ if $n_z<\left\lfloor\frac{L_x}{L_c}\right\rfloor$.
We still set $A_{n_x,n_z}=\gamma_x\gamma_z$. This time we have $\gamma_x=\frac{1}{n_x}$ for $\left\lfloor\frac{L_x}{L_c}\right\rfloor\le n_x\le N_r$
and $\gamma_z=\frac{1}{n_z}$ for $\left\lfloor\frac{L_z}{L_c}\right\rfloor\le n_z\le N_r$.
We then set $\gamma_{n_x}=\frac{N_r}{n_x^2}$ for $N_r\le n_x\le\left\lfloor \frac{N_x}{2}\right\rfloor$
and $\gamma_{n_z}=\frac{N_r}{n_z^2}$ for $N_r\le n_z\le \left\lfloor \frac{N_z}{2}\right\rfloor$.
The Fourier transform of the correlation function $\Gamma_{x,n_x,n_z}=\frac{4\pi^2n_z^2}{L_z^2}A_{n_x,n_z}^2$
is displayed in the $k_x,k_z$ plane in figure~\ref{fact} (a),
$\Gamma_{y,n_x,n_z}=0$ and $\Gamma_{z}=\frac{4\pi^2 n_x^2}{L_x^2}A_{n_x,n_z}^2$ is displayed in the $k_x,k_z$ in figure~\ref{fact} (b).

In the MFU type domain of size $L_x\times L_z=6\times 4$, for the standard resolution, we used the parameters $L_c=1.0$ and $N_r=4$.
In that domain, for the finer resolution, we used the parameters $L_c=1.0$ and $N_r=4$.
In the  domain of intermediate size $L_x\times L_z=18\times 12$, we used the parameters $L_c=1.0$ and $N_r=12$.
In the largest domain, of size $L_x\times L_z=36\times 24$, we used the parameters $L_c=1.0$ and $N_r=24$.
\end{itemize}

\section{Derivation of the kinetic energy budget}\label{abud}

In section~\ref{sdfl},
we have presented the ensemble averaged budget of spatially averaged kinetic energy (Eq.~(\ref{rolland_bud}))
which resembles the Reynolds--Orr energy equation, with the addition of an energy injection rate (Eq.~(\ref{defepsilon})).
We have stated succinctly that this budget, and the similar budget of total energy (Eq.~(\ref{total_bud})),
was obtained by application of the It\^o Lemma for change of variables in stochastic differential equations.
In this appendix, we present the main steps to go from the Stochastically forced Navier--Stokes equations~(\ref{nsf}) to
the energy budgets (Eq.~(\ref{rolland_bud}),~(\ref{total_bud})).

We first present the principle of such a change of variable
for a scalar time dependent variable $f(t)$ following a time discretised Langevin equation of the type
\begin{equation}
f(t+{\rm d}t)-f(t)=df=F(f(t)){\rm d}t+B(f(t)){\rm d}W\,,
\end{equation}
where ${\rm d}W(t)$ is a Wiener process discretised with It\^o convention.
This means that $\langle {\rm d}W(t)\rangle=0$, $\langle {\rm d}W(t){\rm d}W(t')\rangle={\rm d}t\delta(t-t')$,
and the noise covariance $B(f)$ and the Wiener process are uncorrelated $\langle B(f(t)) {\rm d}W(t)\rangle=0$.
The It\^o Lemma (see \citep{gardiner2009stochastic} \S~4) states how a change of variable
towards the Langevin equation for $g(f)$ is performed.
Its application leads to a stochastic differential equation where the noise is still interpreted with an It\^o convention.
It shows that as soon as $g$ is non linear, an additional deterministic term should be included,
on top of the operations that would be naturally performed
to change variables in an ordinary differential equation.
The resulting Langevin equation reads
\begin{equation}
dg=\frac{dg}{df}F(f){\rm d}t+\frac12 B^2\frac{d^2g}{df^2}{\rm d}t+\frac{dg}{df}B(f){\rm d}W\,.\label{itofor}
\end{equation}
In this stochastic differential equation, the terms $\frac{dg}{df}F(f){\rm d}t$ and
$\frac{dg}{df}B(f){\rm d}W$ are what would arise from a change of variable in an ordinary differential equation,
and can be obtained by any standard mean.
The additional term $\frac12 B^2\frac{d^2g}{df^2}{\rm d}t$, sometimes called a \emph{ghost drift},
accounts for the fact that the correlation between the noise and the new variable,
while physically existent, are excluded from the noise term by the It\^o convention  in equation~(\ref{itofor}).
This convention leads to $\left\langle \frac{dg}{df}B(x){\rm d}W\right\rangle=0$ when equation~(\ref{itofor}) is ensemble averaged,
to give
\begin{equation}
\left\langle \frac{dg}{dt} \right\rangle=\left\langle\frac{dg}{df}F \right\rangle +\left\langle \frac12 B^2\frac{d^2g}{df^2} \right\rangle\,.
\end{equation}
This new ghost drift term is absent if $g$ is affine in $f$, and it is an additive constant
if $g$ is quadratic in $f$ and the noise covariance $B$ is independent on the variable $f$.
This last situation represents a one degree of freedom version of our derivation.
We now present the operation in the in more general cases of multidimensional Langevin equations for vector function
$\vec f(t)$ with $N$ components. In that case, $B$ is an $N\times N$ matrix, applied to a vector of independent Wiener processes.
It\^o Lemma then tells us that to perform
a change of variables $g\left(\vec f\right)$, the first part of the resulting equation can be obtained through the classical means,
and that we need to find a convenient mean to calculate the resulting ghost drift that can formally be written as
\begin{equation}\frac{1}{2}\sum_{i=1}^N\sum_{j=1}^N\sum_{k=1}^NB_{ik}B_{kj}\frac{\partial^2g}{\partial f_i\partial f_j}\,, \label{conth}
\end{equation}
where $BB^\dag=\sum_{k=1}^NB_{ik}B_{kj}$ is the covariance matrix of the multidimensional noise, which is contracted against the Hessian matrix of $g$.

In our case, we change variables from the velocity fields $\mathbf{u}$ to the spatially averaged kinetic energy.
This change of variables is quadratic in the variable $\mathbf{u}$, so that we expect a purely additive constant ghost drift.
In order to calculate this drift, we will use two properties:
\begin{enumerate}[i)]
\item the noise covariance is easily written in Fourier--Chebyshev space (Eq.~(\ref{spcor}), \S~\ref{stnoise}),
the vector $\vec f$ that we manipulate then consists of all the Fourier and Chebyshev modes of all the components of $\mathbf{u}$.
\item the spatially averaged  kinetic energy can also be written simply
in Fourier--Chebyshev space using Parseval Theorem
\begin{equation}
E_k(t)=\sum_{\substack{n_x=\frac{N_x}{2}-1 \\ n_y=0 \\ n_z=\frac{N_z}{2}-1 }}^{\substack{n_z=\frac{N_z}{2}+1\\ n_y=N_y-1 \\ nx=\frac{N_x}{2}+1 }}
\frac{\left(|\hat{u}_x|^2+|\hat{u}_y|^2+|\hat{u}_z|^2\right)(n_x,n_y,n_z)}{2}\,
\end{equation}
because $\mathbf{u}$ is written in the orthonormal basis of Fourier and Chebyshev modes in our discretisation.
Moreover, using this spectral representation makes the description of the variable involved in the Langevin equation as a vector with components
labelled by integer much simpler, even in the limit of continuous fields $N_x,N_y,N_z\rightarrow \infty$.
We can still use derivative with respect to the modes instead of differentials in the Hessian.
In that basis, the Hessian of the change of variables is then the identity matrix (the $\frac12$ factor and the factor $2$ coming from the derivative cancel out).
\end{enumerate}
This means that the contraction (Eq.~(\ref{conth})) of the Hessian of the change of variables
with the covariance matrix of the noise in spectral space is just its trace, leading to the energy injection rate of equation~(\ref{defepsilon}).
Note that there is an additional simplification between the factor $\frac{2}{\beta}$ in the noise covariance and the factor $\frac{1}{2}$ in front of the ghost drift.

The same procedure is applied to obtain equation~(\ref{total_bud}).
This leads to the same additive constant as a ghost drift because the difference in
the changes of variables (from $\mathbf{u}$ to $E_k$ (Eq.~(\ref{avek})) and from $\mathbf{u}$ to $E_{\rm tot}$,
(Eq.~(\ref{total_bud}))) only lies in constant or linear terms in components of $\mathbf{u}$:
these differences do not leave a trace in the Hessian of the change of variables.
Meanwhile, after an ensemble average that cancels out the noise, the rest of the differential equation
is the same as in the deterministic case.

\section{Parameter used in definition of reaction coordinate in each set of simulations}\label{aparam}

In section~\ref{dreac}, we presented the set up of the reaction coordinates (Eq.~(\ref{defcoord})),
using three free parameters $E_{\mathcal{A}}$ to define the laminar state $\mathcal{A}$, $\sigma$
to define the hypersurface $\mathcal{C}$ that surrounds it
and $E_{\mathcal{B}}$ to define the turbulent state (Fig.~\ref{rolland_figrare} (a)).
Most AMS computations used a systematic definition of these three parameters with $E_{\mathcal{B}}=0.06$,
and the parameters controlling $E_{\mathcal{A}}$ and $\sigma$ reported in table~\ref{sysparam}.

\begin{table*}
\begin{flushleft}
\begin{tabular}{|c|c|c|c|c|c|c|c|c|c|c|c|}
\hline  $L_x\times L_z$& $N_x\times N_y\times N_z$&$\Delta t$&$\beta_{\rm min}$&$\beta_{\rm max}$&$L_c$&$N_r$&$d_E$&$o_E$&$d_S$&$o_S$&type \\ \hline
$6\times 4$&$32\times 27\times 24$&$0.05$&$1.5\times 10^5$&$6\times 10^5$&1.0&4&-1.05&7.1&-1.11&6.4& $\text{div}(\mathbf{f})=0$ \\ \hline
$6\times 4$&$32\times 27\times 24$&0.05&$1.0\times 10^5$&$1.1\times 10^5$&1.0&4&-1.0&4.52&-1.0&1.44&x \\ \hline
$6\times 4$&$32\times 27\times 24$&0.05&$1.0\times 10^5$&$4.0\times 10^5$&1.0&4&-1.02&6.33&-1.03&4.46&iso \\ \hline
$6\times 4$&$32\times 27\times 24$&0.05&$1.0\times 10^5$&$5.0\times 10^5$&1.0&6&-1.02&6.64&-0.98&3.58&iso \\ \hline
$6\times 4$&$32\times 27\times 24$&0.05&$1.0\times 10^5$&$5.5\times 10^5$&1.0&8&-1.05&7.48&-1.12&6.30&iso \\ \hline
$6\times 4$&$64\times 27\times 48$&0.05&$1.0\times 10^5$&$4.0\times 10^5$&1.0&4&-1.01&6.47&-1.02&4.30&iso \\ \hline
$6\times 4$&$128\times 35\times 96$&0.025&$1.0\times 10^5$&$3.0\times 10^5$&1.0&4&-1.02&6.58&-1.04&4.63&iso \\ \hline
\end{tabular}
\end{flushleft}
\caption{Table of parameters $d_E$, $o_E$, $d_S$, $o_S$ used  in systematic AMS simulations to define the reaction coordinate
for the study of build up at Reynolds number $R$ in a system of size $L_x\times L_z$ using numerical resolution $N_x\times N_y\times N_z$ in space and
time step $\Delta t$ in space, and a forcing spectrum controlled by $\beta$, $L_c$, $N_r$.}
\label{sysparam}
\end{table*}

Some AMS computations did not use a systematic definition for $E_{\mathcal{A}}$, $\sigma$ and $E_{\mathcal{B}}$,
because the value of $\beta$ was too low
for the system to be in the regime where $E$ is linear in $\epsilon$.
For those cases, the control parameters used are reported in table~\ref{tab_param}.

\begin{table*}
\begin{center}
\begin{tabular}{|c|c|c|c|c|c|c|c|c|c|}
\hline $R$& $L_x\times L_z$& $N_x\times N_y\times N_z$&$\Delta t$&$\beta$&$L_c$&$N_r$&$E_{\mathcal{A}}$&$E_{\mathcal{B}}$&$\sigma$ \\ \hline
500 &$6\times 4$&$64\times 47\times 48$&$0.0125$&$5.5\times 10^5$&1.0&4&0.0096&0.06&0.0021 \\ \hline
500&$18\times 12$&$96\times 27\times 72$&$0.05$&62500&1.0&12&0.0117&0.06&0.0034 \\ \hline
500&$18\times 12$&$96\times 27\times 72$&$0.05$&80000&1.0&12&0.0081&0.06&0.0023 \\ \hline																						
500&$36\times 24$&$192\times 27\times 144$&$0.05$&30000&1.0&24&0.0123&0.0477&0.0028 \\ \hline
\end{tabular}
\end{center}
\caption{Table of parameters $E_{\mathcal{A}}$, $E_{\mathcal{B}}$ and $\sigma$ used
in non systematic AMS simulations to define the reaction coordinate
for the study of build up at Reynolds number $R$ in a system of size $L_x\times L_z$ using numerical resolution $N_x\times N_y\times N_z$ in space and
time step $\Delta t$ in space, and a spectrum of the divergence free forcing controlled by $\beta$, $L_c$ and $N_r$.}
\label{tab_param}
\end{table*}

\section{Convergence of AMS computations}\label{aconv}

In this appendix, in complement to the precision tests of section~\ref{sea} (Fig.~\ref{beta_dvf_6_4} (c)),
we will check that the number of clones used in AMS computation was sufficient.
For this matter,
we present convergence results with the number of clones for the two types of forcings in domains of size $L_x\times L_z=6\times 4$ at Reynolds number
$R=500$ using standard resolution $N_x=32$, $N_y=27$, $N_z=24$ and $\Delta t=0.05$.
Isotropic forcing was added using $\beta=4\cdot 10^5$ and $\beta=5\cdot 10^5$ with $N_r=6$, and $L_c=1.0$.  
Divergence free forcing was added using $\beta=2.5\cdot 10^5$ and $3\cdot 10^5$ using $L_c=1$ and $N_r=4$.
In all four cases, the crossing probabilities are not that small,
so that the reactive trajectories and their properties can be computed by DNS with a high enough degree of precision.

\begin{figure}[!htbp]
\centerline{\includegraphics[width=6.5cm]{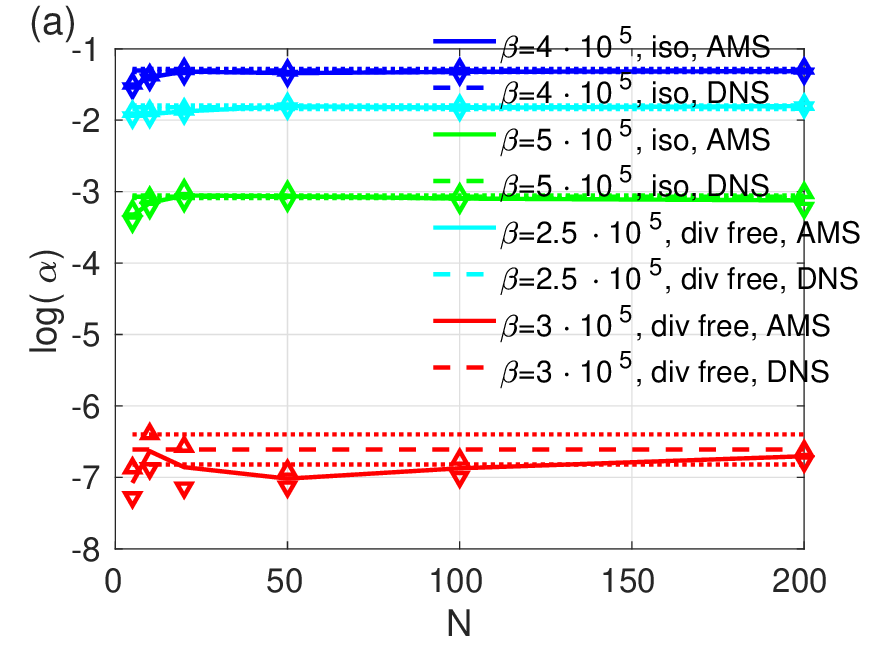}
\includegraphics[width=6.5cm]{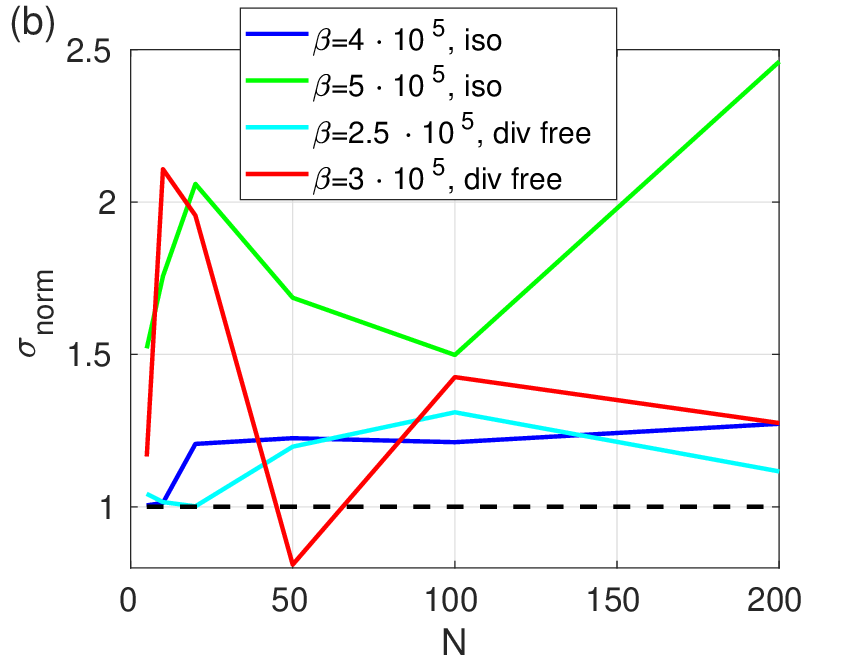}
\includegraphics[width=6.5cm]{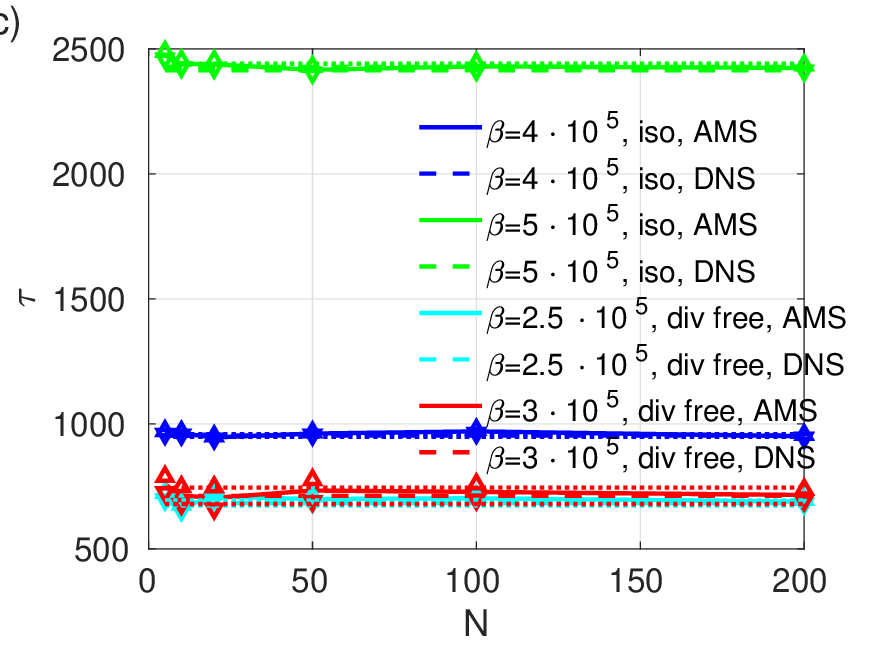}
}
\caption{In a domain of size $L_x\times L_z=6\times 4$ at Reynolds number $R=500$ for four different forcings (see \S~\ref{aconv} for details).
(a) Crossing $\alpha$ probability  as a function of the number of clones,
computed by AMS or DNS.
The error bars correspond to the $66\%$ confidence interval of AMS computations,
the dotted line corresponds to the 66\% interval of confidence of DNS computations.
(b)  normalised standard deviation (Eq.~(\ref{snorm})) of the crossing probability $\alpha$ as a
function of the number of clones, compared to $1$.
(c) Average duration $\tau$ of reactive trajectories as a function of the number of clones,
computed by means of AMS and DNS for the same cases and the same line codes as (a).}
\label{conv_seed}
\end{figure}

We  compare the probability $\alpha$ in logarithmic scale
computed by means of AMS and DNS  in figure~\ref{conv_seed} (a).
For all four examples considered, we observe the overlap of the DNS and AMS $66\%$ intervals of confidence,
obtained using $\pm$ the sample standard deviation,
as soon as $N\ge 20$.
This indicates that no error are performed in these AMS computation if the number of clones is sufficient.
For $N=5$ and $N=10$, we observe underestimates of $\alpha$ which are expected.
Due to the increased rarity of the event for $\beta=3\cdot 10^5$ with divergent free forcing,
the expensive computations means that the dataset is smaller,
which explains the wider confidence intervals for DNS estimates.

We then present the normalised standard deviation of the estimate of $\alpha$ in the case where $N_c=1$ and in the case where $N_c>1$
\begin{equation}
\sigma_{norm}=\frac{\sqrt{\langle \alpha^2\rangle_o-\langle \alpha\rangle_o^2}\sqrt{N}}{\alpha \sqrt{|\log(\alpha)|}}\,,\,
\sigma_{norm,N_c}=\frac{\sqrt{\langle \alpha^2\rangle_o-\langle \alpha\rangle_o^2}\sqrt{N}}{\alpha
\sqrt{\left(\langle \kappa \rangle_o\frac{N_c}{N-N_c}+\frac{1-\langle r\rangle_o}{\langle r\rangle_o} \right)}}\,.
\label{snorm}
\end{equation}
where $\langle \kappa\rangle_o$ is the average number of iterations over AMS runs and $\langle r\rangle_o$ is the average proportion of clones
reaching $\mathcal{B}$ over AMS runs.
It has been demonstrated that this quantity is larger than or equal to one,
and it is one when the committor is used as reaction coordinate \cite{cerou2019asymptotic}.
In parallel, it has also been shown that the closer a reaction coordinate is to the committor,
the most precise the AMS estimates are when a finite number of runs are used.
As a consequence, this quantity is used as an estimate of the quality of the computation because
the better the reaction coordinate, the smaller is $\sigma$. A very large $\sigma$ would be that the estimates of $\alpha$ and $T$ are incorrect,
or even that the paths are incorrectly selected \cite{rolland2015statistical,lucente2022coupling}.
In that case what is typically observed in the selection of shorter though less probable paths.
In the two test cases considered here (Fig.~\ref{conv_seed} (c)), we can see that $\sigma$ is slightly larger than one.
This ensures correct estimates, but also tells us that there is room for improvement of the reaction coordinate.

We finally test the computation of the average duration of reactive trajectories.
The result of the AMS computations of $\tau$ as a function of $N$ in our four cases is presented in figure~\ref{conv_seed} (c).
We add the result of DNS computation and the  66\% interval of confidence to this figure.
We note the overlap of the intervals of confidence in all four cases as soon as $N\ge 10$,
indicating that no discernable bias can be observed in the estimates of $\tau$ by means of AMS.
If those estimates are averaged over enough repetitions, they are precise enough in those cases.
In conclusion to this appendix, we note that we do not discern any irreducible error in the computations
performed by means of AMS, even if the number of clones used is relatively modest (at least 20).
In those case, a larger number of repetitions of the computations is required to compute said estimates.
This confirm the observation performed in section~\ref{sea} for several values of $\beta$
(and the corresponding $\alpha$) with only one clone number per case.

The overlap of intervals of confidences of estimates by means of AMS as well as the closeness of the
rescaled variance to one indicates that AMS computations in domains of size $L_x\times L_z=6 \times 4$ are
performed without errors. This gives us further confidence that the reaction coordinate $\phi$ defined in
section~\ref{dreac} correctly weighs the advance of reactive trajectories by being affine in kinetic energy.
To our knowledge, nothing makes this reaction coordinate irrelevant when the size of the system is increased to $L_x\times L_z=36\times 24$:
the flow transits toward turbulence by near monotonously increasing its kinetic energy, even when the flow displays spanwise localisation during the transition.
Some problem would occur if the flow had to systematically decrease its kinetic energy during transition,
or give some importance to a process that does not leave a trace in total kinetic energy.
We expect these assertions to remain valid if AMS is used to compute reactive trajectories in larger domains,
where the flow displays both spanwise and streamwise localisation during the transition.
We note however that some care should be given in the definition of the thresholds defining sets $\mathcal{A}$
and $\mathcal{C}$ (Fig.~\ref{rolland_figrare} (a)).
Indeed, with spatial localisation of velocity tubes in the flow in the early stages of the trajectory,
the total kinetic energy is reduced as the area in which we find the non trivial flow is reduced.
If said thresholds are too large, the early stages of AMS computation risk being too long
as the algorithm is not used to generate the localised velocity tubes.
Note finally that another effect can impair the quality of AMS computations:
a slow separation of trajectories in phase space.
While this can happen in principle for any type of physical system,
this effect has mostly be seen in system not forced by any noise,
and modifications of AMS have been performed to ensure that AMS computation can generate enough variability
in the obtained trajectories or even function at all \cite{rolland2022collapse}.
No such problem has been observed in the computation of build up trajectories so far.
For such a problem to manifest itself in domains larger than $L_x\times L_z=36\times 24$, where
both spanwise and streamwise localisation of turbulence happens,
the physics of transition would have to change drastically.

\FloatBarrier

\bibliographystyle{apalike}
\bibliography{doc_build_up}

\begin{thebibliography}{}

\bibitem[Arnold et~al., 1995]{arnold1995random}
Arnold, L., Jones, C.~K., Mischaikow, K., Raugel, G., and Arnold, L. (1995).
\newblock {\em Random dynamical systems}.
\newblock Springer.

\bibitem[Baars et~al., 2021]{baars2021application}
Baars, S., Castellana, D., Wubs, F.~W., and Dijkstra, H.~A. (2021).
\newblock Application of adaptive multilevel splitting to high-dimensional
  dynamical systems.
\newblock {\em Journal of Computational Physics}, 424:109876.

\bibitem[Berhanu et~al., 2007]{Berhanu2007}
Berhanu, M., Monchaux, R., Fauve, S., Mordant, N., P\'{e}tr\'{e}lis, F.,
  Chiffaudel, A., Daviaud, F., Dubrulle, B., Mari\'{e}, L., Ravelet, F.,
  Bourgoin, M., Odier, P., Pinton, J.-F., and Volk, R. (2007).
\newblock {Magnetic field reversals in an experimental turbulent dynamo}.
\newblock {\em Eur. Phys. Lett.}, 77(5):59001.

\bibitem[B{\"o}rner et~al., 2023]{borner2023saddle}
B{\"o}rner, R., Deeley, R., R{\"o}mer, R., Grafke, T., Lucarini, V., and
  Feudel, U. (2023).
\newblock Saddle avoidance of noise-induced transitions in multiscale systems.
\newblock {\em arXiv preprint arXiv:2311.10231}.

\bibitem[Bouchet and Reygner, 2016]{bouchet2016generalisation}
Bouchet, F. and Reygner, J. (2016).
\newblock Generalisation of the eyring--kramers transition rate formula to
  irreversible diffusion processes.
\newblock In {\em Annales Henri Poincar{\'e}}, volume~17, pages 3499--3532.
  Springer.

\bibitem[Bouchet et~al., 2019]{prl_jet}
Bouchet, F., Rolland, J., and Simonnet, E. (2019).
\newblock Rare event algorithm links transitions in turbulent flows with
  activated nucleations.
\newblock {\em Physical review letters}, 122(7):074502.

\bibitem[Bouchet and Touchette, 2012]{bouchet2012non}
Bouchet, F. and Touchette, H. (2012).
\newblock Non-classical large deviations for a noisy system with non-isolated
  attractors.
\newblock {\em Journal of Statistical Mechanics: Theory and Experiment},
  2012(05):P05028.

\bibitem[Br{\'e}hier et~al., 2016]{brehier2016unbiasedness}
Br{\'e}hier, C.-E., Gazeau, M., Gouden{\`e}ge, L., Leli{\`e}vre, T., and
  Rousset, M. (2016).
\newblock Unbiasedness of some generalized adaptive multilevel splitting
  algorithms.
\newblock {\em The Annals of Applied Probability}, 26(6):3559--3601.

\bibitem[Brze{\'z}niak et~al., 2015]{brzezniak2015quasipotential}
Brze{\'z}niak, Z., Cerrai, S., and Freidlin, M. (2015).
\newblock Quasipotential and exit time for 2d stochastic navier-stokes
  equations driven by space time white noise.
\newblock {\em Probability Theory and Related Fields}, 162(3-4):739--793.

\bibitem[C{\'e}rou et~al., 2019]{cerou2019asymptotic}
C{\'e}rou, F., Delyon, B., Guyader, A., and Rousset, M. (2019).
\newblock On the asymptotic normality of adaptive multilevel splitting.
\newblock {\em SIAM/ASA Journal on Uncertainty Quantification}, 7(1):1--30.

\bibitem[C{\'e}rou and Guyader, 2007]{rolland_CG07}
C{\'e}rou, F. and Guyader, A. (2007).
\newblock Adaptive multilevel splitting for rare event analysis.
\newblock {\em Stochastic Analysis and Applications}, 25(2):417--443.

\bibitem[C{\'e}rou et~al., 2013]{cerou2013length}
C{\'e}rou, F., Guyader, A., Leli\`evre, T., and Malrieu, F. (2013).
\newblock On the length of one-dimensional reactive paths.
\newblock {\em ALEA}, 10(1):359--389.

\bibitem[C{\'e}rou et~al., 2011]{cerou2011multiple}
C{\'e}rou, F., Guyader, A., Leli\`evre, T., and Pommier, D. (2011).
\newblock A multiple replica approach to simulate reactive trajectories.
\newblock {\em The Journal of chemical physics}, 134(5):054108.

\bibitem[Chantry and Schneider, 2014]{chantry2014studying}
Chantry, M. and Schneider, T.~M. (2014).
\newblock Studying edge geometry in transiently turbulent shear flows.
\newblock {\em Journal of fluid mechanics}, 747:506--517.

\bibitem[Chekroun et~al., 2011]{chekroun2011stochastic}
Chekroun, M.~D., Simonnet, E., and Ghil, M. (2011).
\newblock Stochastic climate dynamics: Random attractors and time-dependent
  invariant measures.
\newblock {\em Physica D: Nonlinear Phenomena}, 240(21):1685--1700.

\bibitem[Cherubini et~al., 2011]{cherubini2011minimal}
Cherubini, S., De~Palma, P., Robinet, J.-C., and Bottaro, A. (2011).
\newblock The minimal seed of turbulent transition in the boundary layer.
\newblock {\em Journal of Fluid Mechanics}, 689:221.

\bibitem[Drazin and Reid, 2004]{drazin2004hydrodynamic}
Drazin, P.~G. and Reid, W.~H. (2004).
\newblock {\em Hydrodynamic stability}.
\newblock Cambridge university press.

\bibitem[Ducimeti{\`e}re et~al., 2022]{ducimetiere2022weakly}
Ducimeti{\`e}re, Y.-M., Boujo, E., and Gallaire, F. (2022).
\newblock Weakly nonlinear evolution of stochastically driven non-normal
  systems.
\newblock {\em Journal of Fluid Mechanics}, 951:R3.

\bibitem[Farrell and Ioannou, 1993a]{farrell1993optimal}
Farrell, B.~F. and Ioannou, P.~J. (1993a).
\newblock Optimal excitation of three-dimensional perturbations in viscous
  constant shear flow.
\newblock {\em Physics of Fluids A: Fluid Dynamics}, 5(6):1390--1400.

\bibitem[Farrell and Ioannou, 1993b]{farrell1993stochastic}
Farrell, B.~F. and Ioannou, P.~J. (1993b).
\newblock Stochastic forcing of the linearized navier--stokes equations.
\newblock {\em Physics of Fluids A: Fluid Dynamics}, 5(11):2600--2609.

\bibitem[Freidlin and Wentzell, 1998]{freidlin1998random}
Freidlin, M.~I. and Wentzell, A.~D. (1998).
\newblock Random perturbations.
\newblock In {\em Random perturbations of dynamical systems}, pages 15--43.
  Springer.

\bibitem[Gardiner, 2009]{gardiner2009stochastic}
Gardiner, C. (2009).
\newblock {\em Stochastic methods}, volume~4.
\newblock Springer Berlin.

\bibitem[Gibson et~al., 2008]{rolland_Gib08}
Gibson, J.~F., Halcrow, J., and Cvitanovi\'c, P. (2008).
\newblock Visualizing the geometry of state space in plane couette flow.
\newblock {\em J. Fluid Mech.}, 611:107--130.

\bibitem[Gom{\'e} et~al., 2022]{gome2022extreme}
Gom{\'e}, S., Tuckerman, L.~S., and Barkley, D. (2022).
\newblock Extreme events in transitional turbulence.
\newblock {\em Philosophical Transactions of the Royal Society A},
  380(2226):20210036.

\bibitem[Grafke and Vanden-Eijnden, 2017]{grafke2017non}
Grafke, T. and Vanden-Eijnden, E. (2017).
\newblock Non-equilibrium transitions in multiscale systems with a bifurcating
  slow manifold.
\newblock {\em Journal of Statistical Mechanics: Theory and Experiment},
  2017(9):093208.

\bibitem[Grafke and Vanden-Eijnden, 2019]{grafke2019numerical}
Grafke, T. and Vanden-Eijnden, E. (2019).
\newblock Numerical computation of rare events via large deviation theory.
\newblock {\em Chaos: An Interdisciplinary Journal of Nonlinear Science},
  29(6):063118.

\bibitem[Grandemange et~al., 2013]{grandemange2013turbulent}
Grandemange, M., Gohlke, M., and Cadot, O. (2013).
\newblock Turbulent wake past a three-dimensional blunt body. part 1. global
  modes and bi-stability.
\newblock {\em Journal of Fluid Mechanics}, 722:51--84.

\bibitem[Hamilton et~al., 1995]{hamilton1995regeneration}
Hamilton, J.~M., Kim, J., and Waleffe, F. (1995).
\newblock Regeneration mechanisms of near-wall turbulence structures.
\newblock {\em Journal of Fluid Mechanics}, 287(1):317--348.

\bibitem[H{\"a}nggi et~al., 1990]{hanggi1990reaction}
H{\"a}nggi, P., Talkner, P., and Borkovec, M. (1990).
\newblock Reaction-rate theory: fifty years after kramers.
\newblock {\em Reviews of modern physics}, 62(2):251.

\bibitem[Herbert et~al., 2020]{herbert2020atmospheric}
Herbert, C., Caballero, R., and Bouchet, F. (2020).
\newblock Atmospheric bistability and abrupt transitions to superrotation:
  wave--jet resonance and hadley cell feedbacks.
\newblock {\em Journal of the Atmospheric Sciences}, 77(1):31--49.

\bibitem[Jim{\'e}nez and Moin, 1991]{jimenez1991minimal}
Jim{\'e}nez, J. and Moin, P. (1991).
\newblock The minimal flow unit in near-wall turbulence.
\newblock {\em Journal of Fluid Mechanics}, 225:213--240.

\bibitem[Kawahara and Kida, 2001]{kawahara2001periodic}
Kawahara, G. and Kida, S. (2001).
\newblock Periodic motion embedded in plane couette turbulence: regeneration
  cycle and burst.
\newblock {\em Journal of Fluid Mechanics}, 449:291--300.

\bibitem[Kim and Durbin, 1988]{kim1988investigation}
Kim, H.-J. and Durbin, P.~A. (1988).
\newblock Investigation of the flow between a pair of circular cylinders in the
  flopping regime.
\newblock {\em Journal of Fluid Mechanics}, 196:431--448.

\bibitem[Kreilos and Eckhardt, 2012]{kreilos2012periodic}
Kreilos, T. and Eckhardt, B. (2012).
\newblock Periodic orbits near onset of chaos in plane couette flow.
\newblock {\em Chaos: An Interdisciplinary Journal of Nonlinear Science},
  22(4).

\bibitem[Lecoanet and Kerswell, 2018]{lecoanet2018connection}
Lecoanet, D. and Kerswell, R.~R. (2018).
\newblock Connection between nonlinear energy optimization and instantons.
\newblock {\em Physical Review E}, 97(1):012212.

\bibitem[Liu et~al., 2021]{liu2021decay}
Liu, T., Semin, B., Klotz, L., Godoy-Diana, R., Wesfreid, J.~E., and Mullin, T.
  (2021).
\newblock Decay of streaks and rolls in plane couette--poiseuille flow.
\newblock {\em Journal of Fluid Mechanics}, 915.

\bibitem[Lopes and Leli{\`e}vre, 2019]{lopes2019analysis}
Lopes, L.~J. and Leli{\`e}vre, T. (2019).
\newblock Analysis of the adaptive multilevel splitting method on the
  isomerization of alanine dipeptide.
\newblock {\em Journal of computational chemistry}, 40(11):1198--1208.

\bibitem[Lucarini and B{\'o}dai, 2017]{lucarini2017edge}
Lucarini, V. and B{\'o}dai, T. (2017).
\newblock Edge states in the climate system: exploring global instabilities and
  critical transitions.
\newblock {\em Nonlinearity}, 30(7):R32.

\bibitem[Lucarini and B{\'o}dai, 2019]{lucarini2019transitions}
Lucarini, V. and B{\'o}dai, T. (2019).
\newblock Transitions across melancholia states in a climate model: Reconciling
  the deterministic and stochastic points of view.
\newblock {\em Physical review letters}, 122(15):158701.

\bibitem[Lucente et~al., 2022]{lucente2022coupling}
Lucente, D., Rolland, J., Herbert, C., and Bouchet, F. (2022).
\newblock Coupling rare event algorithms with data-based learned committor
  functions using the analogue markov chain.
\newblock {\em Journal of Statistical Mechanics: Theory and Experiment},
  2022(8):083201.

\bibitem[Metzner et~al., 2006]{metzner2006illustration}
Metzner, P., Sch{\"u}tte, C., and Vanden-Eijnden, E. (2006).
\newblock Illustration of transition path theory on a collection of simple
  examples.
\newblock {\em The Journal of chemical physics}, 125(8):084110.

\bibitem[Monokrousos et~al., 2011]{monokrousos2011nonequilibrium}
Monokrousos, A., Bottaro, A., Brandt, L., Di~Vita, A., and Henningson, D.~S.
  (2011).
\newblock Nonequilibrium thermodynamics and the optimal path to turbulence in
  shear flows.
\newblock {\em Physical review letters}, 106(13):134502.

\bibitem[Nusse and Yorke, 1989]{nusse1989procedure}
Nusse, H.~E. and Yorke, J.~A. (1989).
\newblock A procedure for finding numerical trajectories on chaotic saddles.
\newblock {\em Physica D: Nonlinear Phenomena}, 36(1-2):137--156.

\bibitem[Onsager, 1938]{onsager1938initial}
Onsager, L. (1938).
\newblock Initial recombination of ions.
\newblock {\em Physical Review}, 54(8):554.

\bibitem[Onsager and Machlup, 1953]{onsager1953fluctuations}
Onsager, L. and Machlup, S. (1953).
\newblock Fluctuations and irreversible processes.
\newblock {\em Physical Review}, 91(6):1505.

\bibitem[Philip and Manneville, 2011]{philip2011temporal}
Philip, J. and Manneville, P. (2011).
\newblock From temporal to spatiotemporal dynamics in transitional plane
  couette flow.
\newblock {\em Physical Review E}, 83(3):036308.

\bibitem[Podvin and Sergent, 2017]{podvin2017precursor}
Podvin, B. and Sergent, A. (2017).
\newblock Precursor for wind reversal in a square rayleigh-b{\'e}nard cell.
\newblock {\em Physical Review E}, 95(1):013112.

\bibitem[Rabin et~al., 2012]{rabin2012triggering}
Rabin, S. M.~E., Caulfield, C.-C.~P., and Kerswell, R.~R. (2012).
\newblock Triggering turbulence efficiently in plane couette flow.
\newblock {\em Journal of Fluid Mechanics}, 712:244.

\bibitem[Rigas et~al., 2021]{rigas2021nonlinear}
Rigas, G., Sipp, D., and Colonius, T. (2021).
\newblock Nonlinear input/output analysis: application to boundary layer
  transition.
\newblock {\em Journal of Fluid Mechanics}, 911.

\bibitem[Rolland, 2015]{rolland2015mechanical}
Rolland, J. (2015).
\newblock Mechanical and statistical study of the laminar hole formation in
  transitional plane couette flow.
\newblock {\em The European Physical Journal B}, 88(3):66.

\bibitem[Rolland, 2018]{rolland_pre18}
Rolland, J. (2018).
\newblock Extremely rare collapse and build-up of turbulence in stochastic
  models of transitional wall flows.
\newblock {\em Physical Review E}, 97(2):023109.

\bibitem[Rolland, 2022]{rolland2022collapse}
Rolland, J. (2022).
\newblock Collapse of transitional wall turbulence captured using a rare events
  algorithm.
\newblock {\em Journal of Fluid Mechanics}, 931.

\bibitem[Rolland et~al., 2016]{rolland2016computing}
Rolland, J., Bouchet, F., and Simonnet, E. (2016).
\newblock Computing transition rates for the 1-d stochastic
  ginzburg--landau--allen--cahn equation for finite-amplitude noise with a rare
  event algorithm.
\newblock {\em Journal of Statistical Physics}, 162(2):277--311.

\bibitem[Rolland and Simonnet, 2015]{rolland2015statistical}
Rolland, J. and Simonnet, E. (2015).
\newblock Statistical behaviour of adaptive multilevel splitting algorithms in
  simple models.
\newblock {\em Journal of Computational Physics}, 283:541--558.

\bibitem[Romanov, 1973]{romanov1973stability}
Romanov, V.~A. (1973).
\newblock Stability of plane-parallel couette flow.
\newblock {\em Functional analysis and its applications}, 7(2):137--146.

\bibitem[Schmid and Henningson, 1994]{schmid1994optimal}
Schmid, P.~J. and Henningson, D.~S. (1994).
\newblock Optimal energy density growth in hagen--poiseuille flow.
\newblock {\em Journal of Fluid Mechanics}, 277:197--225.

\bibitem[Schmiegel and Eckhardt, 1997]{schmiegel1997fractal}
Schmiegel, A. and Eckhardt, B. (1997).
\newblock Fractal stability border in plane couette flow.
\newblock {\em Physical review letters}, 79(26):5250.

\bibitem[Schneider et~al., 2007]{schneider2007turbulence}
Schneider, T.~M., Eckhardt, B., and Yorke, J.~A. (2007).
\newblock Turbulence transition and the edge of chaos in pipe flow.
\newblock {\em Physical review letters}, 99(3):034502.

\bibitem[Schneider et~al., 2008]{schneider2008laminar}
Schneider, T.~M., Gibson, J.~F., Lagha, M., De~Lillo, F., and Eckhardt, B.
  (2008).
\newblock Laminar-turbulent boundary in plane couette flow.
\newblock {\em Physical Review E}, 78(3):037301.

\bibitem[Schneider et~al., 2010]{schneider2010localized}
Schneider, T.~M., Marinc, D., and Eckhardt, B. (2010).
\newblock Localized edge states nucleate turbulence in extended plane couette
  cells.
\newblock {\em Journal of Fluid Mechanics}, 646:441--451.

\bibitem[Simonnet, 2016]{simonnet2016combinatorial}
Simonnet, E. (2016).
\newblock Combinatorial analysis of the adaptive last particle method.
\newblock {\em Statistics and Computing}, 26(1):211--230.

\bibitem[Simonnet et~al., 2021]{simonnet2020multistability}
Simonnet, E., Rolland, J., and Bouchet, F. (2021).
\newblock Multistability and rare spontaneous transitions between climate and
  jet configurations in a barotropic model of the jovian mid-latitude
  troposphere.
\newblock {\em Journal of the Atmospheric Sciences}, 78:1889--1911.

\bibitem[Spangler and Wells~Jr, 1968]{spangler1968effects}
Spangler, J.~G. and Wells~Jr, C.~S. (1968).
\newblock Effects of freestream disturbances on boundary-layer transition.
\newblock {\em AIAA Journal}, 6(3):543--545.

\bibitem[Stuart, 1958]{stuart1958non}
Stuart, J.~T. (1958).
\newblock On the non-linear mechanics of hydrodynamic stability.
\newblock {\em Journal of Fluid Mechanics}, 4(1):1--21.

\bibitem[Toh and Itano, 2003]{toh2003periodic}
Toh, S. and Itano, T. (2003).
\newblock A periodic-like solution in channel flow.
\newblock {\em Journal of Fluid Mechanics}, 481:67--76.

\bibitem[Touchette, 2009]{touchette2009large}
Touchette, H. (2009).
\newblock The large deviation approach to statistical mechanics.
\newblock {\em Physics Reports}, 478(1-3):1--69.

\bibitem[Trefethen et~al., 1993]{trefethen1993hydrodynamic}
Trefethen, L.~N., Trefethen, A.~E., Reddy, S.~C., and Driscoll, T.~A. (1993).
\newblock Hydrodynamic stability without eigenvalues.
\newblock {\em Science}, 261(5121):578--584.

\bibitem[Van~Erp et~al., 2003]{van2003novel}
Van~Erp, T.~S., Moroni, D., and Bolhuis, P.~G. (2003).
\newblock A novel path sampling method for the calculation of rate constants.
\newblock {\em The Journal of chemical physics}, 118(17):7762--7774.

\bibitem[Waleffe, 1997]{waleffe1997self}
Waleffe, F. (1997).
\newblock On a self-sustaining process in shear flows.
\newblock {\em Physics of Fluids}, 9(4):883--900.

\bibitem[Wan, 2013]{wan2013minimum}
Wan, X. (2013).
\newblock A minimum action method for small random perturbations of
  two-dimensional parallel shear flows.
\newblock {\em Journal of Computational Physics}, 235:497--514.

\bibitem[Wan and Yu, 2017]{mam_pois}
Wan, X. and Yu, H. (2017).
\newblock A dynamic-solver-consistent minimum action method: With an
  application to 2d navier-stokes equations.
\newblock {\em Journal of Computational Physics}, 331:209--226.

\bibitem[Wan et~al., 2015]{wan2015model}
Wan, X., Yu, H., and Weinan, E. (2015).
\newblock Model the nonlinear instability of wall-bounded shear flows as a rare
  event: a study on two-dimensional poiseuille flow.
\newblock {\em Nonlinearity}, 28(5):1409.

\bibitem[Willis and Kerswell, 2009]{willis2009turbulent}
Willis, A.~P. and Kerswell, R.~R. (2009).
\newblock Turbulent dynamics of pipe flow captured in a reduced model: puff
  relaminarization and localized 'edge' states.
\newblock {\em Journal of Fluid Mechanics}, 619:213--233.

\end{thebibliography}

\end{document}